\documentclass[twocolumn]{aastex631}
\usepackage{amsmath,amstext}
\usepackage{bm}

\received{April 29, 2022}
\accepted{October, 28, 2022}

\shorttitle{3D MHD simulations of magnetospheric accretion}
\shortauthors{Takasao et al.}
\graphicspath{{./}{figures/}}
\begin{document}

\title{Three-dimensional Simulations of Magnetospheric Accretion in a T Tauri Star: \\
Accretion and Wind Structures Just Around Star}

\correspondingauthor{Shinsuke Takasao}
\email{takasao@astro-osaka.jp}

\author[0000-0003-3882-3945]{Shinsuke Takasao}
\affiliation{Department of Earth and Space Science, Graduate School of Science, Osaka University, Toyonaka, Osaka 560-0043, Japan}

\author[0000-0001-8105-8113]{Kengo Tomida}
\affiliation{Astronomical Institute, Tohoku University, Sendai, Miyagi 980-8578, Japan}

\author[0000-0002-2707-7548]{Kazunari Iwasaki}
\affiliation{Center for Computational Astrophysics, National Astronomical Observatory of Japan, Mitaka, Tokyo 181-8588, Japan}

\author[0000-0001-9734-9601]{Takeru K. Suzuki}
\affiliation{School of Arts \& Sciences, The University of Tokyo, 3-8-1, Komaba, Meguro, Tokyo , 153-8902, Japan; Department of Astronomy, The University of Tokyo, 7-3-1, Hongo, Bunkyo, Tokyo, 113-0033, Japan}



%% 250 words limit
\begin{abstract}
We perform three-dimensional magnetohydrodynamic simulations of magnetospheric accretion in a T Tauri star to study the accretion and wind structures in the close vicinity of the star. The gas accreting onto the star consists of the gas from the magnetospheric boundary and the failed disk winds. The accreting gas is commonly found as a multi-column accretion, which is consistent with observations. A significant fraction of the angular momentum of the accreting flows is removed by the magnetic fields of conical disk winds and turbulent failed winds inside and near the magnetosphere. As a result, the accretion torque is significantly reduced compared to the simple estimation based on the mass accretion rate. The stellar spin affects the time variability of the conical disk wind by changing the stability condition of the magnetospheric boundary. However, the time-averaged magnetospheric radius only weakly depends on the stellar spin, which is unlike the prediction of classical theories that the stellar spin controls the magnetospheric radius through the magnetic torque. The ratio of the toroidal to the poloidal field strengths at the magnetospheric boundary, which is a key parameter for the magnetic torque, is also insensitive to the spin; it is rather determined by the disk dynamics. Considering newly found three-dimensional effects, we obtain a scaling relation of the magnetospheric radius very similar to the Ghosh \& Lamb relation from the steady angular momentum transport equation.
\end{abstract}



\section{Introduction} \label{sec:intro}
Dynamics in the close vicinity of young stellar objects have significant impacts on the stellar and disk evolution. Observations have suggested that magnetospheric accretion generally occurs in low mass pre-main-sequence (pre-MS) stars or T Tauri stars; inner disks are truncated by stellar magnetospheres, and accretion is funneled by stellar magnetic fields \citep[see reviews by][]{Romanova2015SSRv,Hartmann2016ARA&A}. 
Evolution of stellar spin depends on star-disk interaction via the magnetosphere. The accreting flows bring angular momentum to the star, while the angular momentum can be extracted from the star by magnetic fields connected to the disk beyond the corotation radius \citep[e.g.][]{Ghosh_Lamb1979_paperIII,Wang_YM1987} and by outflows/stellar winds \citep[e.g.][]{Matt&Pudritz2005ApJ,Zanni2013A&A}. The funnel accretion flows move at nearly free-fall velocities of a few 100 km~s$^{-1}$ and form accretion shocks when they hit the stellar surface \citep{Koenigl1991ApJ,Calvet1998ApJ}. FUV, EUV and X-rays emitted from the accretion shocks are considered to play important roles in dissipating disk gas via photoevaporation \citep[e.g.][]{Ercolano2017RSOS}. 
The inner disk structure can affect the orbits of short-period planets and therefore their formation process \citep[e.g.][]{Lee_Chiang2017ApJ,Liu_Ormel_Lin2017A&A}.
The innermost disk structure smaller than 0.1 au has been investigated using near-infrared interferometric observations \citep[e.g.][]{Gravity_Collaboration2017A&A}.
Currently, {\it Hubble Space Telescope} is being devoted to a large survey toward low-mass pre-MS stars at UV wavelengths, accelerating comprehensive observational studies about accretion and ejections \citep{Manara2021A&A,Espaillat2022_ODYSSEUS}.

Mass ejections such as jets, outflows, and winds are commonly found in classical T Tauri stars \citep[e.g.][]{Ray2007prpl,Frank2014prpl,Pascucci2022arXiv}. As mass ejections extract angular momentum from the launching regions, it is important to reveal how and where they are driven for understanding of the stellar spin evolution and the accretion mechanism. Different types of ejections have been proposed from theoretical studies. Strong stellar winds may be powered by accretion \citep{Matt&Pudritz2005ApJ,Cranmer2008ApJ,Matt2012ApJ}. Episodic magnetospheric ejections may occur in response to disk-magnetosphere interaction \citep{Lynden-Bell1994MNRAS,Hayashi_etal_1996,Zanni2013A&A}. Conical disk winds can emanate from the disk-magnetosphere boundary either by magnetocentrifugal force \citep{Shu1994ApJ,Hirose1997PASJ,Ferreira2013MNRAS} or by magnetic pressure gradient force \citep{Romanova2009MNRAS,Lii2012MNRAS}. 
These magnetohydrodynamic processes can also work in the outer region \citep[e.g.][]{Blandford1982MNRAS,Shibata_Uchida1985PASJ}. When magnetic fields sufficiently couple with the disk plasma, magnetorotational instability \citep[MRI; ][]{Balbus_Hawley1991} can make the disk turbulent. Recent three-dimensional (3D) magnetohydrodynamic (MHD) simulations found that highly fluctuating winds are launched by MRI-driven turbulence \citep{Suzuki_Inutsuka2009ApJ,Bai_Stone2013ApJ,Suzuki_etal_2014ApJ}. \citet[][hereafter, ST18]{Takasao_Tomida_Iwasaki_Suzuki_2018} pointed out that a part of the MRI-driven winds fails to escape from the stellar gravity and falls onto the star in the form of funnel accretion flows, suggesting a new relation between disk winds and accretion flows.

Stellar spin has been considered as a key parameter to characterize accretion and ejections around the star.
If the star enters the propeller regime in which the magnetospheric radius is larger than the corotation radius, strong winds may blow from the disk-magnetosphere boundary \citep{Lovelace1999ApJ,Miller_Stone1997ApJ,Romanova2009MNRAS,Lii2014MNRAS}. As a result, the accretion onto the star is considerably suppressed. However, the efficiency of the propeller (the ratio of the ejection rate to the accretion rate) remains unclear. \citet{Ustyugova2006ApJ} pointed out that the efficiency in 2D models strongly depends on effective viscosity and magnetic diffusivity. Since the effective viscosity and magnetic diffusivity arise in response to turbulence in three-dimension, 3D analysis is required. Stellar spin also affects the stability of the accretion structure. The disk-magnetosphere boundary can be unstable to the magnetic interchange instability \citep[e.g.][]{Spruit1995MNRAS}, depending on the stellar spin and other parameters \citep[see also][]{Blinova2016MNRAS}. Magnetospheric accretion can produce multiple accretion columns when the interchange instability occurs \citep{Kulkarni2008MNRAS,Romanova2012MNRAS}, which may be relevant to spectroscopic observations suggesting that multiple accretion columns with different energy fluxes should be hitting the stellar surface \citep[e.g.][]{Ingleby2013ApJ,Johnstone2014MNRAS,Robinson2019ApJ}.

The spin evolution of pre-MS stars remains a puzzle. Observations have suggested that the angular momentum should be efficiently removed from stars during the pre-MS phase \citep[e.g.][]{Edwards1993AJ,Bouvier1993A&A}. The classical model by \citet{Ghosh_Lamb1979_paperII} proposes that the star can spin-down by a torque of the magnetic field lines connecting the star and the disk beyond the corotation radius. However, as pointed out by many authors \citep[e.g.][]{Ireland2021ApJ}, the assumption that the stellar magnetic field is mostly closed and threading a large portion of the disk is not supported by our modern picture based on both analytic and numerical studies \citep[e.g.][]{Agapitou2000MNRAS,Zanni2013A&A}. Opening the stellar field flux via the star-disk interaction has been considered as a key to efficient angular momentum removal because it can enhance the spin-down torque by the stellar winds \citep{Matt2005bApJ}. 
Recent axisymmetric simulations have been investigating this possibility quantitatively. A current major challenge is that massive stellar winds with a mass loss rate of a few 10\% of the accretion rate would be required to balance the spin-up torque due to accretion \citep{Pantolmos2020A&A,Ireland2021ApJ}. How such strong stellar winds can be realized remains unresolved.

Many pre-MS stars show bursts and dips in their lightcurves \citep[e.g.][]{Stauffer2014AJ,Cody2018AJ}. Recently, such stars have been regarded as good targets for studying the gas and dust distributions at the innermost scale \citep[e.g.][]{Bodman2017MNRAS}. While periodic variability is usually attributed to the stellar rotation, aperiodic variability could be caused by fluctuations in the accreting or ejected flows around the stars. 
Pre-MS stars called ``dippers" show sudden (typically day scale) drops in their optical lightcurves because of transient partial occultation \citep[e.g.][]{Ansdell2016ApJ}. Considering the short timescale, dynamical events in the vicinity of the stellar magnetosphere are responsible for dippers at least  in some stars. To reveal the detailed mechanisms of the variability, we need to understand the magnetospheric dynamics.

As mentioned earlier, stellar radiation such as FUV, EUV and X-rays will drive photoevaporative winds and contribute to disk gas dissipation. However, the amount of radiation that can reach the outer disk depends on the density structure around the star. The screening hydrogen column densities required for the optical depth of unity are approximately $10^{22}$ cm$^{-2}$ for X-rays and $10^{19}$ cm$^{-2}$ for EUV \citep[][]{Ercolano2009ApJ,Owen2010MNRAS}. Accreting and ejected flows can veil the stellar radiation. The stability of magnetospheric accretion will affect the degree of veiling, as fragmented accretion flows will veil a smaller amount of the starlight than smooth accretion flows without gaps.
All the current photoevaporation scenarios face uncertainty about the assumption of the density structure around the star. To establish a robust disk evolutionary scenario, the complex density structure in the central region needs to be studied with 3D MHD simulations.

To reveal the inner sub-au scale dynamics, we have been performing 3D MHD simulations of magnetospheric accretion onto a T Tauri star. This paper reports the initial results about the issues raised above. We study the magnetospheric accretion for three models with different stellar spins. The remainder of this paper is structured as follows. Section~\ref{sec:method} describes our models and assumptions. The results are shown in Section~\ref{sec:result}, where the accretion and wind structures are described in detail. Section~\ref{sec:discussion} discusses the mechanism that determines the magnetospheric radius. We will also compare our results with previous models. In Section~\ref{sec:summary}, we summarize our key findings.

\section{Methods} \label{sec:method}
\subsection{Numerical Approach}
We simulate the accretion onto a rotating magnetized star in the following manner.
Our numerical methods and models are similar to those in ST18 and \citet{Takasao_Tomida_Iwasaki_Suzuki_2019}, although the stars in these previous studies do not have stellar magnetospheres. 

The basic equations are 3D resistive MHD equations in a conservative form in spherical coordinates $(r,\theta,\varphi)$. The equations are solved with Athena++ \citep{Stone_Tomida_White_Felker_2020}.
The dual energy formalism (the time evolution of the internal energy is also solved in parallel) is adopted to avoid negative pressure in very low plasma $\beta$ regions \citep{Takasao_Fan_Cheung_Shibata_2015,Iijima_2016}. When the MHD solver returns unphysically small or negative internal energy density, it is replaced with the internal energy density calculated from the equation of the internal energy density.
We use the Harten-Lax-van Leer Discontinuities (HLLD) approximate Riemann solver \citep{Miyoshi_Kusano_2005} and the constrained transport method \citep{Stone_Gardiner_2009} to update the equations. The second-order piecewise linear reconstruction is used, and the third-order Runge-Kutta time integration is adopted. The equation of state is for an ideal gas with specific heat ratio $\gamma=5/3$.
As in ST18, we include a simpliﬁed radiative cooling for the disk gas in the energy equation so that the disk temperature will not continue to rise in response to accretion heating (the method will be described below).

The inner boundary of the numerical domain is regarded as the magnetized rotating stellar surface, which is assumed to be approximately the bottom of the stellar corona.
The central star is rotating with the constant angular velocity $\Omega_*$. The corresponding corotation radius $r_{\rm cor}$ is $(GM_*/\Omega_*^2)^{1/3}$, where $G$ is the gravitational constant, and $M_*$ and $R_*$ are the stellar mass and radius, respectively. 
The source of gravity is the central star only.
The dipole magnetosphere of the field strength of $\sim 160$~G is imposed to the star at $2.3$~days after the simulations start in all the models.
In this paper, we assume that $R_*=2R_\odot$ and $M_*=0.5M_\odot$. For later use, we define the Keplerian velocity at the stellar radius of $v_{\rm K0}=\sqrt{GM_*/R_*}\approx 220~{\rm km~s^{-1}}$ and the reference density of $\rho_0\approx 1.7\times 10^{-9}~{\rm g~cm^{-3}}$.
This paper presents results of three models (Model A, B, and C) with different $\Omega_*$. The corotation radii $r_{\rm cor}$ of Model A, B, and C are $1.5R_*$, $3R_*$, and $5R_*$, respectively.

\subsection{Initial Conditions}
The initial gas distribution is a cold dense torus embedded in a warm atmosphere. We construct the initial gas distribution by combining the two hydrostatic solutions for the cold torus and the warm atmosphere.

The warm atmosphere is obtained by solving the following hydrostatic balance equations:
\begin{align}
\frac{\partial p}{\partial r} -\frac{\rho v_{\varphi}^2}{r}&= -\frac{\rho G M_*}{r^2},\label{eq:hydrostatic-r}\\
\frac{\partial p}{\partial \theta} & = \rho v_{\varphi}^2 \frac{\cos{\theta}}{\sin{\theta}},\label{eq:hydrostatic-theta}
\end{align}
where $\rho$ and $p$ are the gas density and pressure, respectively. $v_\varphi$ is the azimuthal component of the velocity. 
The temperature profile is 
\begin{align}
    T_{\rm warm}(r,\theta)=T_{\rm warm,0}\left(\frac{r}{R_*}\right)^{-1}.
\end{align}
$T_{\rm warm,0}$ is determined by 
\begin{align}
    c_{\rm iso,w0}^2 = 0.1 v_{\rm K0}^2,
\end{align}
where $c_{\rm iso,w0}$ is the isothermal sound speed for the temperature of $T_{\rm warm,0}$. Therefore, the initial atmosphere is warm in the sense that the thermal energy density is non-negligible compared to the gravitational energy density. 
The density in the equatorial plane is
\begin{align}
    \rho_{\rm warm}(r,\theta=\pi/2)=\rho_{\rm warm,0}\left(\frac{r}{R_*}\right)^{-2},
\end{align}
where $\rho_{\rm warm,0}=3\times 10^{-4}\rho_0$.

The torus is initially magnetized and rotating, and it becomes an accretion disk after the simulations start. For the cold torus gas we also utilize a hydrostatic equilibrium solution obtained under the assumption that the specific angular momentum $j_{\rm torus}$ and the gas pressure $p_{\rm torus}$ have the functional forms of
\begin{align}
    j_{\rm torus} &\propto \left(\frac{R}{R_{\rm c}}\right)^{a_{\rm j}}\\
    p_{\rm torus} & \propto \rho_{\rm torus}^{\gamma},
\end{align}
respectively, where $R=r\sin\theta$ is the cylindrical radius, and the torus density takes its maximum at $R=R_{\rm c}$ in the midplane \citep[see, e.g.,][]{Hayashi_etal_1996}. $\rho_{\rm torus}$ is the torus density. We set $a_{\rm j}=0.46$ and $R_{\rm c}=7R_*$. The maximum value of the torus density $\rho_{\rm torus,p}$ is $0.01\rho_0$. The ratio of the thermal energy density to the gravitational energy density at the density peak is 0.01.
The torus has a purely poloidal magnetic field which is constructed by giving the $\varphi$ component of the vector potential $A_{\varphi,{\rm torus}}(r,\theta)$, where $A_{\varphi,{\rm torus}}(r,\theta)\propto \rho_{\rm torus}(r,\theta)-\rho_{\rm cutoff}$. This vector potential produces magnetic loops embedded in the torus. The torus magnetic field is only given to the region where $\rho > \rho_{\rm cutoff}$. The coefficient is chosen such that the plasma $\beta$ at $(r,\theta)=(R_{\rm c},\pi/2)$ is 100. In our models, $\rho_{\rm cutoff}=0.01\rho_{\rm torus,p}$.

The torus solution is used as the initial gas profile in the region with $p_{\rm torus}>p_{\rm warm}$, otherwise the warm atmospheric solution is adopted.
The resulting density becomes very small around the poles. To avoid numerical problems, we slightly increase the initial density only around the poles.

The simulation domain spans $0.9\le r/R_*\le 30$, $0\le \theta \le \pi$ and $0\le \varphi < 2\pi$, and we used one level of static mesh refinement with twice higher resolution within the range of $r\lesssim 13R_*$ and $0.52 \lesssim \theta \lesssim 2.6$. The root level consists of $120\times 120\times 112$ cells. The cells are uniformly spaced in the $\theta$ and $\varphi$ directions, while the radial cell size is proportional to the radius ($\Delta r_{\rm i+1}/\Delta r_{\rm i}=1.03$). We resolve one local pressure scale height of the disk with approximately 20 cells, which is similar to the resolution sufficient to capture MRI dynamics \citep{Hawley2013ApJ}.
In the analysis, we denote the azimuthal average of a physical quantity $Q$ as $\langle Q \rangle$.

The typical time step after the insertion of the stellar magnetic fields is approximately 0.3 s. Approximately 200 day scale simulations required $> 4\times 10^7$ integration steps.

\subsection{Stellar Surface Model and Boundary Conditions}
The stellar surface model is based on ST18 with some modifications. We put a damping layer as a thin spherical shell around the actual inner boundary. The thickness is $w_{\rm d}=0.1R_*$. We define the stellar coronal density and temperature as $\rho_*$ and $T_*$, respectively. The stellar coronal pressure is accordingly defined as $p_{*}=\rho_* T_*$ in the nondimensional form. Only in the damping layer, we additionally solve the following equations in an operator splitting manner:
\begin{align}
    \frac{\partial \rho}{\partial t}&= -\frac{\rho-\rho_*}{t_{\rm d}(r,\rho)}\\
    \frac{\partial p}{\partial t}&= -\frac{p-p_*}{t_{\rm d}(r,\rho)},
\end{align}
where $t_{\rm d}(r,\rho)$ is the damping timescale. It is defined as follows:
\begin{align}
    t_{\rm d}(r,\rho)^{-1} &= f_{\rm rad}(r)t_{\rm d0}(\rho)^{-1}\\
t_{\rm d0}(\rho) &= \max\left[ \min\left(f_{\rm d,min}\frac{\rho}{\rho_{*}}, f_{\rm d,max} \right), f_{\rm d,min}\right] \times t_{\rm cross} \label{eq:td0} \\ 
f_{\rm rad}(r) & = \frac{1}{2}\left[1-\tanh{\left( \frac{R_{*}-r_{\rm d}}{w_{\rm d2}} \right)} \right]
\end{align}
where $w_{\rm d2} = 0.25w_{\rm d}$, $f_{\rm d,min}=0.1$, and $f_{\rm d,max}=1$. $t_{\rm cross}=w_{\rm d}/c_{s*}$ ($c_{s*}=\sqrt{T_*}$ in the numerical unit) is the sound crossing timescale for the damping layer. 
Therefore, the density and the pressure in the damping layer are controlled to approach the coronal values in a spatially and temporally smooth way. The stellar corona and wind are kept as a result of this treatment.
Compared to ST18, we remove the constraint for the radial velocity component ($f_{\rm v}(v_r)$ in our previous paper) from $t_{\rm d}(r,\rho)$. Also, we adopt a much smaller $f_{\rm d,max}$ to quickly remove the mass of accreting flows hitting the stellar surface.
We set $\rho_*=3\times 10^{-4}\rho_0$, and $T_{*}$ is determined by requiring that the corresponding isothermal sound speed is $0.5 v_{\rm K0}^2$ (i.e., $T_*\approx 0.87$ MK).

Our damping layer method enables us to realize a self-regulated corona.
In reality, if the stellar field is closed, the magneto-hydrostatic corona will be established. If the stellar field is open and the coronal plasma is allowed to flow out, then the thermally driven stellar wind will be driven.  No mass injection will occur when the magneto-hydrostatic corona is established above the stellar surface.
In our simulations, such a stellar coronal condition is automatically realized by the damping layer method.
In our numerical setting, the stellar wind continuously blows from the polar regions as the magnetic field is open there. The stellar wind behaves as an adiabatically expanding stellar wind because there is no artificial heat source outside the thin damping layer. The velocity and magnetic field components are not artificially modified in the damping layer. In the ghost cells (where we set the boundary conditions), the gas is assumed to be rigidly rotating at the stellar angular velocity $\Omega_*$. The poloidal velocity in the ghost cells is set to zero. Namely, $v_r=v_\theta = 0$ and $v_\varphi=r\Omega_*\sin\theta $. The magnetic field components in the ghost cells are unchanged during the simulations. Therefore, the stellar dipole magnetosphere is anchored to the rigidly rotating stellar surface.

We model the stellar wind to study the general structure of outflows around the star, and the detailed modeling of the stellar wind is beyond the scope of this study.
We note that the velocity structure and the mass loss rate of the stellar wind in our models cannot be directly compared to observations.
We are aware that the adiabatic wind is unrealistic because it is a decelerating wind \citep[e.g.][]{Lamers1999isw}. However, this setting is acceptable because it escapes from the numerical domain before it stops (the speeds near the stellar surface and at the outer boundary are $\sim 200$~km~s$^{-1}$ and $\sim 90$~km~s$^{-1}$, respectively).
In addition, the mass outflow rate depends on the assumed stellar temperature and density. We choose the coronal density so that the numerical time step, which is determined by the Alfv\'en speed in the polar regions, is achievable for the available computational resources.
We consider that our wind model does not significantly affect the general accretion and wind structures. See Appendix~\ref{appendix:stellar-wind} for more details about the influences of the wind model.

An outgoing boundary is used at the outer boundary, where the flows coming into the numerical domain from the outside are prohibited. For the outgoing flows, the zero-gradient boundary condition is applied.

\subsection{Cooling and Resistivity}
We include a simpliﬁed radiative cooling for the disk gas in the energy equation so that the disk temperature will not continue to rise in response to accretion heating. The reference disk temperature profile is 
\begin{align}
    T_{\rm disk,ref}(r,\theta)=T_{\rm disk,0}\left(\frac{r}{R_*}\right)^{-1},
\end{align}
where $T_{\rm disk,0}$ is determined such that the corresponding isothermal sound speed is $0.01v_{\rm K0}^2$ ($\sim 1.7\times 10^4$~K).
The radiative cooling is considered via operator splitting by solving the following equation:
\begin{align}
    \frac{\partial T(t,r,\theta)}{\partial t}=-\frac{T(r,\theta)-T_{\rm disk,ref}(r,\theta)}{\tau_{\rm cool}(r,\theta)},
\end{align}
where $T$ is the gas temperature.
The radiative cooling timescale $\tau_{\rm cool}$ at each radius is set to 40\% of one orbital period. The cooling is switched-on only for the gas with a temperature in the range of $T_{\rm disk,ref}<T<0.3T_*$ in $|\theta-90^\circ|<20^\circ$.

A resistivity is included to model magnetic reconnection which can drive magnetospheric ejections. Our resistivity $\eta_{\rm anom}$ is a type of anomalous resistivity which is a function of the density and the electric current density \citep[e.g.][]{Ugai1992,Yokoyama_Shibata_2001} and is also used in our previous work \citep{Takasao_Tomida_Iwasaki_Suzuki_2019}. 
This type of resistivity is useful to realize magnetic reconnection with a realistic reconnection rate using a limited numerical resolution. 
The resistivity $\eta$ in this study is written as
\begin{align}
    \eta &= \begin{cases}{\rm max}\left[\eta_0 \left(\frac{v_{\rm drift}}{v_{\rm crit}} -1\right),\eta_{\rm max}\right] \\
    \hspace{1cm}{\rm if}\phantom{a} \rho<\rho_{*} \phantom{a} {\rm and} \phantom{a} v_{\rm drift} > v_{\rm crit} \\
    0 \hspace{1cm} {\rm otherwise}\end{cases}
\end{align}
where $J$ is the absolute value of the electric current density ($\bm{J}=\nabla\times \bm{B}$ in the simulation units). $v_{\rm drift}$ ($=J/\rho$ in the simulation units) is the so-called ion-electron drift velocity. The resistivity works only when the drift velocity exceeds the critical value $v_{\rm crit}$ outside the disk. The density threshold is set so that the resistivity will not operate in the accretion disk. 
For the normalized units of $L_0=R_*$ and $t_0=L_0/v_{\rm K0}$, we choose $\eta_0=0.01L_0^2/t_0$, $\eta_{\rm max}=0.1L_0^2/t_0$, and $v_{\rm drift}=10^3L_0/t_0$.
It has been known that the reconnection rate depends only weakly on the details of the functional form of $\eta$ as long as the resistivity is localized in the regions where strong electric currents exist \citep{Ugai1992}.

\section{Results} \label{sec:result}

\subsection{Overview of Accretion and Wind Structures}\label{subsec:accretion_structure}

\begin{figure*}
    \centering
    \includegraphics[width=2.1\columnwidth]{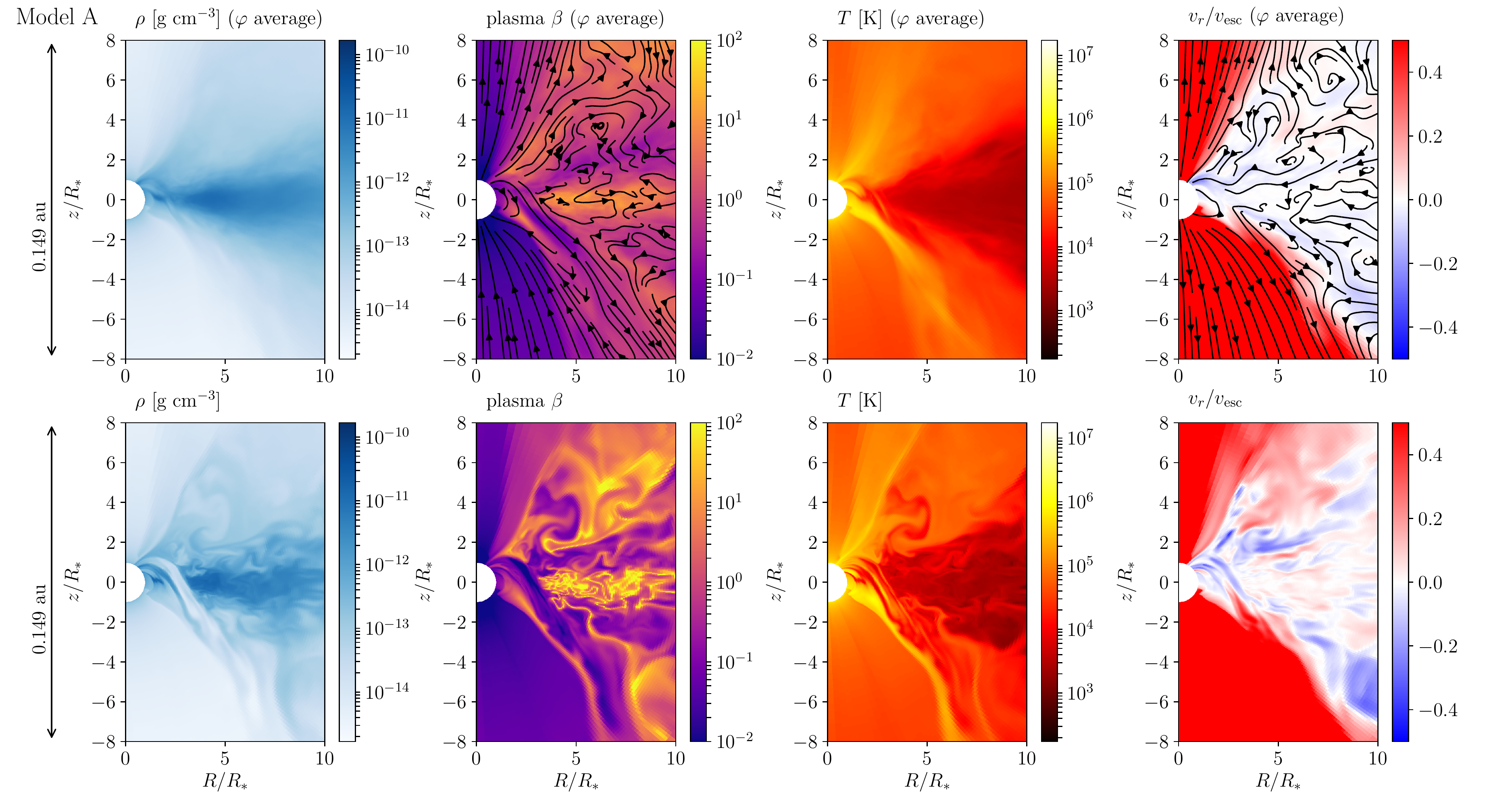}
    \caption{Accretion and ejection structures of Model A at $t=194.7$ day. Top row: azimuthally averaged data. Solid lines with arrows in the plasma $\beta$ map denote poloidal magnetic field lines, while those in the $v_r/v_{\rm esc}$ map show streamlines. Bottom row: data sliced at $\varphi=0$. From left to right, the density, the plasma $\beta$, the temperature, and the radial component of the velocity normalized by the local escape velocity, $v_r/v_{\rm esc}$.}
    \label{fig:4phys_modelA}
\end{figure*}

\begin{figure*}
    \centering
    \includegraphics[width=2.1\columnwidth]{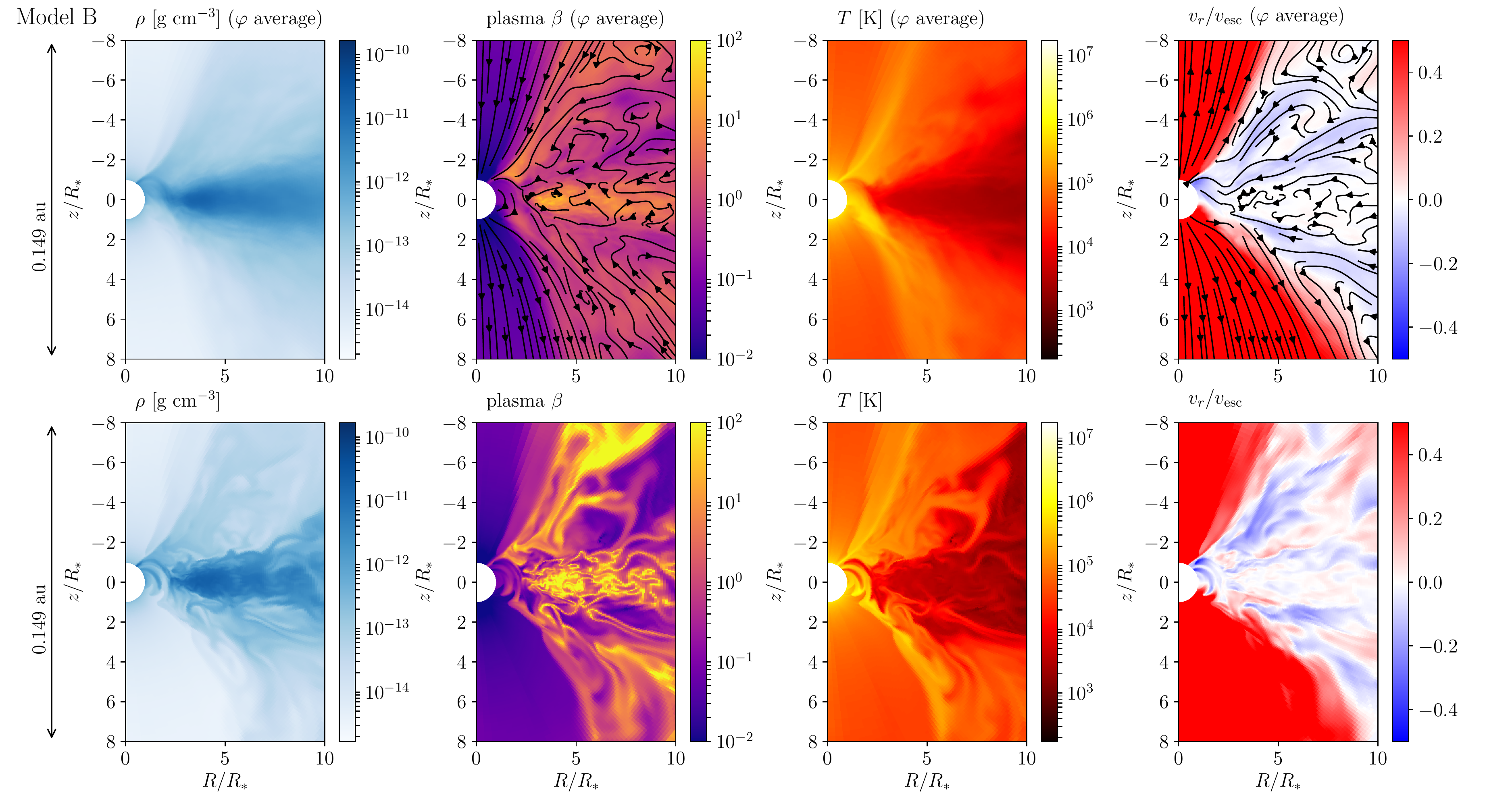}
    \caption{Same as Figure \ref{fig:4phys_modelA}, but for Model B at $t=194.7$ day. The images are flipped vertically so that we can easily compare Model B and the other two models.}
    \label{fig:4phys_modelB}
\end{figure*}

\begin{figure*}
    \centering
    \includegraphics[width=2.1\columnwidth]{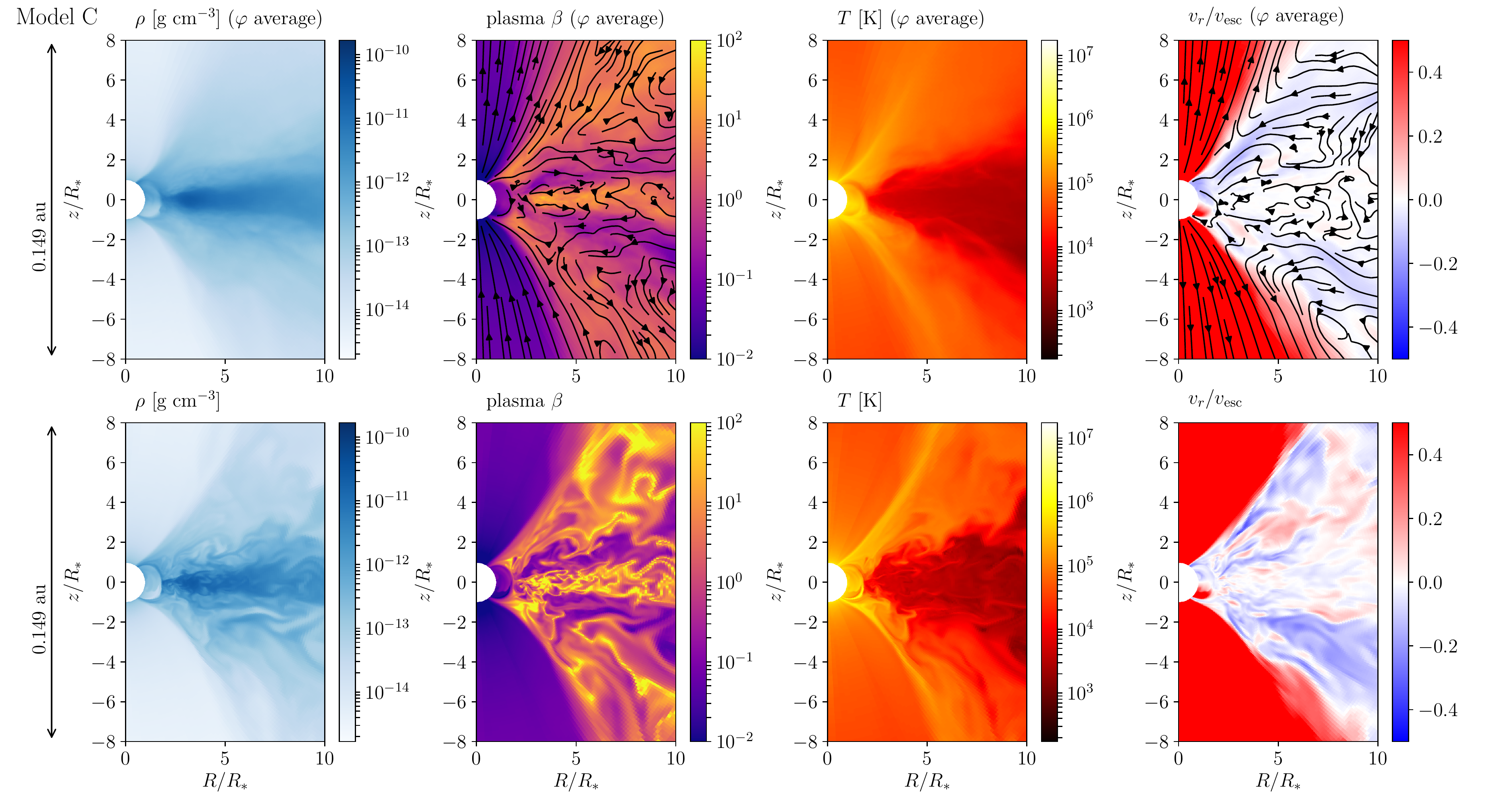}
    \caption{Same as Figure \ref{fig:4phys_modelA}, but for Model C at $t=194.7$ day (before a stable conical wind is established).}
    \label{fig:4phys_modelC}
\end{figure*}

\begin{figure*}
    \centering
    \includegraphics[width=2.1\columnwidth]{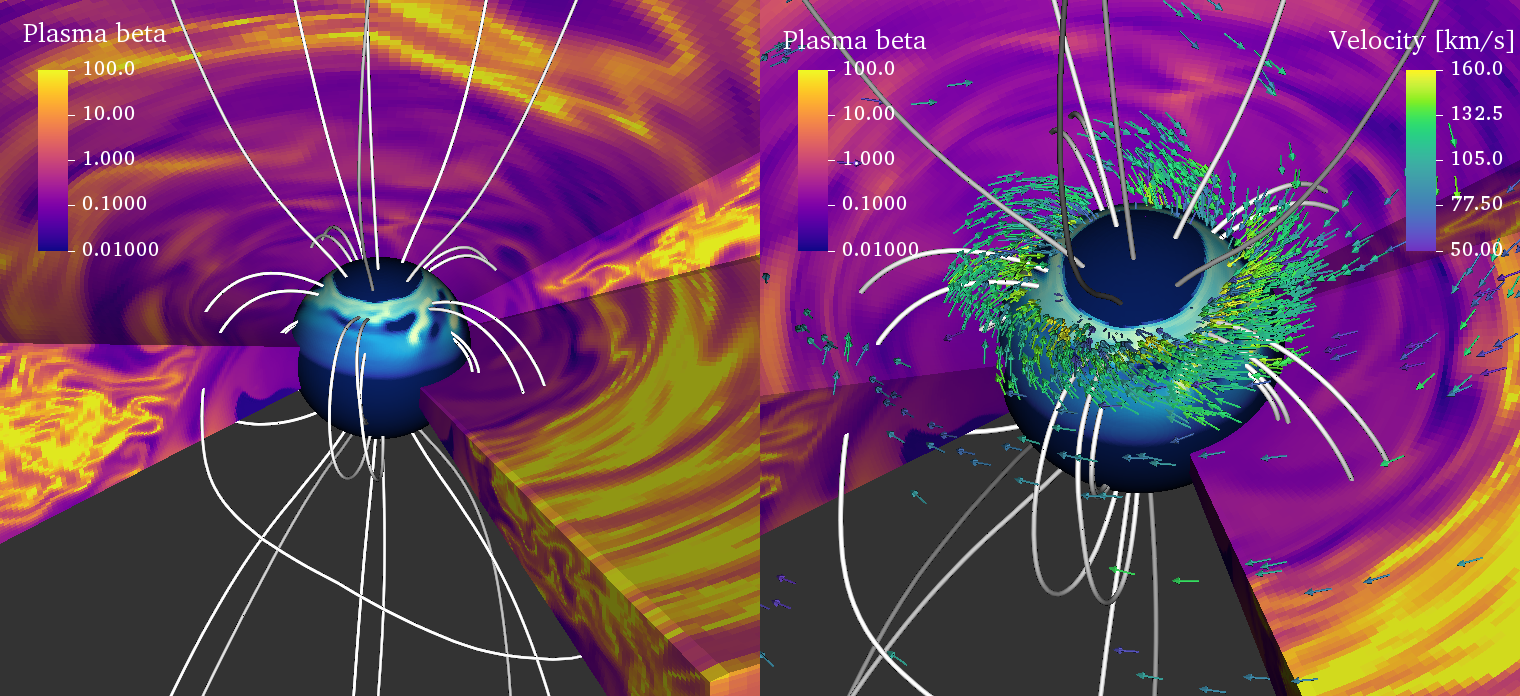}
    \caption{3D accretion structure for Model B at $t=190.1$ day (top-down view in the southern hemisphere). The southern hemisphere is highlighted as the funnel accretion occurs there. The central star is shown as the central sphere. The sphere is colored with the value of $-\rho v_r^3$ (only the regions with negative $v_r$ are colored). The disk is colored with the value of plasma $\beta$. Lines denote magnetic field lines. In the right panel, arrows indicate velocity vectors.}
    \label{fig:3D_modelB}
\end{figure*}

Figure \ref{fig:4phys_modelA} displays the snapshot of the accretion structure around the star in Model A (fast rotator). The top row shows the azimuthally averaged data, while the bottom row exhibits the data sliced at $\varphi=0$. From left to right, the density, the plasma $\beta$, the temperature, and the $r$ component of the velocity normalized by the local escape velocity, $v_r/v_{\rm esc}$ are displayed. In the top row, solid lines with arrows in the plasma $\beta$ map denote poloidal magnetic field lines, while those in the $v_r/v_{\rm esc}$ map show streamlines. Figures \ref{fig:4phys_modelB} and \ref{fig:4phys_modelC} show the same figures but for Model B and C, respectively.

We first overview common accretion structures among the three models by referring to Figures \ref{fig:4phys_modelA}, \ref{fig:4phys_modelB} and \ref{fig:4phys_modelC}. 
All models show time-variable magnetospheric accretion. Their accretion rates are $\sim 10^{-8}~{M_\odot~{\rm yr^{-1}}}$ (the accretion rates will be shown in Figure~\ref{fig:mdot_angmomdot}). In Model A, accretion mainly occurs in the northern hemisphere, and the stellar magnetosphere expands and quenches stellar accretion in the southern hemisphere. The azimuthally averaged accretion structure in Model C shows a transition from a quasi-symmetric to an asymmetric structure at $t\sim 250$ day. After the transition, accretion mainly occurs in the northern hemisphere, as in Model A. In Model B, stellar accretion takes place in the southern hemisphere. For easy comparison among the three models, hereafter the 2D images of Model B are flipped in the $z$ and $\theta$ directions. Another noticeable structure is fast accretion flows above the disk (approximately 10 to 100\% of $v_{\rm esc}$). They develop well outside the magnetosphere. They are evident in the $v_r/v_{\rm esc}$ maps of the sliced data in Figures \ref{fig:4phys_modelA}, \ref{fig:4phys_modelB} and \ref{fig:4phys_modelC}. The fast accretion flows are highly inhomogeneous. ST18 also found such fast accretion in a model without a stellar magnetosphere. ST18 showed that the failed MRI-driven disk wind becomes the fast accretion as a result of efficient angular momentum loss by magnetic fields well above the disk surfaces. The fluctuating disk wind can be discerned in both the averaged and slice maps of $v_r/v_{\rm esc}$. These simulations demonstrate that such accretion also occurs even around the star with a magnetosphere. Therefore, the accretion flows consist of the gas from the disk-magnetosphere boundary and the failed disk winds.

In the ST18 model without a magnetosphere, fast accretion develops around the boundary between the stellar wind and the disk atmosphere and occurs as a funnel accretion. However, fast accretion in the present models are found in a wider range of the latitudinal angle than in our previous model. 
We attribute this difference to the magnetic field strength in the disk atmospheres.
In the previous model, the plasma $\beta$ in the MRI turbulent disk is 30-100 around the midplane and approximately unity in the disk atmosphere. The initial disk is weakly magnetized (the initial plasma $\beta$ of the disk is set to $10^4$).
In the present models, the plasma $\beta$ is 1-10 in the inner disk and smaller than unity in a large volume around the disk surfaces (see Figures~\ref{fig:4phys_modelA},\ref{fig:4phys_modelB} and \ref{fig:4phys_modelC}). 
As the disk surfaces in the present models are more strongly magnetized, the fast accretion is driven around the disk surfaces as well as at higher latitudes. The efficient amplification of the magnetic field around the disk-magnetosphere boundary is a possible reason why the stronger magnetization is realized in the present simulations (Section \ref{subsec:magnetosphere-boundary}). However, direct comparison is not straightforward as the initial condition of the present models largely differs from the previous one (e.g. the presence of a large-scale magnetic field and the initial disk model).

The 3D accretion structure for Model B is shown in Figure \ref{fig:3D_modelB}. The southern hemisphere is highlighted as the accretion mainly occurs there. In the right panel, fast accretion flows are indicated by arrows of velocity vectors. The inner disk is turbulent as a result of MRI. The turbulent disk is truncated by the stellar magnetosphere approximately at $r=2 R_*$. As we will show in Section \ref{subsec:magnetosphere-boundary}, the magnetospheric boundary is highly perturbed. The accretion flows originating from the failed disk wind are also indicated by arrows well outside the magnetosphere. 
The free-fall velocity of the gas accreting from the magnetospheric radius $r_{\rm m}$ is
\begin{align}
    v_{\rm ff}&=\sqrt{\frac{2GM_*}{R_*}}\zeta^{1/2} \nonumber \\
    & \approx 217~{\rm km~s^{-1}} \left(\frac{M_*}{0.5M_\odot}\right)^{1/2}\left(\frac{R_*}{2R_\odot}\right)^{-1/2}\left(\frac{\zeta}{0.5}\right)^{1/2},
\end{align}
at the stellar radius for $r_{\rm m}=2R_*$, where 
\begin{equation}
    \zeta=(1-R_*/r_{\rm m})^{1/2}.
\end{equation}
The figure shows that the accretion velocity reaches $\sim 150~{\rm km~s^{-1}}$, which is 70\% of $v_{\rm ff}$.
The magnetospheric accretion flows are inhomogeneous in the azimuthal direction, which generally agrees with the multi-column accretion picture \citep[e.g.][]{Ingleby2013ApJ}. The inhomogeneous accretion structure originates from both the perturbed magnetospheric boundary and turbulent failed disk wind.
We find similar structures in the three models, although there are quantitative differences.

\begin{figure}
    \centering
    \includegraphics[width=0.9\columnwidth]{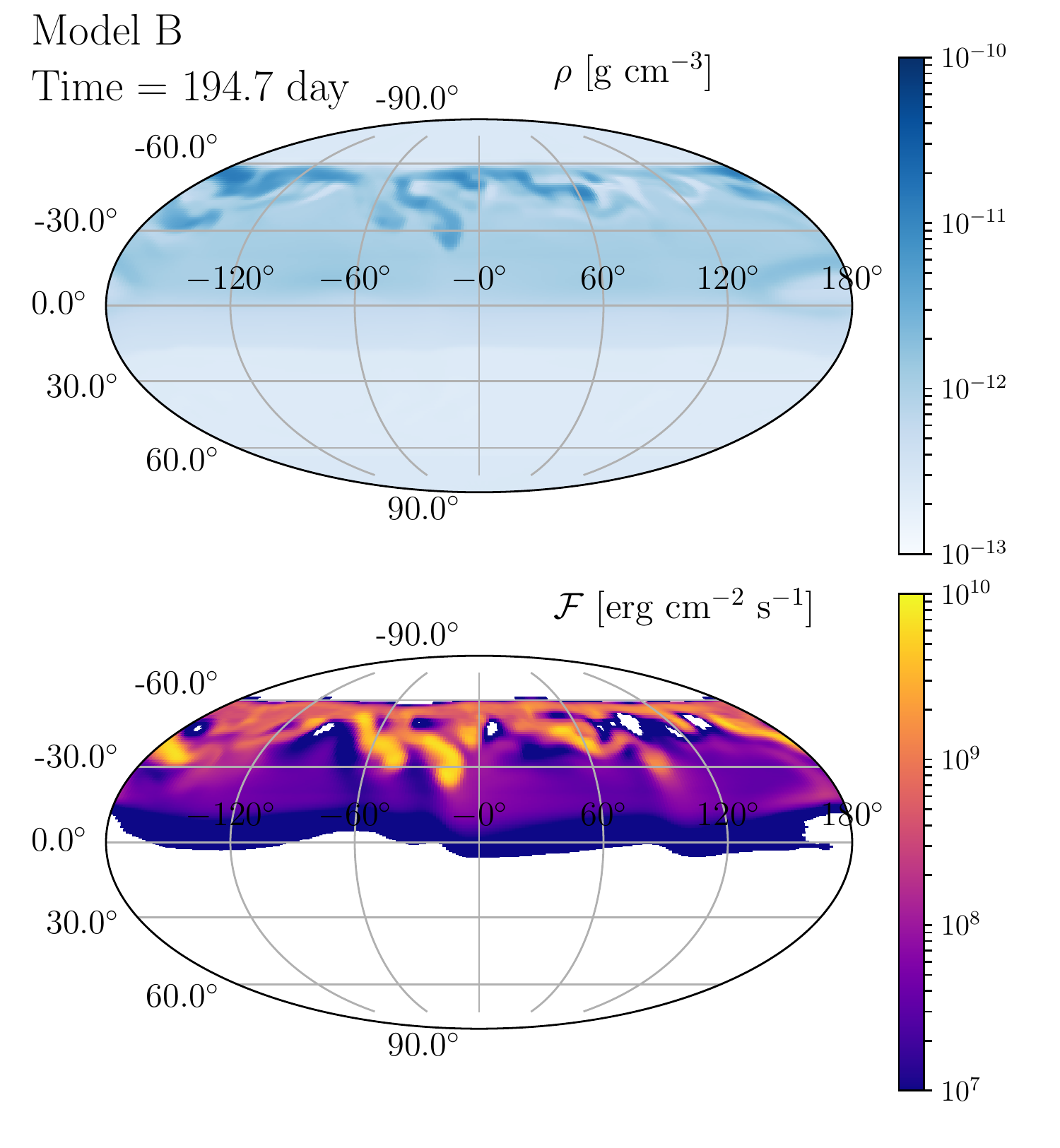}
    \caption{The density (top) and kinetic energy flux (bottom) distributions at $r=R_*$ in the Mollweide projection for Model B. The images are flipped in the $\theta$ direction so that they are shown in the same way as for the other images of Model B.}
    \label{fig:rho_ekinflux_mollweide}
\end{figure}

Figure \ref{fig:rho_ekinflux_mollweide} demonstrates how accreting flows impact the stellar surface in Model B. The density map in the top panel highlights the spotted accretion streams. The bottom panel displays the kinetic energy flux $\mathcal{F}$ defined as 
\begin{align}
    \mathcal{F}&=-\rho \left[v_r^2+v_\theta^2 + (v_\varphi-v_{\varphi,*}(\theta))^2\right]v_r,\\
    v_{\varphi,*}(\theta) &= R_* \sin\theta \Omega_*,
\end{align}
where $v_{\varphi,*}(\theta)$ is the rotational speed of the stellar surface. We find multiple locations with a high kinetic energy flux, which may be consistent with the observations suggesting the multiple accretion components. The kinetic energy flux found in our model ($\sim 10^{10}$ erg~cm$^{-2}$~s$^{-1}$) generally shows smaller values than observationally inferred values \citep[e.g., $10^{10}$-$10^{12}$ erg~cm$^{-2}$~s$^{-1}$ in ][]{Ingleby2013ApJ}. One reason could be that the accretion velocity at the stellar surface in our model ($\sim 150$~km~s$^{-1}$) is smaller than the typical value in observations ($\sim 300$~km~s$^{-1}$, see, e.g., \citet{Hartmann2016ARA&A}). 
The stellar field strength of our models is 160~G, but there are T Tauri stars showing kG fields \citep[e.g.][]{Johnstone2014MNRAS}. If the stellar field is stronger, the magnetospheric radius and the free-fall velocity will be larger. Another reason could be numerical. The damping layer near the inner boundary may reduce the kinetic energy of the accretion because the accreted gas is removed there over a finite period instead of directly falling onto the star. We will investigate this point in future papers.

\begin{figure*}
    \centering
    \includegraphics[width=2.2\columnwidth]{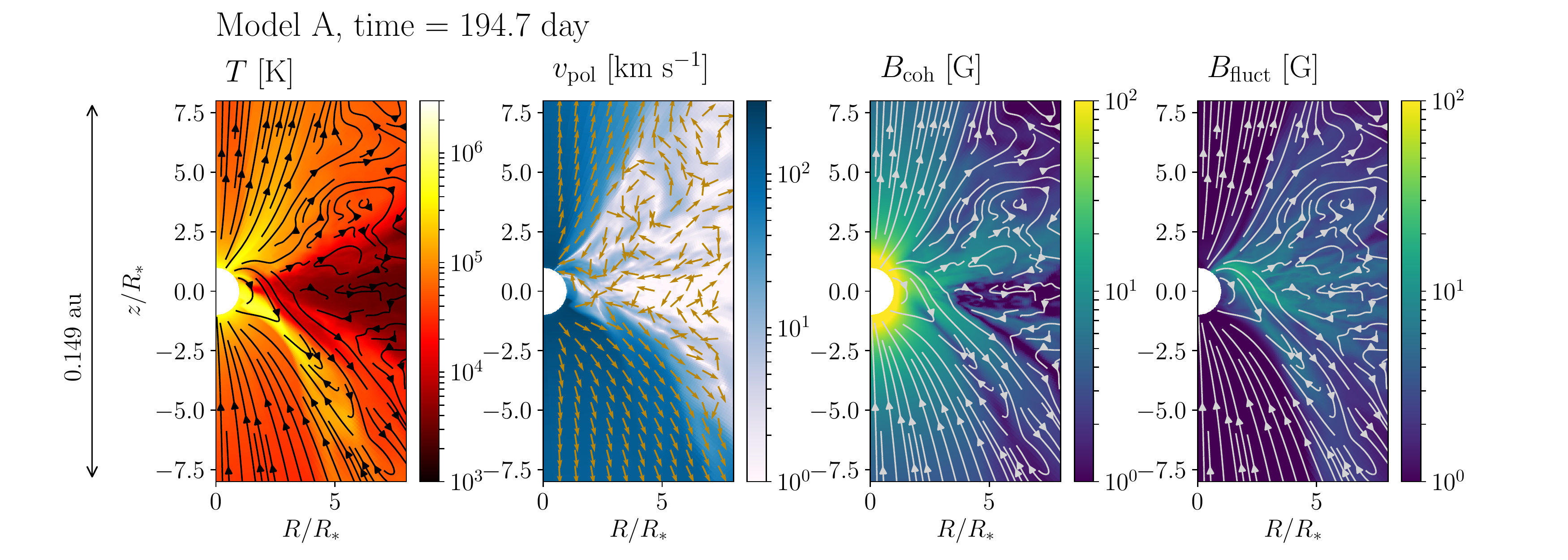}
    \includegraphics[width=2.2\columnwidth]{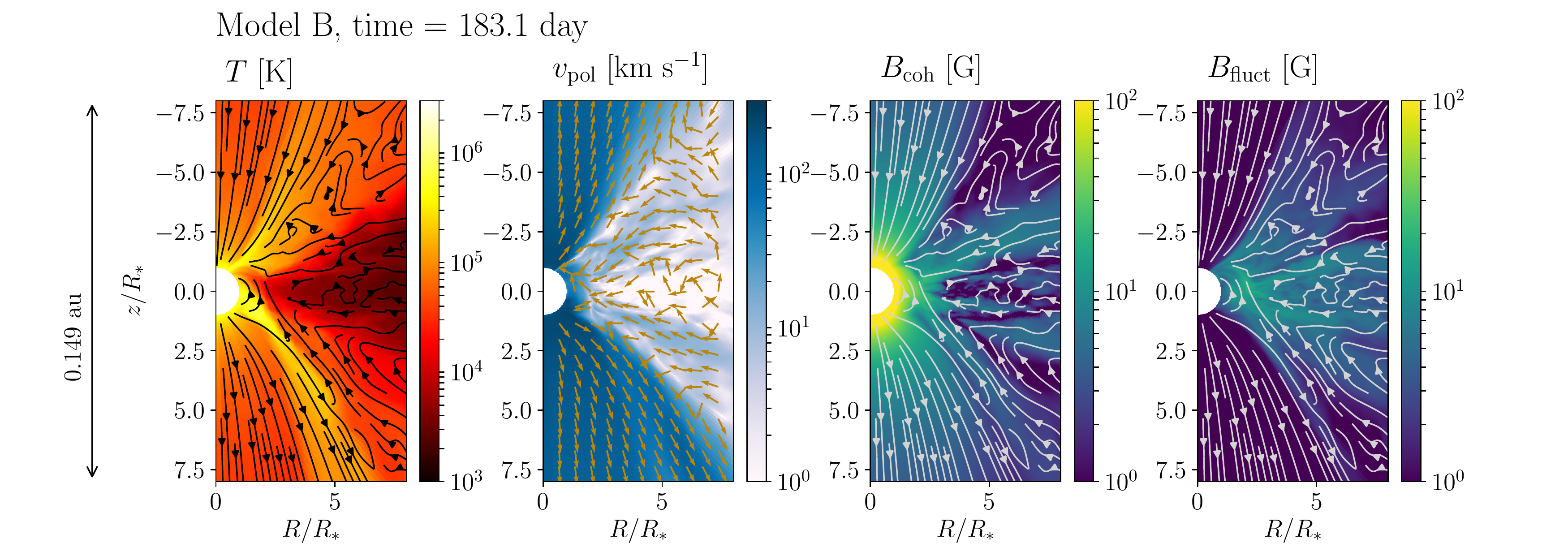}
    \caption{The wind structures in Model A (top) and B (bottom). From left to right, the temperature, the poloidal speed $v_{\rm pol}$, the coherent component of the magnetic field, and the fluctuating component of the magnetic field are shown. Lines with arrows display the azimuthally averaged magnetic structures. Arrows in the $v_{\rm pol}$ map denote the directions of the poloidal velocity vectors. Note that the image for Model B is vertically flipped.}
    \label{fig:conical_wind}
\end{figure*}

\begin{figure}
    \centering
    \includegraphics[width=1\columnwidth]{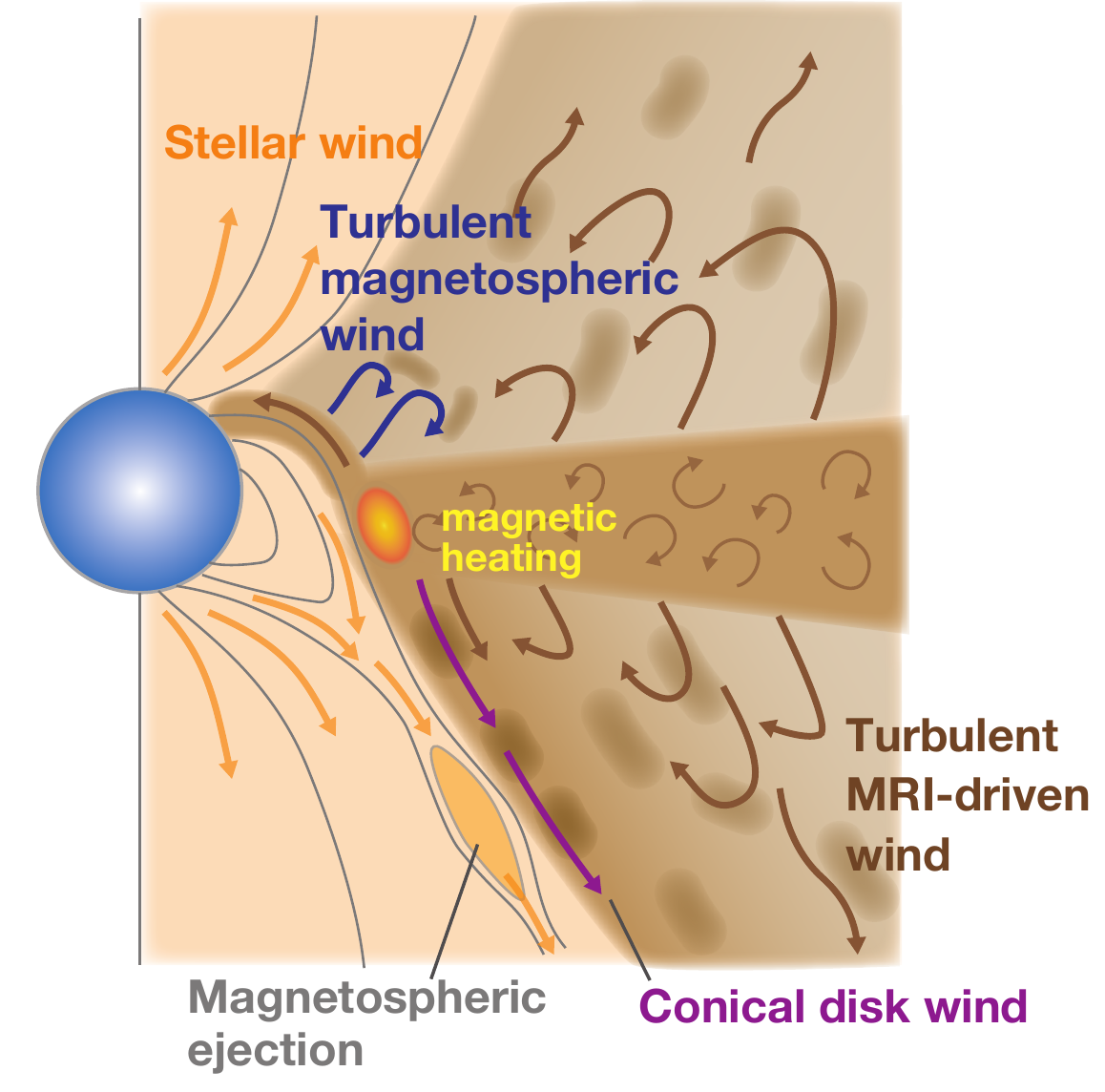}
    \caption{Schematic illustration of the wind structure. Gray lines denote the structure of the stellar poloidal magnetic field.}
    \label{fig:schematic_diagram}
\end{figure}

Figure~\ref{fig:conical_wind} displays the wind structures in Model A (top) and B (bottom). 
The strength of the coherent and fluctuating magnetic fields are denoted as $B_{\rm coh}$ and $B_{\rm fluct}$, respectively. They are defined as
\begin{align}
    B_{\rm coh}&=\sqrt{\langle B_r\rangle^2 + \langle B_\theta\rangle^2+\langle B_\varphi\rangle^2}\\
    B_{\rm fluct}&=\sqrt{\langle B_r^2 \rangle + \langle B_\theta^2 \rangle + \langle B_\varphi^2 \rangle - B_{\rm coh}^2}.
\end{align}
The two models show similar wind structures, which is summarized in the schematic diagram of Figure~\ref{fig:schematic_diagram}. In the following, we describe the wind structure using Figures~\ref{fig:conical_wind} and \ref{fig:schematic_diagram}.
We use several terminologies to indicate different types of winds. The classification in this study is as follows (Figure~\ref{fig:schematic_diagram}):
\begin{itemize}
  \setlength\itemsep{0.2em}
    \item stellar winds
    \item magnetospheric ejections
    \item disk winds
    \begin{itemize}
    \setlength\itemsep{0.2em}
        \item conical disk winds
        \item turbulent MRI-driven winds (failed disk winds)
        \item turbulent magnetospheric winds (failed winds emanating from the magnetosphere)
    \end{itemize}
\end{itemize}

The stellar coronal plasma blows in different forms. The stellar winds flow along the open magnetic field in the polar regions.
The magnetospheres inflate in the bottom half domain of the images (note that the images of Model B are flipped in the $z$ direction).
The stellar coronal plasma also flows out in the inflating magnetospheres, which is similar to a helmet streamer seen around the Sun \citep[e.g.][]{Abbo2016SSRv}. 
We find magnetospheric ejections associated with magnetic reconnection. For instance, we can find the ejection of closed magnetic fields (plasmoid ejections) in Figure~\ref{fig:4phys_modelB} (see also Figure~\ref{fig:schematic_diagram}). Magnetospheric ejections are commonly found in 2D models \citep[e.g.][]{Hayashi_etal_1996,Zanni2013A&A}. However, as we will show in Section~\ref{subsec:conical-wind} and discuss in Section \ref{subsec:comparison_2D}, the magnetospheric ejections do not introduce a significant time variability in the accretion rate, unlike 2D models. In addition, the magnetospheric ejections in our models are not accompanied by superhot plasma.

The disk winds around the magnetospheres show complicated structures.
The conical disk winds, which are collimated, outgoing flows, surround the coronal plasma. The conical winds have the poloidal velocity of several 10 to 100 km~s$^{-1}$ at the 0.1 au scale. They are time-variable and sometimes disappear, which causes the density fluctuation. The time variability is weaker in Model A than in the other two models, which suggests that the stellar spin affects the time variability. Figure~\ref{fig:conical_wind} indicates that $B_{\rm coh}$ is larger than $B_{\rm fluct}$, suggesting that the conical winds are driven by the coherent magnetic field. They are accelerated by the magnetic pressure gradient force of the coherent toroidal field. The acceleration is further investigated in Section~\ref{subsec:conical-wind}.

In addition to the conical winds, we also find failed winds emanating from the magnetosphere. They are turbulent and trapped by the stellar gravity. The velocity structure is a mixture of inflow and outflow. They are always present regardless of the development of the conical disk winds.
As we will see in Section \ref{subsec:ang}, the turbulent magnetospheric winds also play crucial roles in reducing the accretion torque exerting on the star.

The MRI-turbulent disks blow the turbulent MRI-driven winds.
The wind density structure is inhomogeneous. The inflow and outflow coexist because they are turbulent (ST18). One will find that $B_{\rm fluct}$ is larger than $B_{\rm coh}$ in the disk atmospheres, suggesting that the wind is driven by the fluctuating magnetic field (see also \citet{Suzuki_Inutsuka2009ApJ} and ST18). However, a large part of the wind gas fails to escape from the stellar gravity because of their small velocity. 
The low plasma $\beta$ region in the stellar wind is separated by the relatively high $\beta$ disk atmosphere (including the disk wind). See the plasma $\beta$ maps in Figures \ref{fig:4phys_modelA}, \ref{fig:4phys_modelB} and \ref{fig:4phys_modelC}. 
The averaged plasma $\beta$ in a large volume of the disk atmosphere is close to or larger than unity. ST18 found that mass loading and heating by the MRI-driven wind results in such a moderate ($\gtrsim 1$) plasma $\beta$ even in the high latitudinal region of the disk atmosphere.

The turbulent MRI-driven winds play a role in confining the stellar winds in the polar regions by the gas pressure. The effect of the confinement by the disk winds should be underestimated in 2D models, because the MRI-driven winds blow only in 3D.
We present an additional analysis about the dynamical effects of the stellar winds in Appendix~\ref{appendix:stellar-wind}.

\subsection{Magnetospheric 
boundary}\label{subsec:magnetosphere-boundary}

\begin{figure*}
    \centering
    \includegraphics[width=1.9\columnwidth]{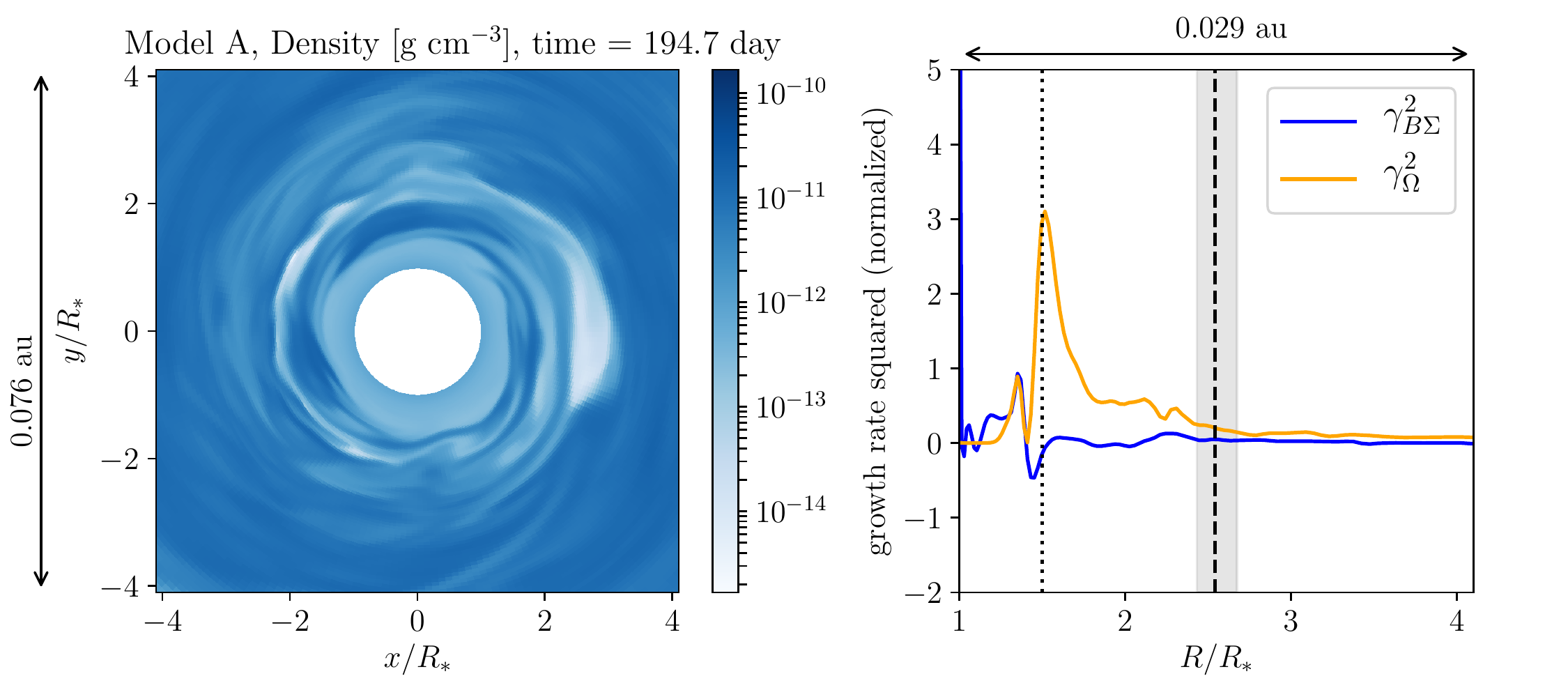}
    \includegraphics[width=1.9\columnwidth]{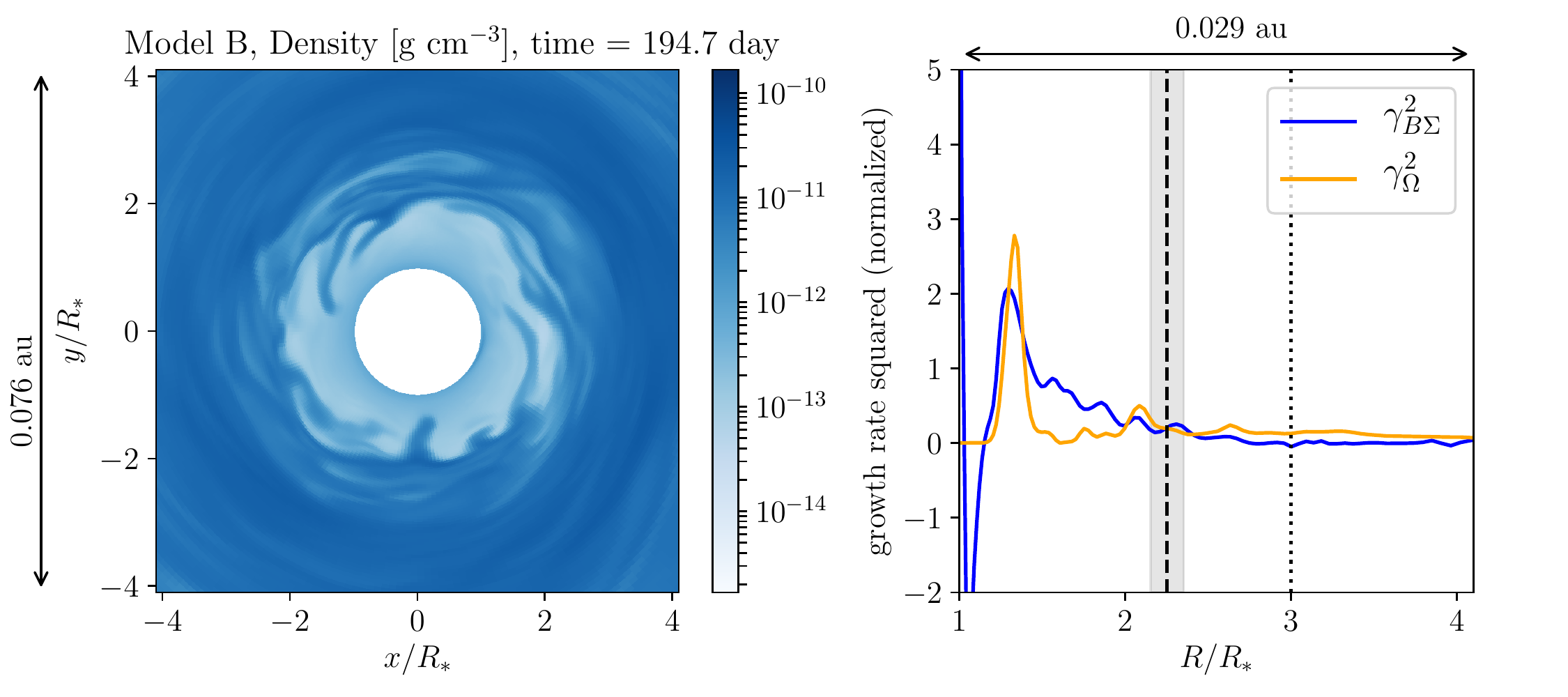}
    \includegraphics[width=1.9\columnwidth]{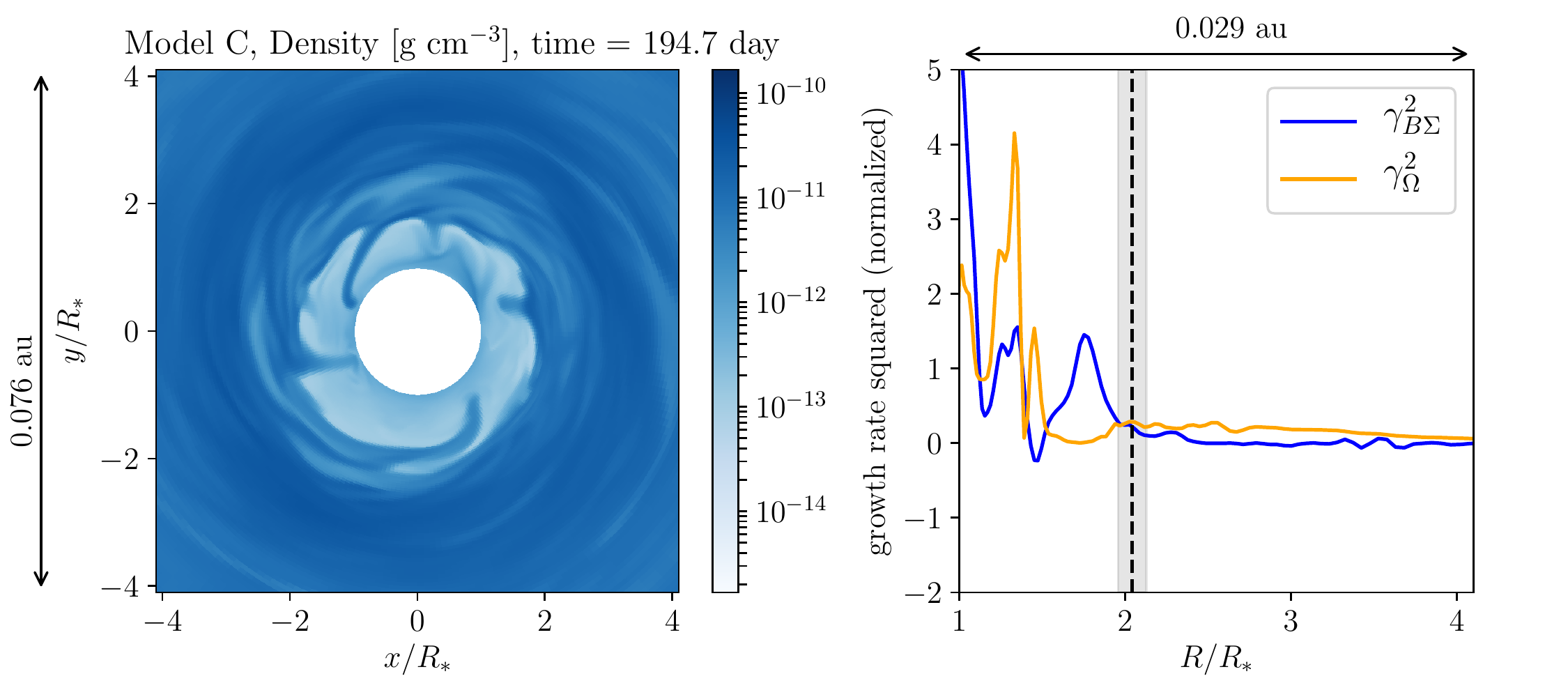}
    \caption{The density structures at the midplane (left) and the radial profiles of $\gamma_{B\Sigma}^2$ and $\gamma_{\Omega}^2$ normalized by $\Omega_{\rm K*}^2=GM_*/R_*^3$. The top, middle, and bottom panels are for Model A, B, and C, respectively. See the text for the definitions of $\gamma_{B\Sigma}^2$ and $\gamma_{\Omega}^2$. In the right panels, the vertical dashed lines denote the time averaged values of the magnetospheric radii during the period of $t=$190.1-199.4~day. The gray bands indicate the ranges between the minimum and maximum values of the magnetospheric radii. The vertical dotted lines show the corotation radii.}
    \label{fig:interchange}
\end{figure*}

\begin{figure}
    \centering
    \includegraphics[width=0.9\columnwidth]{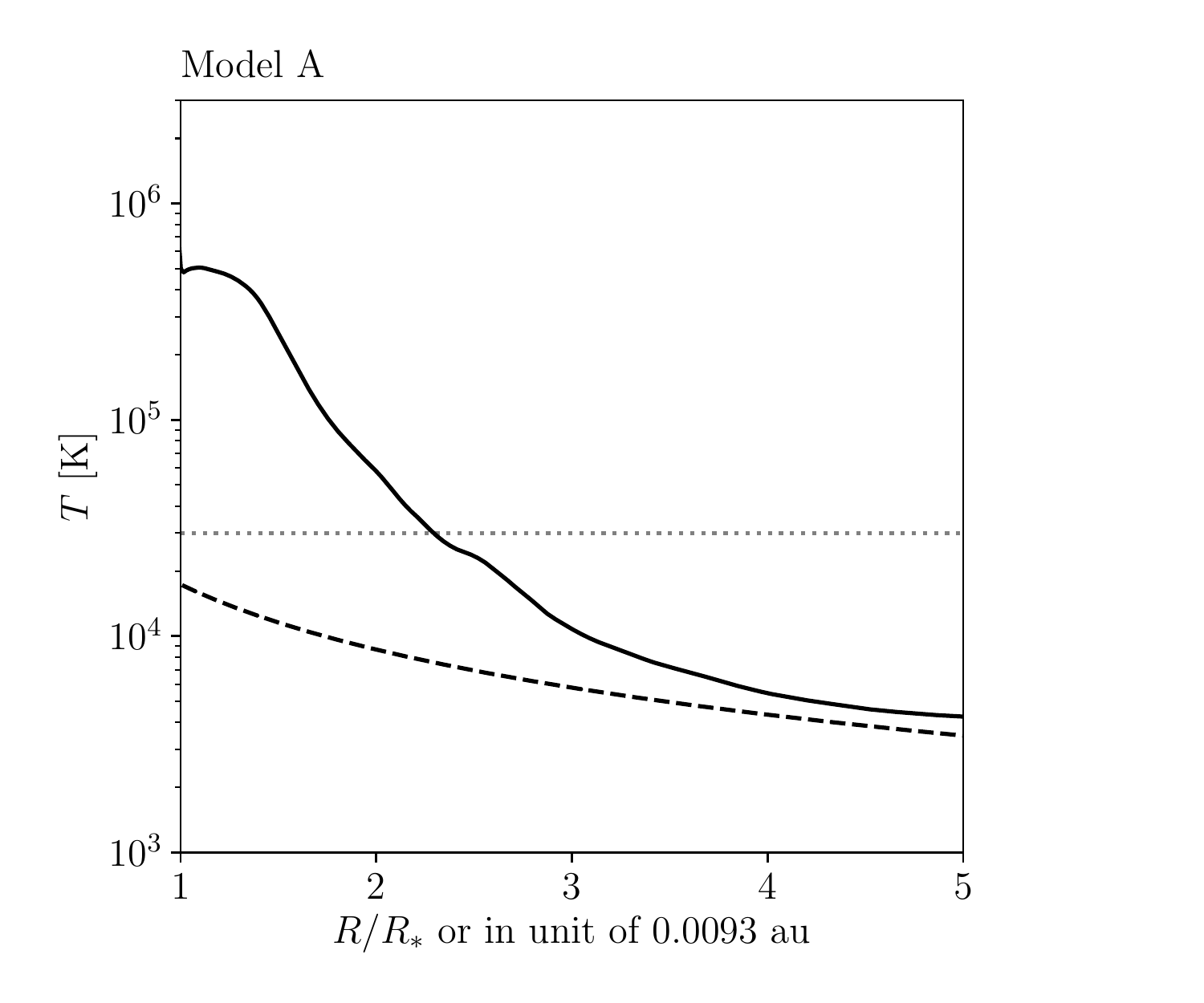}
    \includegraphics[width=0.9\columnwidth]{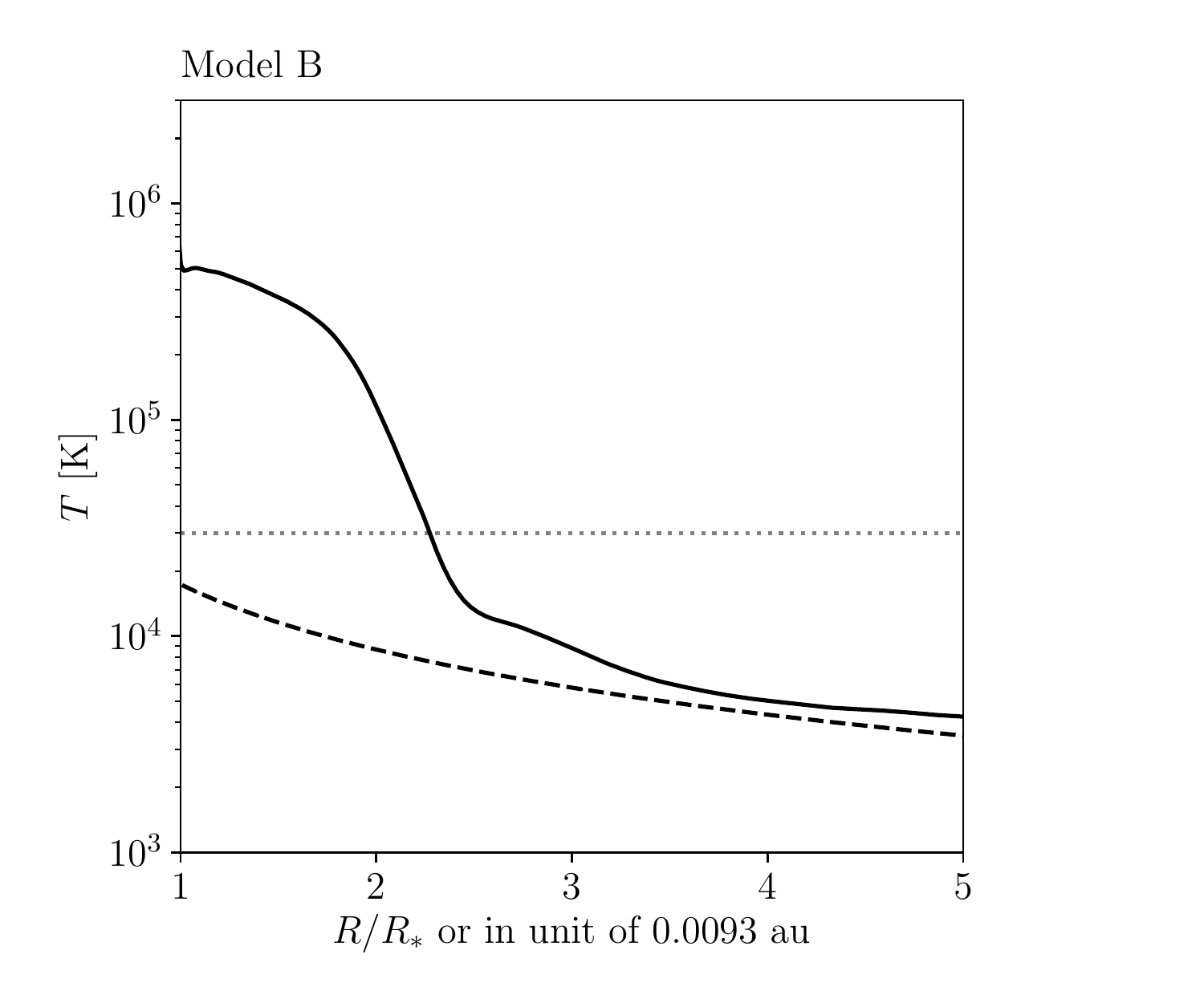}
    \includegraphics[width=0.9\columnwidth]{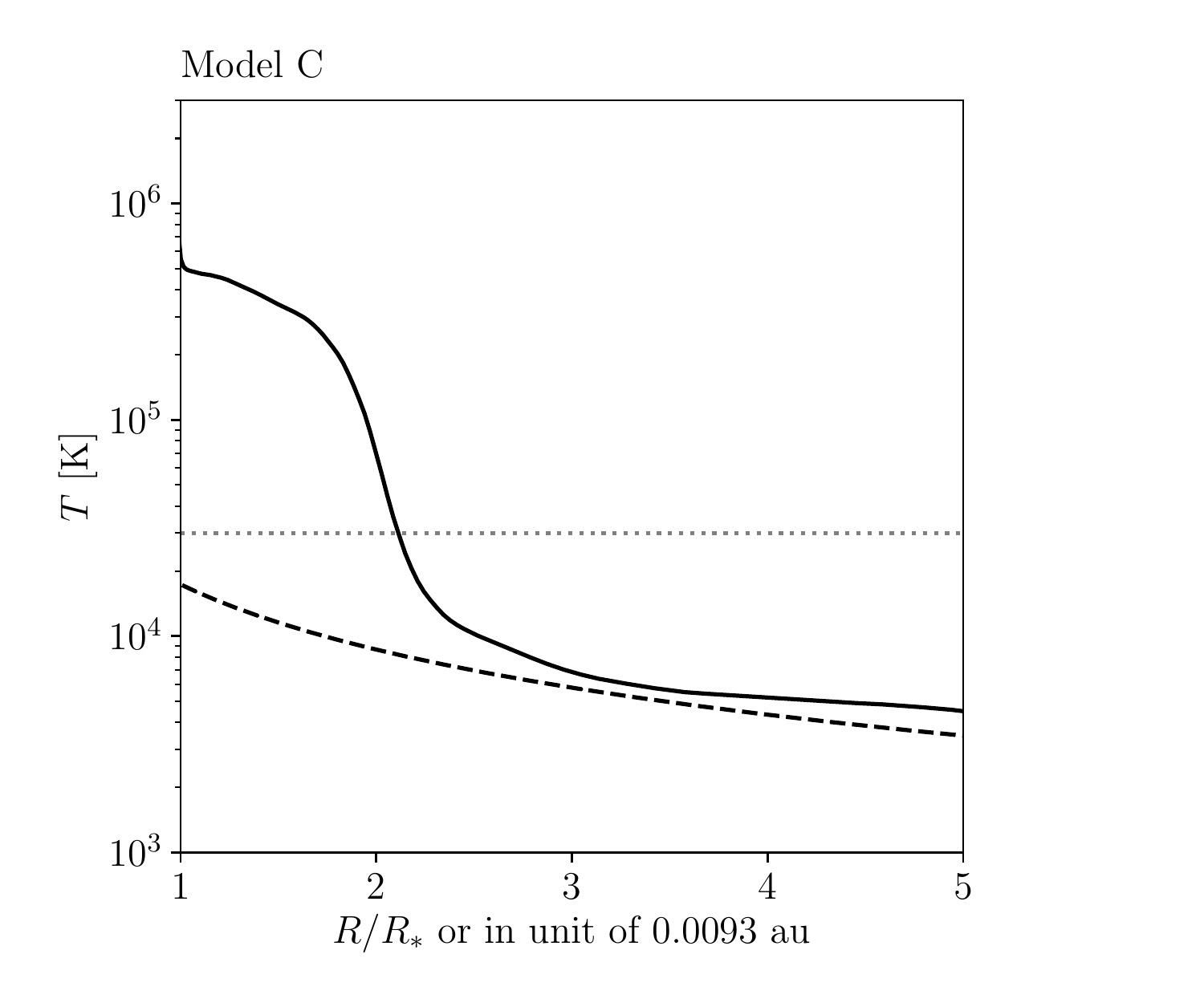}
    \caption{The equatorial temperature profiles for Model A, B, and C. The solid line is the numerical results. The dashed lines denote $T_{\rm disk,ref}(R,\pi/2)$. The horizontal dotted lines indicate the threshold temperature, $T_{\rm mag}=3\times 10^4$~K.}
    \label{fig:temp_midplane}
\end{figure}

\begin{figure}
    \centering
    \includegraphics[width=1.0\columnwidth]{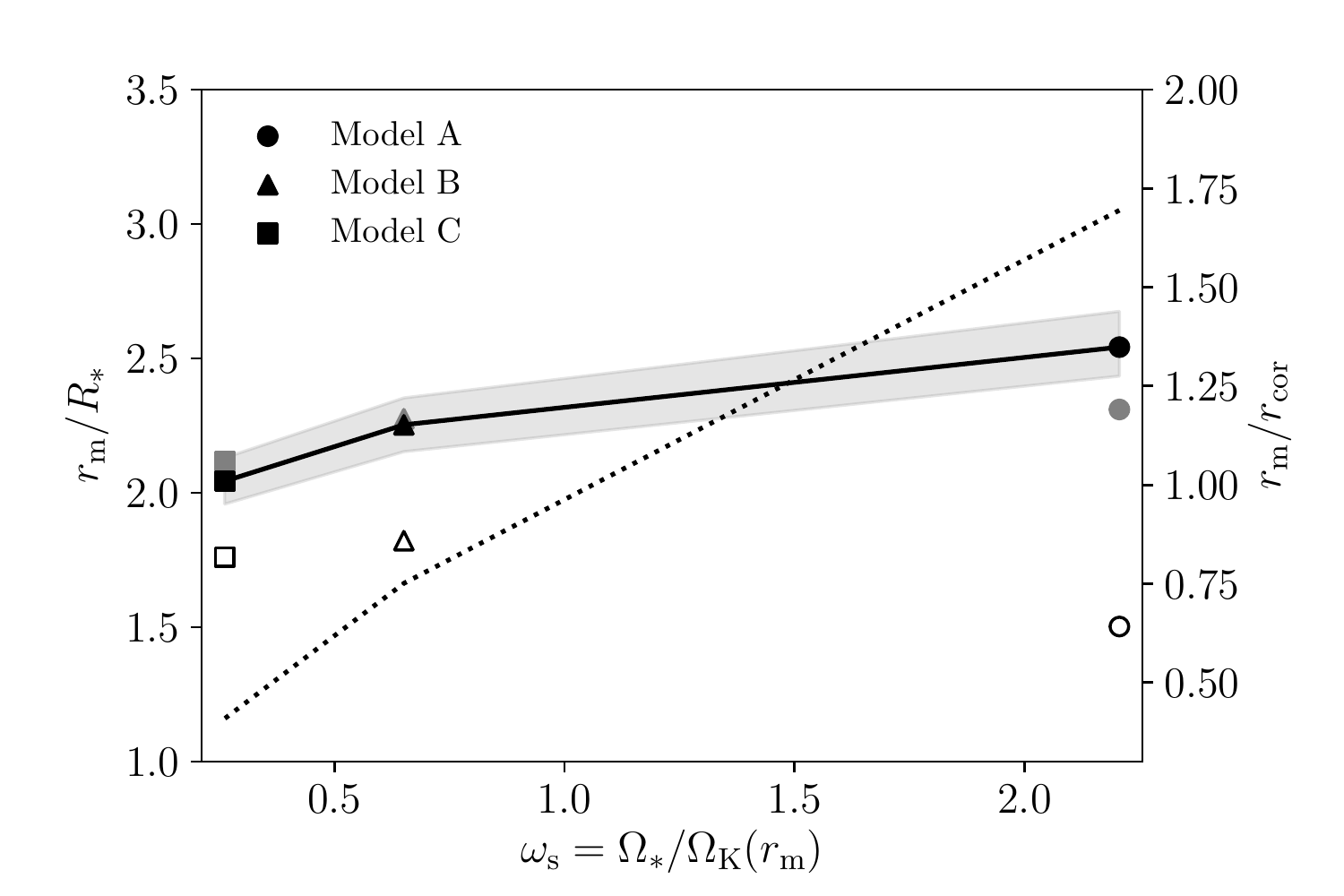}
    \caption{The magnetospheric radii normalized by the stellar radius (the left vertical axis) are shown against the fastness parameter $\omega_{\rm s}$. The circles, triangles, and squares are the results for Model A, B, and C, respectively. The black-filled, open, and gray-filled symbols denote $r_{\rm m,1}$, $r_{\rm m,2}$, and $r_{\rm m,3}$, respectively. The solid line with the black-filled symbols shows $r_{\rm m,1}$. The minimum and maximum values of $r_{\rm m,1}$ are indicated by the shaded region. The dotted line shows the ratio of the magnetospheric radius to the corotation radius (the right vertical axis). The data are taken during the period of $t=$190.1-199.4~day.}
    \label{fig:rmag}
\end{figure}

The left panels of Figure \ref{fig:interchange} display the density structures at the midplane. The low-density regions generally correspond to the magnetospheric plasma. The fluctuation of the boundary between the dense and tenuous regions is prominent in all the models.

We estimate the magnetospheric radii in three different ways. 
The first method is to define the magnetospheric radius as the radius where the plasma $\beta$ is unity \citep[e.g.][]{Bessolaz2008A&A,Kulkarni_Romanova2013MNRAS}. Previous simulations suggest that this radius corresponds to the radius where the funnel accretion flows start to fall onto the star \citep{Bessolaz2008A&A,Kulkarni_Romanova2013MNRAS}.
For each azimuthal angle, we search for the minimum radius where $\beta=1$ at the equatorial plane, and then we take the azimuthal average. When we search for the minimum radius, we ignore the regions where $\beta>1$ if their radial size is smaller than three cells. We also take the time average of the radius during the period of $t=$190.1-199.4~day. The resulting value is defined as $r_{\rm m,1}$.

The second method is to use the ratio of the total matter pressure (the gas pressure plus the ram pressure) to the magnetic pressure, $\beta_{\rm t}$:
\begin{equation}
\beta_{\rm t} \equiv (p+\rho v^2)/(B^2/8\pi).
\end{equation}
The condition $\beta_{\rm t}=1$ is used to find the magnetospheric radius in some previous studies \citep[e.g.][]{Romanova2002ApJ}. We calculate the azimuthally and temporally averaged radius in the same manner as for $r_{\rm m,1}$, but we use the condition $\beta_{\rm t}=1$. The calculated radius is defined as $r_{\rm m,2}$.

The third estimation, $r_{\rm m,3}$, is based on the stellar coronal structure. As the stellar coronal plasma that fills the magnetosphere is much hotter ($\sim 10^6$ K in this study) than the disk gas, the magnetosphere can be characterized by the temperature. For this reason, we define $r_{\rm m,3}$ as the radius where $\langle T \rangle$ is equal to a threshold temperature, $T_{\rm mag}$, at the equatorial plane. We also perform the time average for $r_{\rm m,3}$ for the same time span as the other two.
In our models, the disk temperature is maintained near the reference disk temperature $T_{\rm disk,ref}(r,\theta)$ and is approximately $1\times 10^4$ K at $R=2R_*$. Considering this, we set $T_{\rm mag}=3\times 10^4$ K in this study. Figure \ref{fig:temp_midplane} displays the temperature profiles, where one will find that the temperature deviates from the reference disk temperature and drastically rises toward the center. As this estimation depends on the chosen value for $T_{\rm mag}$, this definition may be somewhat specific for T Tauri stars. However, this estimation is more compatible with observations than the others because neither $\beta_{\rm t}$ nor $\beta$ can be measured but the size of the innermost disk with a temperature of a few $10^4$ K can \citep{Gravity_Collaboration2020Natur}. For this reason, we use $r_{\rm m,3}$ to evaluate the other two estimation methods ($r_{\rm m,1}$ and $r_{\rm m,2}$).

The results of the magnetospheric radius measurement are summarized in Figure~\ref{fig:rmag}. The horizontal axis is the fastness parameter of the models, $\omega_{\rm s}=\Omega_*/\Omega_{\rm K}(r_{\rm m})$. 
$r_{\rm m,1}$ and $r_{\rm m,3}$ are very similar, but $r_{\rm m,2}$ is systematically smaller than the others. For Model A, $r_{\rm m,2}$ is approximately 60\% of either $r_{\rm m,1}$ or $r_{\rm m,3}$. This result suggests that $r_{\rm m,1}$ based on the plasma $\beta$ represents the hot magnetospheric size more accurately than $r_{\rm m,2}$ in our simulations. Considering this, we use $r_{\rm m,1}$ as the representative value for the magnetospheric radius and denote $r_{\rm m,1}$ just as $r_{\rm m}$ in the following.
The solid line with black-filled symbols in Figure~\ref{fig:rmag} shows the time-averaged $r_{\rm m}$. The minimum and maximum values are indicated by the shaded region. The amplitude of the variability is found to be roughly several to 10\% in all the models.

Our simulations indicate that the magnetospheric radius weakly depends on the stellar spin.
The dotted line denotes the $r_{\rm m}/r_{\rm cor}$ (the right vertical axis), which shows that our models cover a wide range of the ratio. We note that many previous theories are based on the assumption that $r_{\rm m}/r_{\rm cor}=1$ \citep[e.g.][]{Shu1994ApJ}.

We investigate the destabilization mechanisms of the magnetospheric boundary.
\citet{Spruit1995MNRAS} derived the criterion for development of the magnetic interchange instability as follows:
\begin{equation}
    \gamma_{B \Sigma}^2\equiv -g_{\rm eff}\frac{d}{dr}\ln \frac{\Sigma}{B_z} > 2\left( r \frac{d\Omega}{dr}\right)^2 \equiv \gamma_{\Omega}^2,
\end{equation}
where $\Sigma$ is the surface density, $B_z$ is the absolute value of the vertical component of the magnetic field, and $\Omega$ is the angular velocity of the gas. When we calculate these values, we use the temporally and azimuthally averaged quantities. $\gamma_{\Omega}$ expresses the suppression of the interchange instability by velocity shear.
$g_{\rm eff}$ is the effective gravity and is written as
\begin{equation}
    g_{\rm eff} \equiv -\left[\Omega_{\rm K}(r)^2  - \langle\Omega(r)\rangle^2\right] r,\label{eq:geff}
\end{equation}
where $\Omega_{\rm K}(r)=\sqrt{GM_*/r^3}$.
We calculate the angular velocity $\langle\Omega\rangle$ as
\begin{align}
    \langle \Omega \rangle = \frac{\langle\rho v_\varphi \rangle}{R\langle \rho \rangle}.
\end{align}
\citet{Blinova2016MNRAS} also performed the stability analysis on the basis of this criterion.

The right panels of Figure \ref{fig:interchange} display the profiles of $\gamma_{B \Sigma}^2$ and $\gamma_{\Omega}^2$ for Model A, B, and C. In Model B and C, the unstable condition $\gamma_{B \Sigma}^2 > \gamma_{\Omega}^2$ is satisfied around the magnetospheric boundary (shown as the vertical dashed lines with the shaded regions). Therefore, we consider that the finger-like structures seen in the density maps for both models are results of the interchange instability. The plasma seems marginally stable to the interchange instability just at the magnetospheric boundaries probably because of the convection mixing in response to the instability. The boundary in Model A is highly disturbed and the accreting flows break into spiraling filaments. However, the unstable condition is not satisfied for Model A. As we will discuss in Section~\ref{subsec:intability_modelA}, the instability seems to be relevant to the magneto-gradient driven instability proposed by \citet{Hirabayashi2016ApJ}.

\subsection{Magnetic fields around the magnetospheric boundary}\label{subsec:Bfield_magnetosphere}
\begin{figure} 
    \centering
    \includegraphics[width=0.9\columnwidth]{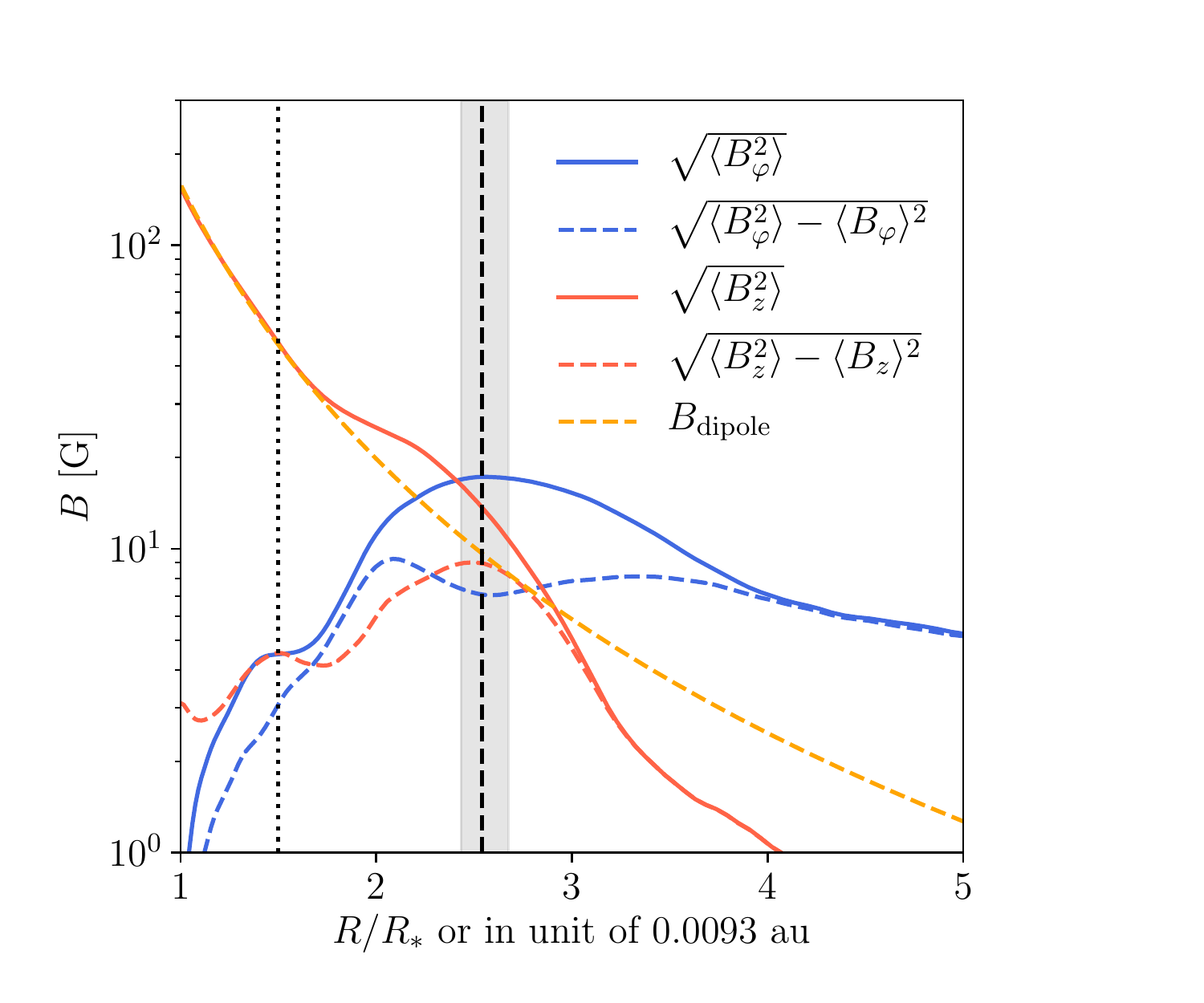}
    \includegraphics[width=0.9\columnwidth]{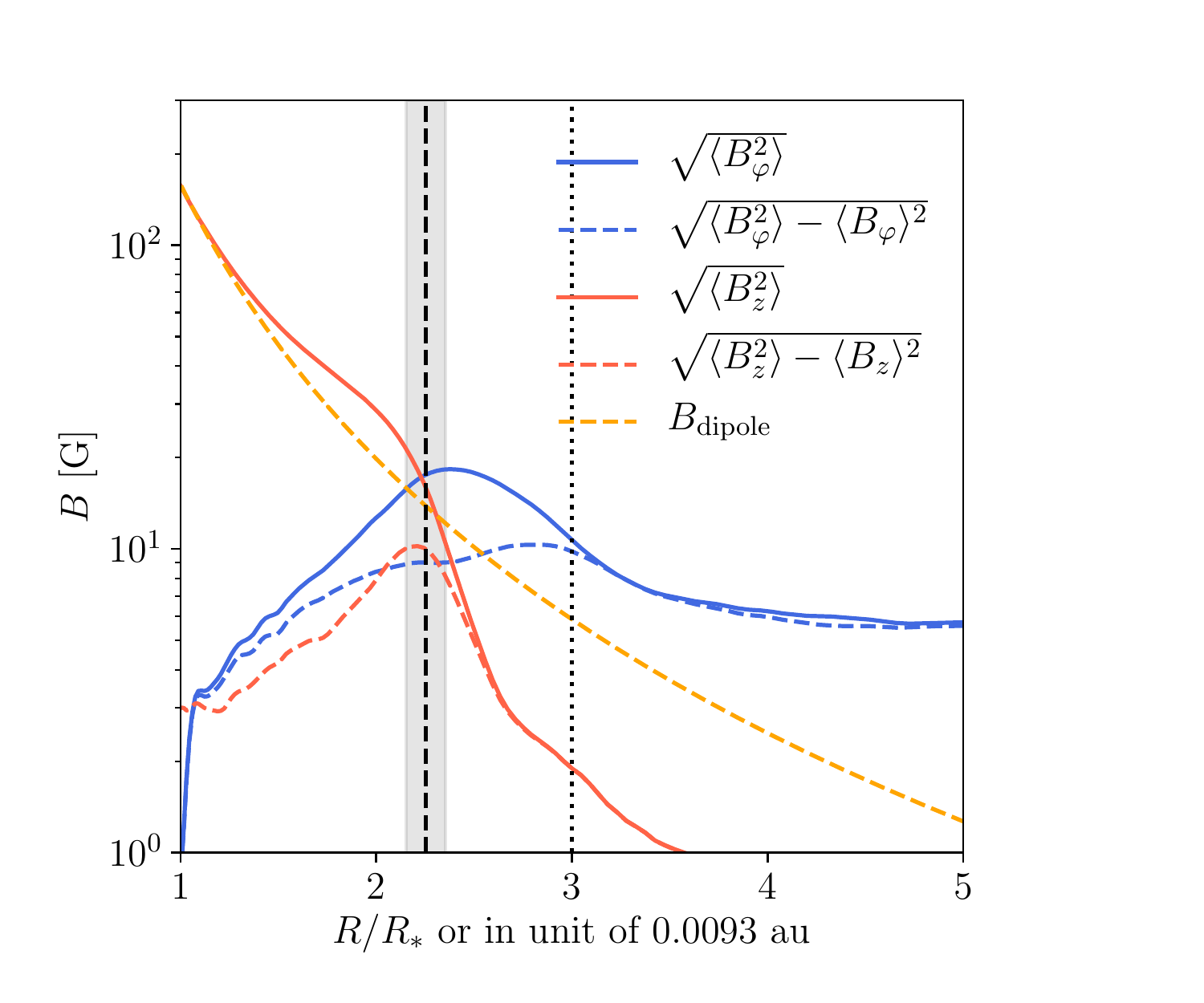}
    \includegraphics[width=0.9\columnwidth]{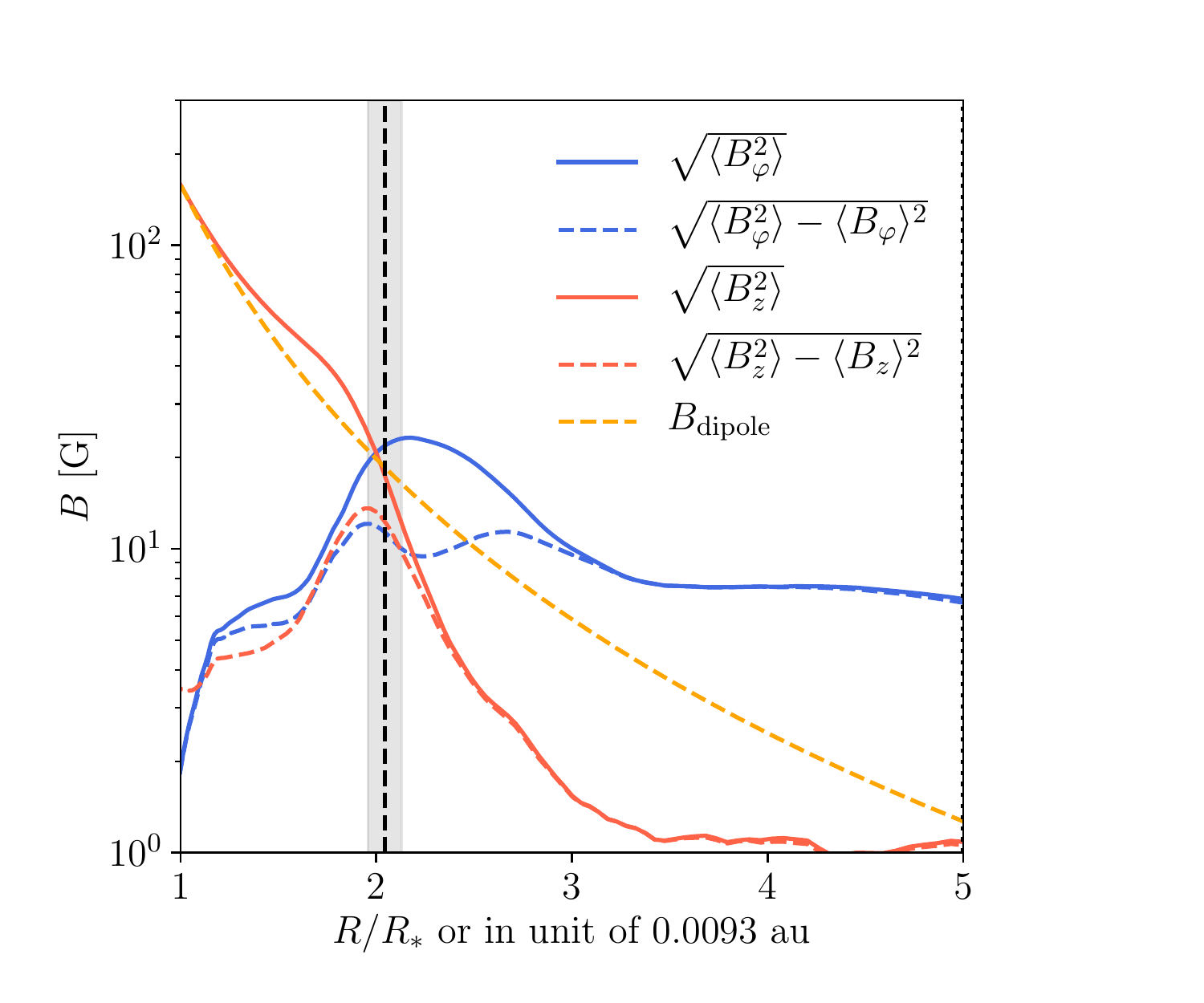}
    \caption{Magnetic field profiles at the midplane. The top, middle and bottom panels show the results for Model A, B, and C, respectively. Blue and orange solid lines show the vertical and azimuthal components of magnetic fields, respectively. The pure dipole field strength is also indicated by dashed orange lines as a reference. The vertical dashed lines denote $r_{\rm m}/R_{*}$ during the period of $t=$190.1-199.4~day. The gray bands indicate the ranges between the minimum and maximum values of $r_{\rm m}/R_*$. The vertical dotted lines show $r_{\rm cor}/R_*$.}
    \label{fig:Bfield}
\end{figure}

The magnetic field profiles of the three models are shown in Figure \ref{fig:Bfield}. The data are measured around the equatorial plane. We separately measure the total components ($\sqrt{\langle B_z^2 \rangle}$ and $\sqrt{\langle B_\varphi^2 \rangle}$), the coherent components ($\langle B_z \rangle$ and $\langle B_\varphi \rangle$), and the fluctuation components ($\sqrt{\langle B_z^2 \rangle-\langle B_z\rangle^2}$ and $\sqrt{\langle B_\varphi^2 \rangle-\langle B_\varphi\rangle^2}$). The instability at the boundary increases the fluctuation components, but the enhancement of the total $B_z$ is insignificant. Regarding the toroidal component, the total $B_\varphi$ (blue solid lines) has a peak around the magnetospheric radius. The coherent component dominates around the magnetospheric boundary, suggesting the formation of the coherent toroidal magnetic flux bundle. Inside the magnetospheric boundary, the strength of the toroidal field rapidly decreases toward the center, which means that the accreting gas does not strongly twist the stellar field. Outside the boundary, the fluctuation component (blue dashed lines) is the largest. In the disk, the fluctuation components are mainly produced by MRI.

An important difference between the previous axisymmetric models and our 3D models is the presence of additional mechanisms to amplify the toroidal field.
Previous axisymmetric models treat the radial transport of the stellar field and gas around the magnetospheric boundary as a diffusion process by using effective magnetic diffusivity and viscosity. In the steady state, the stellar magnetic field is confined by the axisymmetric disk gas. In other words, the outward transport of the stellar field does not occur.
However, the stellar field in three-dimension is continuously extruded to the disk as a result of instabilities of the magnetospheric boundaries (Figure \ref{fig:interchange}). 
As a result of the continuous injection of the stellar field to the inner disk, the strong toroidal field can be efficiently generated by the disk differential rotation. This process has been ignored in the previous axisymmetric models. In addition, axisymmetric models generally ignore the toroidal field generation via MRI. The lack of this mechanism will lead to the reduction in the toroidal field strength away from the boundary, which will make it difficult that the inner disk keeps the strong toroidal field against diffusion from the boundary toward the outer disk.

\begin{figure*} 
    \centering
    \includegraphics[width=1.7\columnwidth]{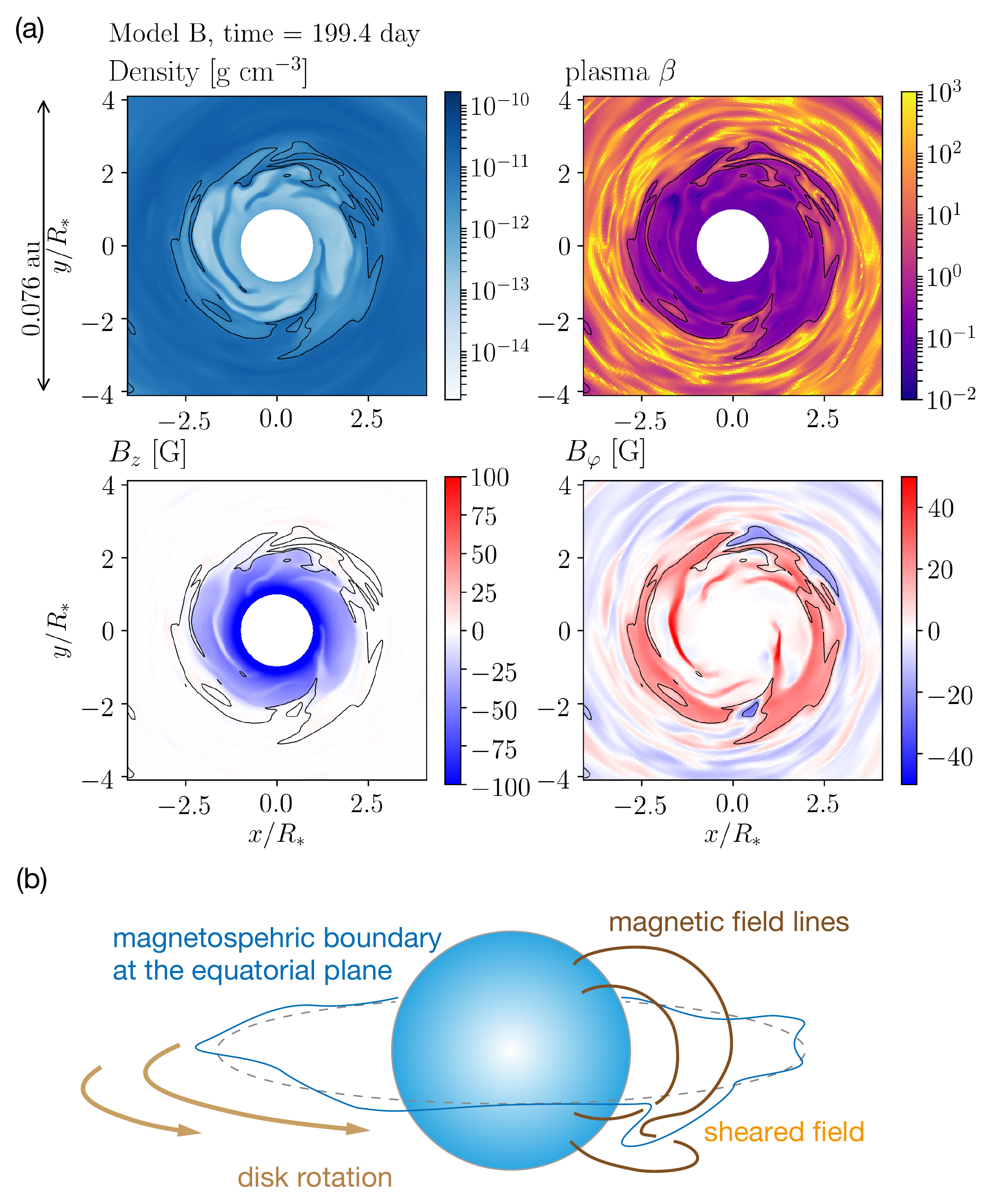}
    \caption{(a) The equatorial plane distributions of the density (top left), the plasma $\beta$ (top right), $B_z$ (bottom left), and $B_{\varphi}$ (bottom right). The locations of $\beta=1$ is indicated by the solid lines. The data is for Model B. (b) Schematic diagram about the toroidal field generation after a part of the magnetospheric field is extruded toward the disk.}
    \label{fig:midplane_bfield_amplification}
\end{figure*}

Our simulations show that the ratio $|B_\varphi/B_z|$ at the magnetospheric boundary is close to unity at the equator regardless of the stellar spin in the parameter range investigated. Figure~\ref{fig:midplane_bfield_amplification} describes the generation process of the toroidal field. Figure~\ref{fig:midplane_bfield_amplification}(a) shows the equatorial plane distributions of the density, the plasma $\beta$, $B_z$, and $B_\varphi$ of Model B. 
The low-density and low-$\beta$ regions denote magnetospheric plasma.
The figures show that a part of the magnetospheric plasma is extruded to the disk via the interchange instability. In the extruded regions, the vertical field is weakened, but the toroidal field is amplified. This result indicates the generation of the toroidal field from the poloidal field via the velocity shear. The panel (b) shows a schematic diagram about the toroidal field generation. We note that the plasma $\beta$ in the regions with a strong toroidal field is close to or smaller than unity.

The magnetic field structure seems to be regulated by the force balance.
The amplified toroidal field around the magnetosphere produces the magnetic tension force toward the center (so-called the hoop stress). The magnetic pressure gradient force of the poloidal field should balance with the tension force when the plasma $\beta$ is close to or smaller than unity (if $\beta\gg 1$, the gas pressure can be important). As $\beta\lesssim 1$ around the magnetospheric radius, the force balance results in the condition that $\langle B_{\varphi}^2 \rangle\approx \langle B_{z}^2 \rangle$.

Our simulations indicate that the temporally and azimuthally averaged plasma $\beta$ for the amplified toroidal field does not become much smaller than unity and approximately remains $\mathcal{O}(1)$. This is a result of the balance between the field amplification and the field escape by magnetic buoyancy. When the plasma $\beta$ for the toroidal field is close to unity, the toroidal field can escape from the disk on the Keplerian orbital timescale because of magnetic buoyancy such as the Parker instability \citep{Paker1955ApJ,Paker1966ApJ}. Our simulations show rising magnetic flux bundles from the inner disk, suggesting that magnetic buoyancy plays a role. ST18 also reported such eruptions of buoyantly rising flux tubes. \citet{Wang_YM1987} argued that the strength of the toroidal field around the magnetospheric boundary will be limited by the effect of magnetic buoyancy \citep[see also][]{Campbell1992GApFD}. Our simulations support this idea.
The inflation of the twisted magnetospheric field may give a similar result \citep[e.g.][]{Lovelace1995MNRAS,Agapitou2000MNRAS,Matt&Pudritz2005ApJ}. However, we may have to reexamine the argument. As we will show in Section \ref{subsec:magnetic_twist}, the picture about the twisting of the stellar field needs to be improved.

\subsection{Rotation profile}\label{subsec:rotation}
\begin{figure} 
    \centering
    \includegraphics[width=0.9\columnwidth]{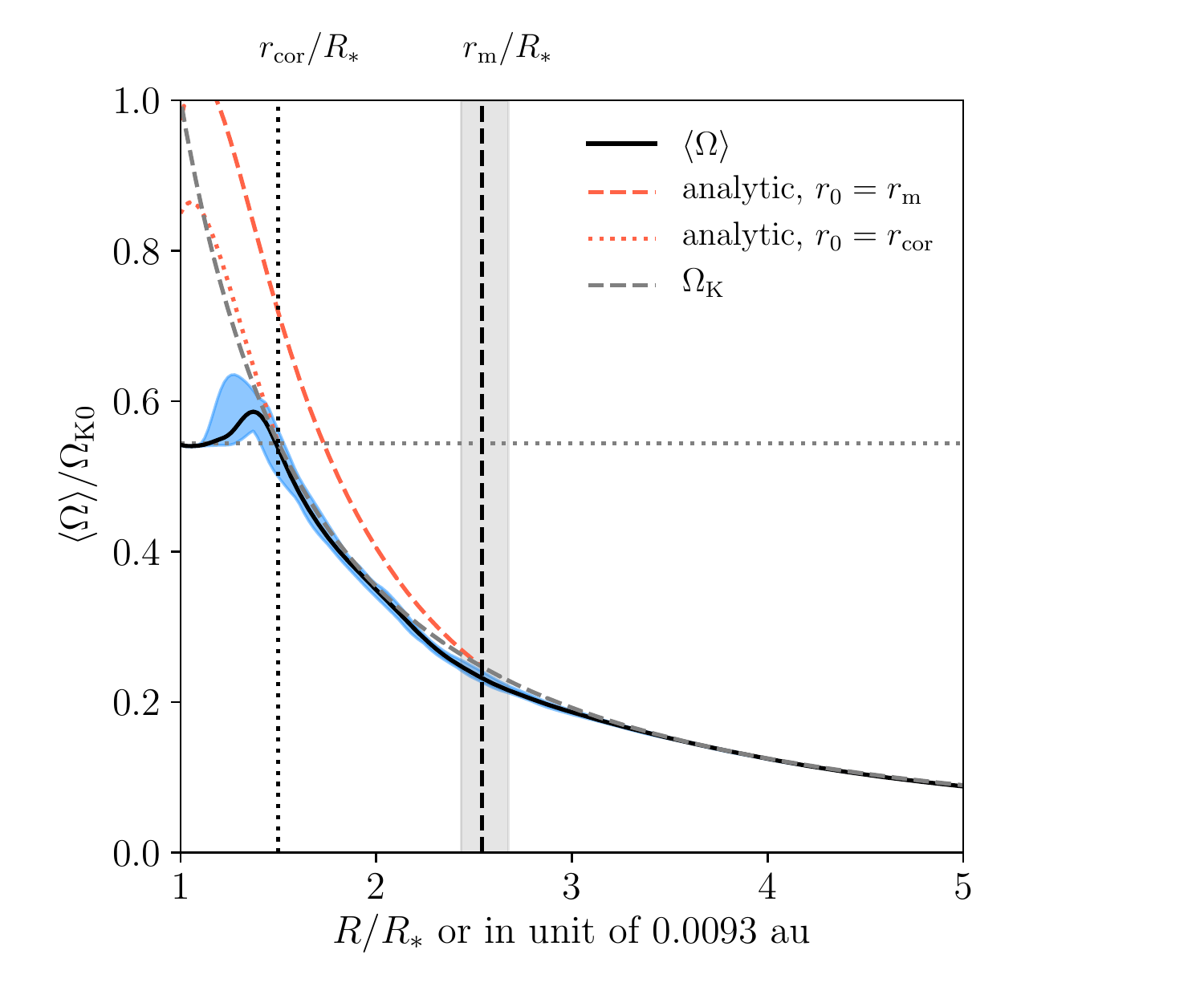}
    \includegraphics[width=0.9\columnwidth]{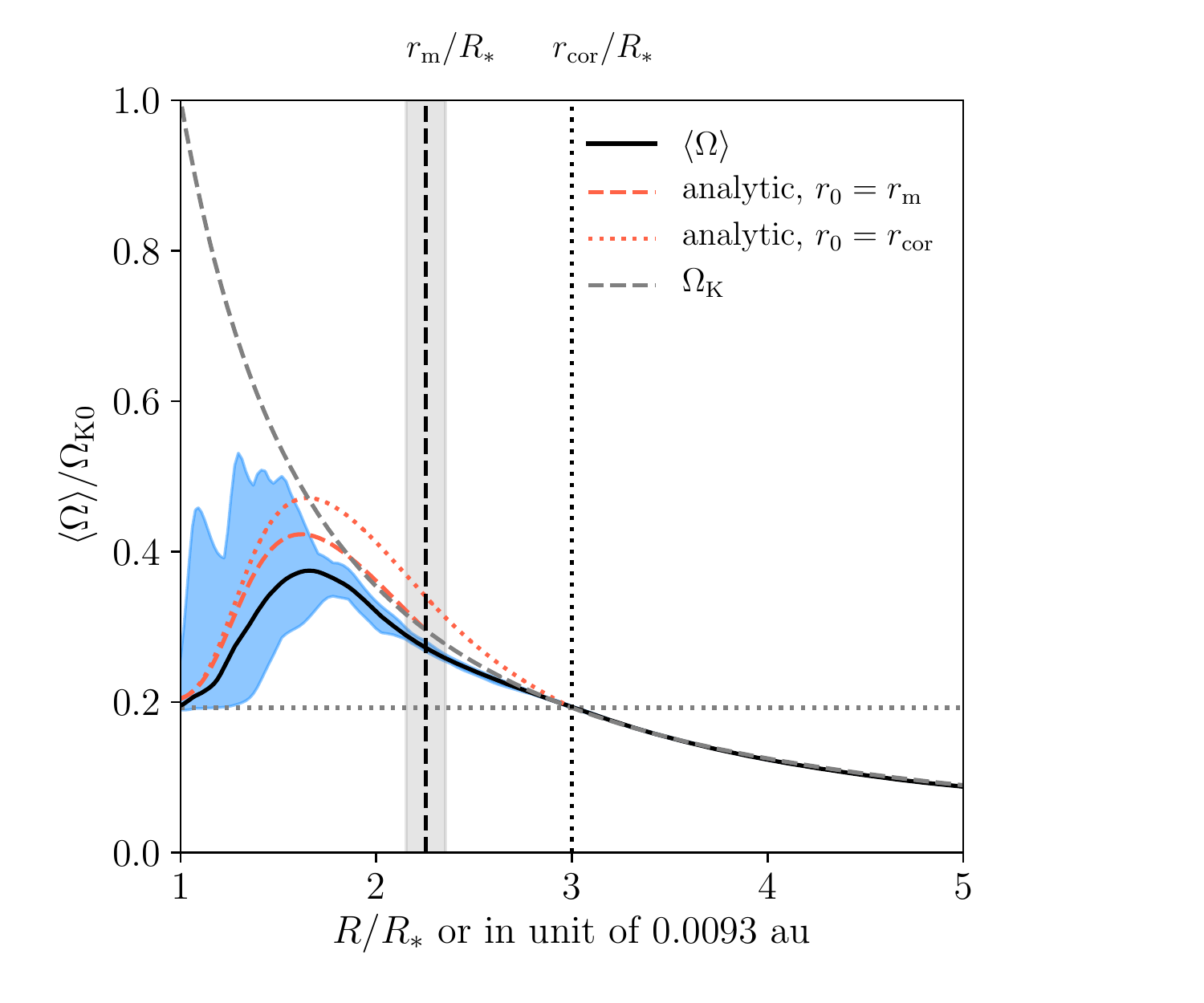}
    \includegraphics[width=0.9\columnwidth]{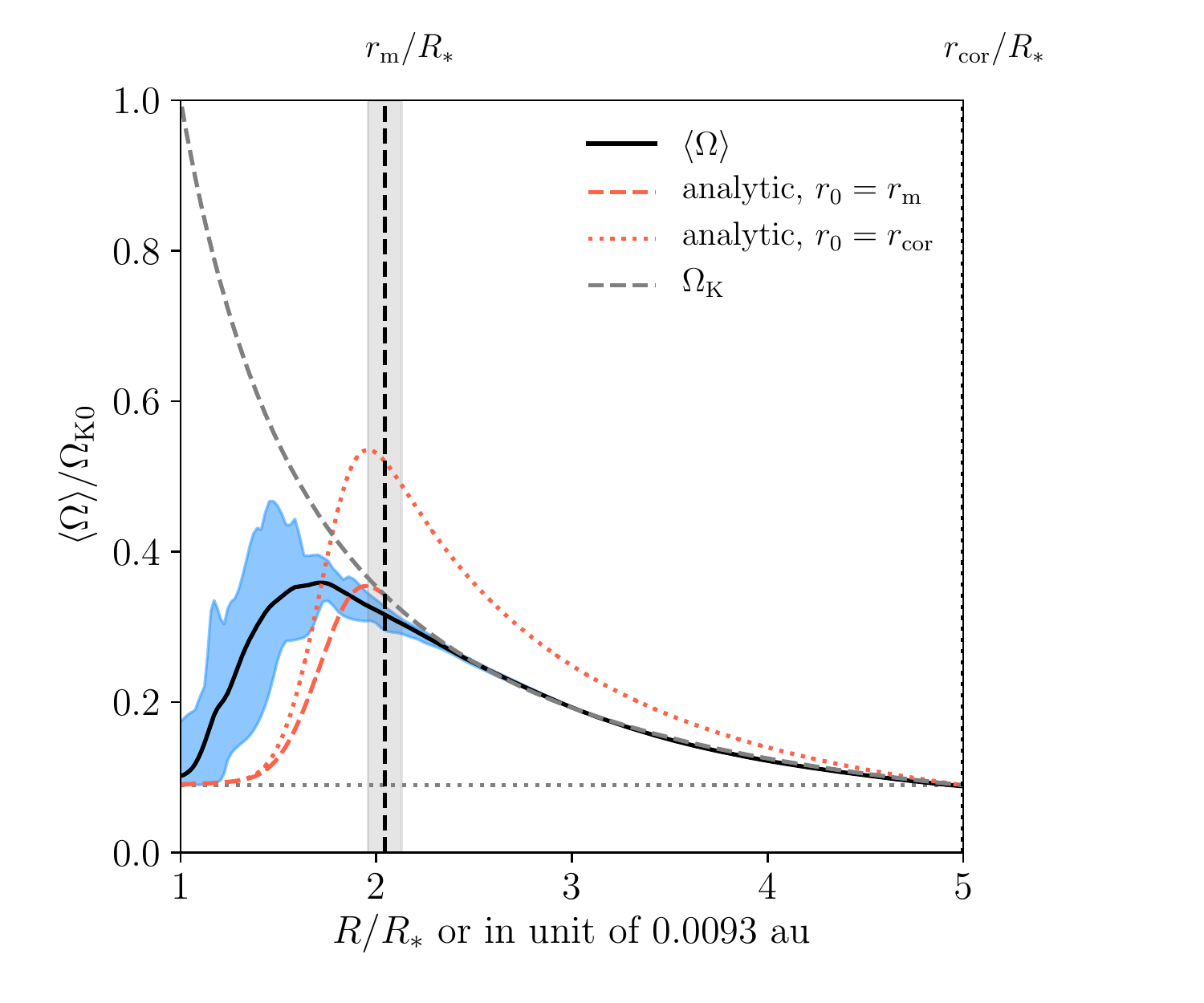}
    \caption{Angular velocity profiles at the midplane. The solid lines show the angular velocity. The variation range during the measurement period is shown with the blue bands. The gray dashed lines indicate the Keplerian angular velocity $\Omega_{\rm K}$. The red dashed and dotted lines denote the predictions of a previous theory (see text). These are normalized by $\Omega_{\rm K0}$. The horizontal dotted lines denote $\Omega_*/\Omega_{\rm K0}$. The vertical dashed lines denote $r_{\rm m}/R_{*}$ during the period of $t=$190.1-199.4~day. The gray bands indicate the ranges between the minimum and maximum values of $r_{\rm m}/R_*$. The vertical dotted lines show $r_{\rm cor}/R_*$.}
    \label{fig:Omega}
\end{figure}

It has been often assumed that the angular velocity within the magnetospheric radius is equal to the stellar spin. However, our results demonstrate that the gas penetrating into the magnetosphere forces to rotate the magnetosphere nearly at the Keplerian velocity. Figure \ref{fig:Omega} shows the angular velocity profiles at the midplane for the three models. 
We find a non-negligible (several \%) deviation of $\langle \Omega \rangle$ from the Keplerian profile around $R=r_{\rm m}$. The increase in the variation amplitude (indicated by the blue bands) is also prominent within this radius.
These results indicate that the definition of the magnetospheric radius based on the plasma $\beta$ is reasonable.

For Model A, the deviation from the Keplerian profile is less prominent within $r_{\rm cor}<R<r_{\rm m}$ than those in the other two models, although we do see a small deviation. The reason why the deviation is small is related to the presence of the spiral density structure (see the density map of Figure \ref{fig:interchange}). This pattern starts to form around $R=r_{\rm m}$ as a result of the instability and is spiraling toward the center. This high-density spiral is less susceptible from the magnetic braking due to its large inertia. As a result, the spiral pattern is found within $R\approx 1.5$-$2.5 R_*$, which produces a nearly Keplerian profile down to $R=1.5R_*$. Even though the rotation profile is near Keplerian, the disk structure of $R\approx 1.5$-$2.5 R_*$ is largely modified by the magnetic field to produce such a spiral pattern. As shown in the density map of Figure \ref{fig:interchange}, the low-density coronal region extends outside $R=2R_*$. From these reasons, we consider that the estimated $r_{\rm m}$ for Model A ($\approx 2.5 R_*$) is reasonable.

In all the models, a large volume of the magnetosphere rotates nearly at the Keplerian velocity. Unlike the classical expectation, we find no clear jump in the angular velocity at the magnetospheric radius \citep[compare our numerical result with the schematic diagram in Figure 4 of][]{Romanova2015SSRv}. That is, the spin transition occurs over a large width ($\sim R_*$). This result indicates that around the magnetospheric radius, the accreting gas virtually maintains the magnetospheric spin at the Keplerian value, regardless of the actual stellar spin.

The sophisticated analytic model by \citet{Kluzniak2007ApJ} also reproduces a smooth transition of the angular velocity.
Although the angular velocity profiles of our simulations are qualitatively similar to their results, quantitative comparison reveals significant differences. 
They consider a height-averaged angular momentum equation. 
Their model assumes that the viscous torque or the $r\varphi$ component of the Maxwell stress can be ignored within the radius ($r_0$ according to their notation) where the rotation profile starts to deviate from the Keplerian one. 
In addition, the model adopts an ad hoc prescription about the magnetic twist. 
Under the assumption that the magnetic twist is produced as a result that the stellar magnetic field is sheared by the deferentially rotating disk, the model adopts one of the following ad hoc analytic prescriptions \citep[see also e.g.][]{Livio1992MNRAS,Wang1995ApJ}:
\begin{align}
    (\langle B_\varphi \rangle / \langle B_z \rangle)_{\rm s}=1-\frac{\langle \Omega \rangle}{\Omega_*},\label{eq:twist_an1}
\end{align}
or
\begin{align}
    (\langle B_\varphi \rangle / \langle B_z \rangle)_{\rm s}=\frac{\Omega_*}{\langle \Omega \rangle}-1,\label{eq:twist_an2}
\end{align}
where the subscript $s$ denotes that the value is measured at the disk surface. 
Note that the magnetic twist is zero at the corotation radius in both expressions.
We examine both functional forms, but the function of Equation~(\ref{eq:twist_an1}) is mainly focused, as in \citet{Kluzniak2007ApJ}.

Using the above two simplifications, \citet{Kluzniak2007ApJ} obtained the following equation:
\begin{align}
    \frac{\dot{M}}{R}\frac{dl}{dR}=-B_z^2 R \left( 1 - \frac{l}{\Omega_* R^2}\right),\label{eq:kluzniak}
\end{align}
where $l=\Omega(R)R^2$ is the specific angular momentum of the matter flow. From a mathematical point of view, neglecting the viscous torque allows us to easily solve the equation. Solving this equation inside $R=r_0$ with the boundary condition that $l(r_0)=\Omega_{\rm K}(r_0)r_0^2$ gives the angular velocity profile. Using the numerically obtained $\dot{M}$ and $B_z (= \langle B_z \rangle)$, we numerically integrate Equation (\ref{eq:kluzniak}). The solutions with $r_0=r_{\rm m}$ and $r_{0}=r_{\rm cor}$ are both examined. In all the models, we adopt $\dot{M}=10^{-8}~M_\odot~{\rm yr^{-1}}$ (Figure \ref{fig:mdot_angmomdot} will show that this value is reasonable).

The red dashed and dotted lines in Figure \ref{fig:Omega} denote the analytic predictions for the cases with $r_0=r_{\rm m}$ and $r_{0}=r_{\rm cor}$, respectively. For Model B, the analytic solution with $r_{\rm 0}=r_{\rm m}$ may be similar to the time-averaged numerical solution. However, the analytic solutions of Model A and C significantly deviate from the numerical solutions. The analytic solutions of Model A do not show the transition to the stellar spin. For Model C, the peak locations of the analytic solutions are shifted from the peak of the numerical solution.

We find the breakdown of the following assumptions in the theory: 1. the magnetic twist around the magnetospheric radius is generated via the twisting of the stellar magnetic field, and 2. the $r\varphi$ component of the Maxwell stress can be ignored around the magnetosphere. In Section \ref{subsec:magnetic_twist}, we will analyze the magnetic twist at the disk surfaces to clarify the first point.
In Sections \ref{subsec:ang} and \ref{subsec:ang_flux}, we will demonstrate that the $r\varphi$ component of the Maxwell stress is important. The main reason why the analytic solutions systematically predict larger angular velocities seems to be because the theory ignores this important stress component.

\subsection{Magnetic twist at the disk surfaces}\label{subsec:magnetic_twist}
In Section \ref{subsec:Bfield_magnetosphere}, we have seen the profile of the magnetic field at the equator. It is also important to investigate $\langle B_\varphi\rangle /\langle B_z\rangle$ (the magnetic twist) at the disk surfaces, as it determines the magnetic torque exerting on the disk. 
\footnote{The magnetic twist in the main text is defined by the coherent components of the magnetic field. If we define it by using the magnetic torque in 3D, we can also evaluate it as $\langle B_\varphi B_z\rangle / \langle B_z \rangle^2$. However, we confirmed that the magnetic twists based on the two definitions are very similar inside the magnetospheres, although $\langle B_\varphi B_z\rangle / \langle B_z \rangle^2$ gives smoother profiles away from the magnetospheric boundaries. Therefore, we only show $\langle B_\varphi\rangle /\langle B_z\rangle$.}
As described in Section \ref{subsec:rotation}, previous models often adopt the ad hoc prescription of the magnetic twist, Equations (\ref{eq:twist_an1}) and (\ref{eq:twist_an2}) \citep[e.g.][]{Livio1992MNRAS,Wang1995ApJ,Kluzniak2007ApJ}.
The validity of the prescriptions should be investigated.

\begin{figure*} 
    \centering
    \includegraphics[width=0.9\columnwidth]{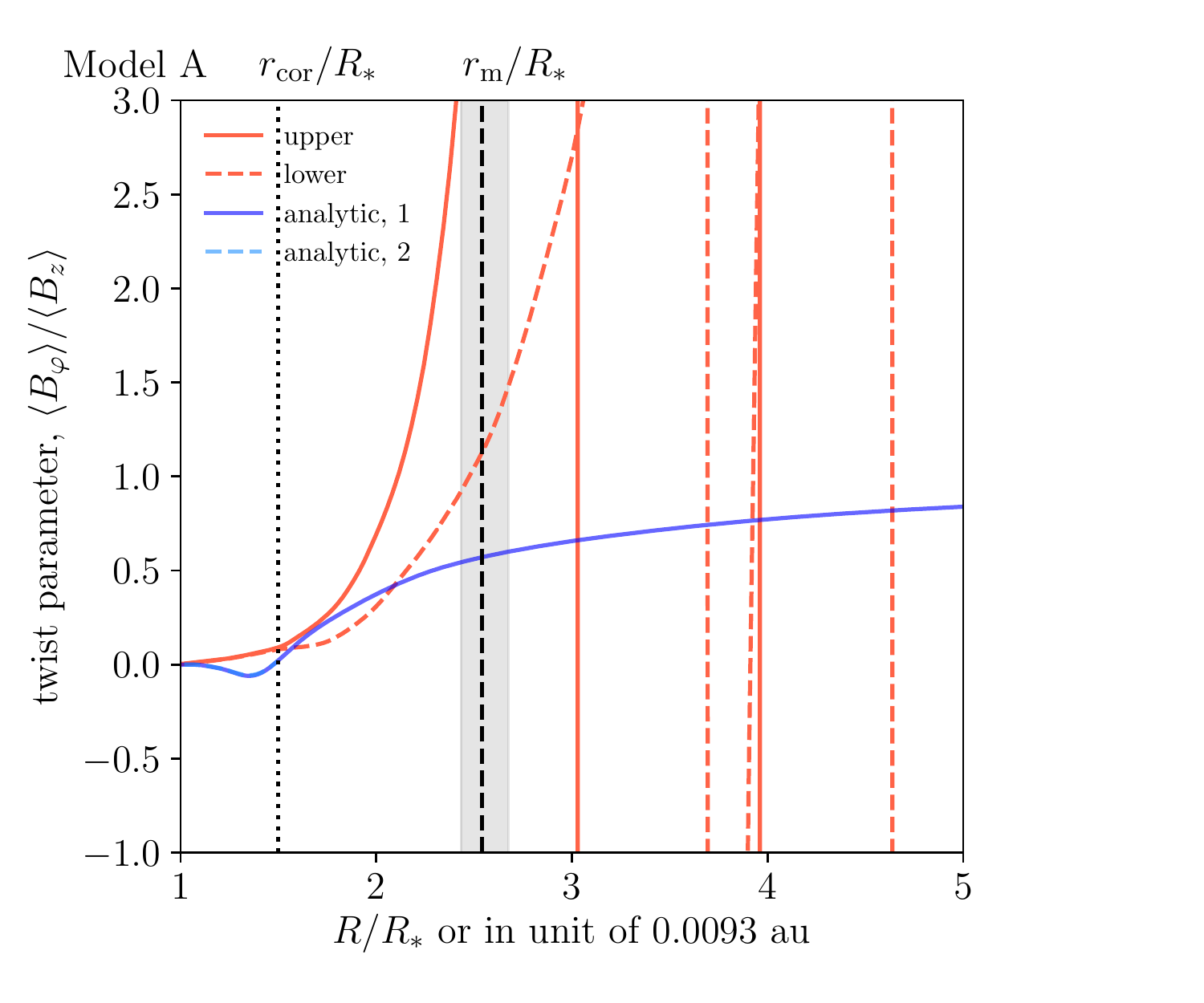}
    \includegraphics[width=0.9\columnwidth]{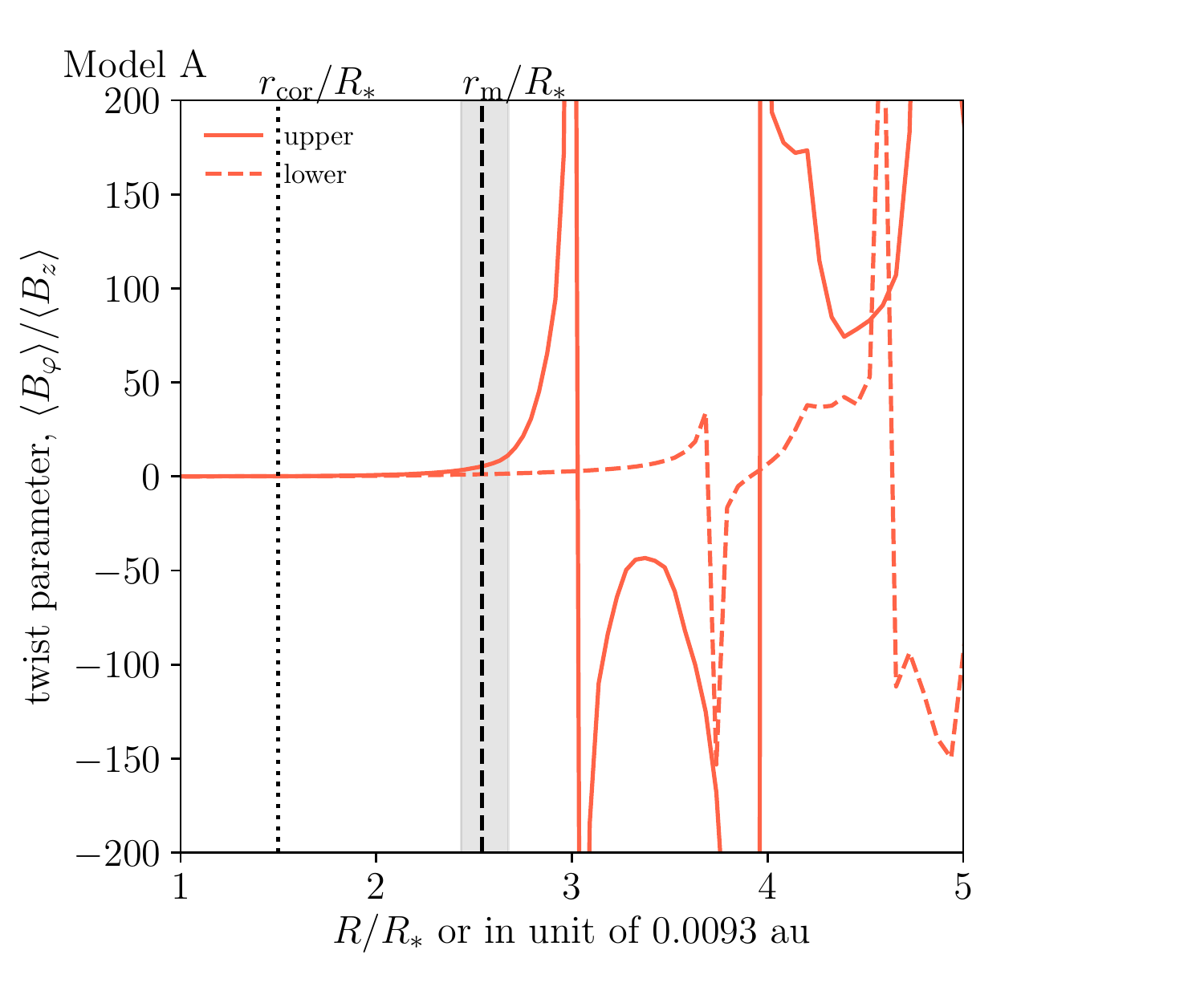}
    \includegraphics[width=0.9\columnwidth]{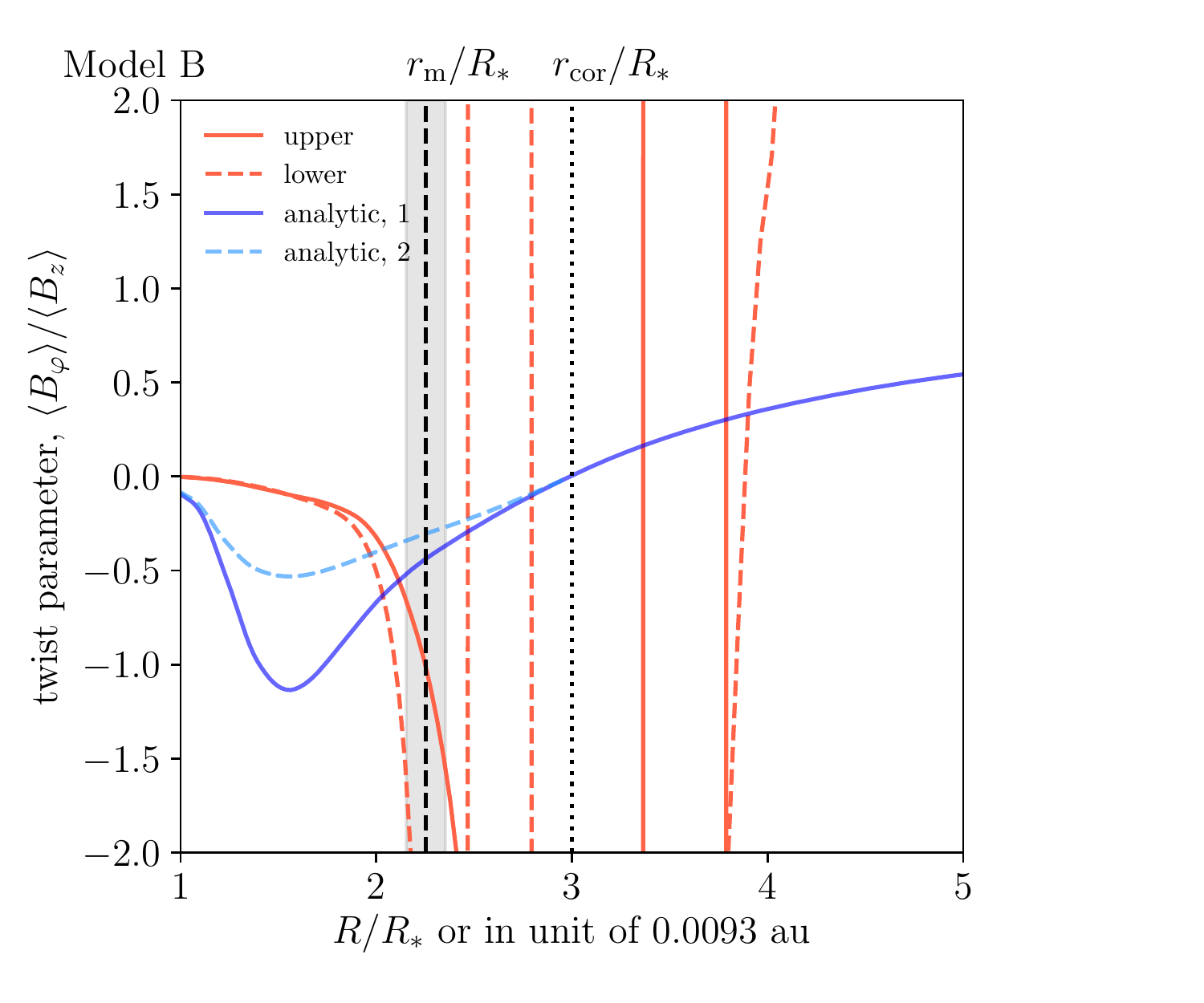}
    \includegraphics[width=0.9\columnwidth]{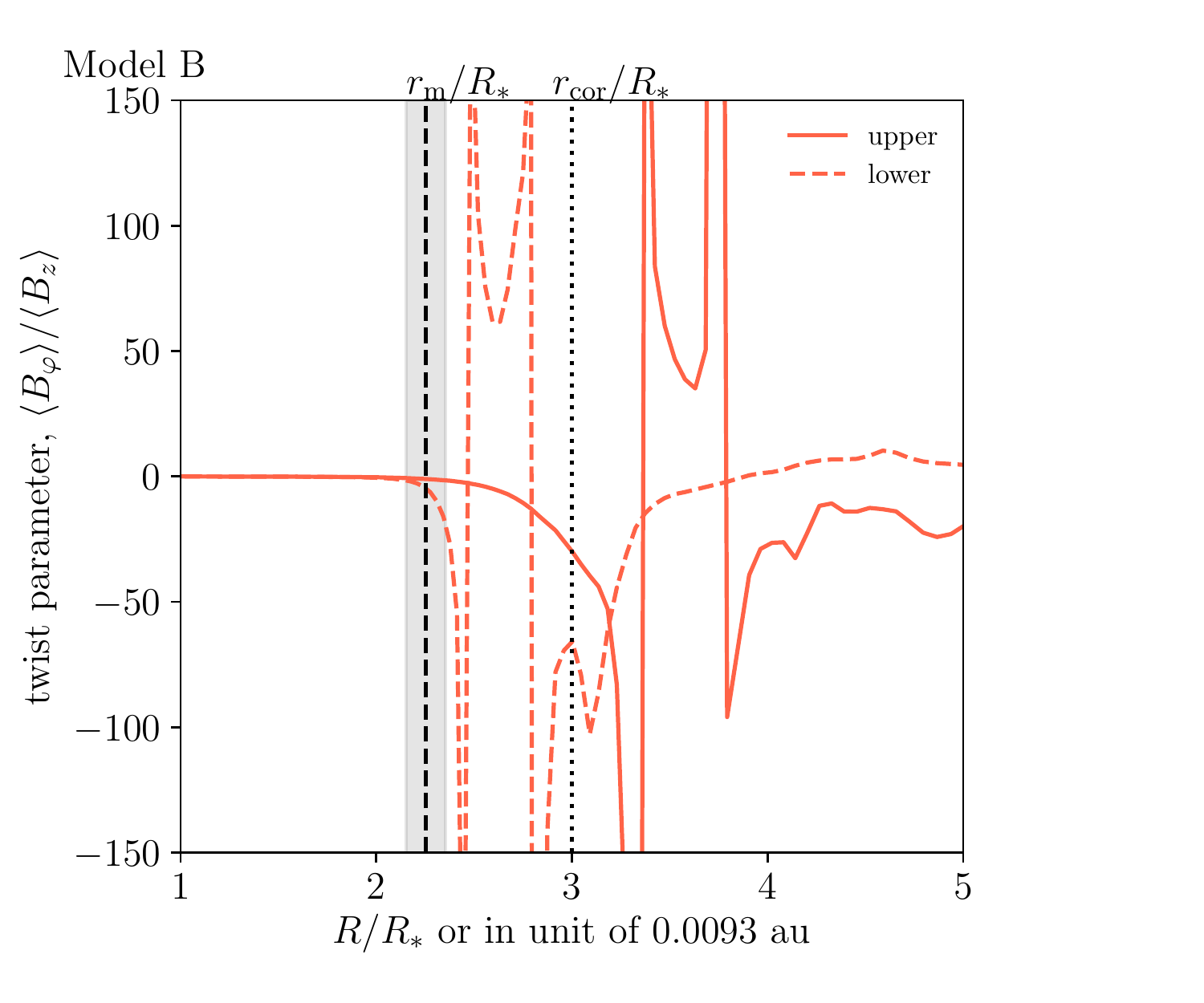}
    \includegraphics[width=0.9\columnwidth]{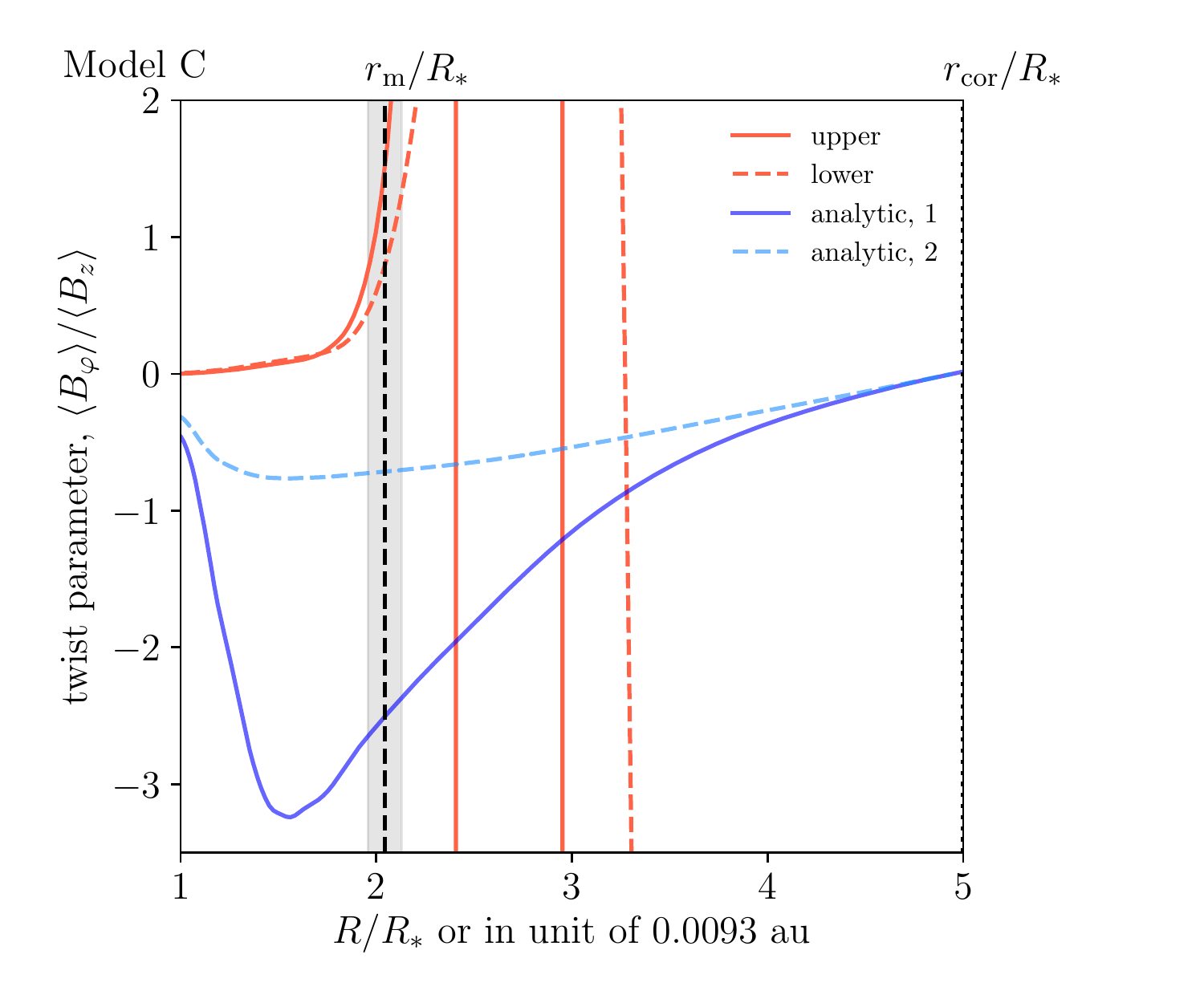}
    \includegraphics[width=0.9\columnwidth]{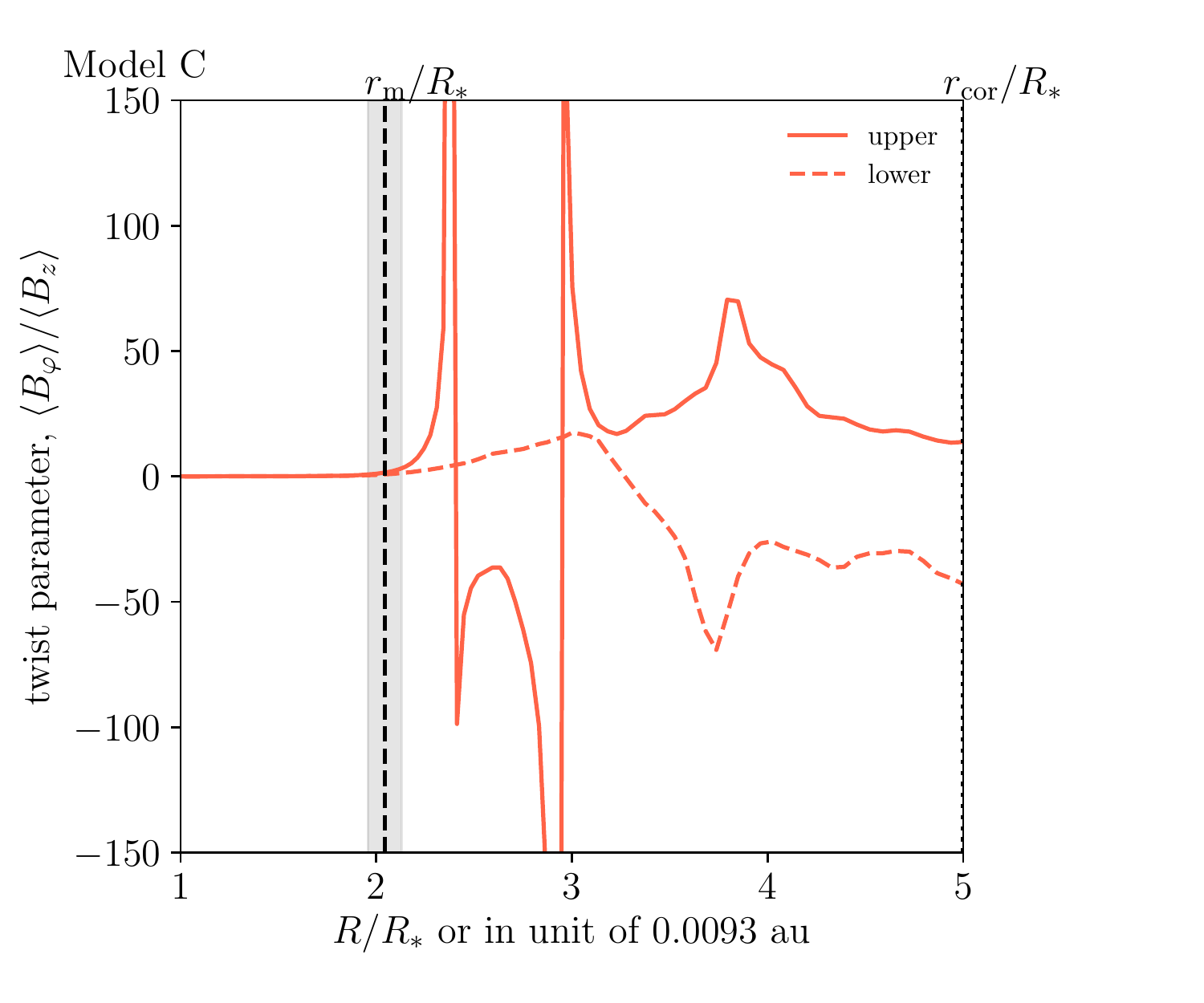}
    \caption{The radial profiles of the magnetic twist, $\langle B_\varphi\rangle / \langle B_z\rangle$, measured at the disk surfaces. The left and right panels show the same radial range, but the vertical ranges are different. The red solid and dashed lines respectively indicate the data for the upper and lower disk surfaces. The time averaged data during the period of $t=157.6$-$204.0$ day are used.
    The vertical dashed lines denote $r_{\rm m}/R_{*}$ during the period of $t=$190.1-199.4~day. The gray bands indicate the ranges between the minimum and maximum values of $r_{\rm m}/R_*$. The vertical dotted lines show $r_{\rm cor}/R_*$. 
    The solid and dashed blue lines in the left panels denote the analytical functions used in previous studies (Equations (\ref{eq:twist_an1}) and (\ref{eq:twist_an2}), respectively).}
    \label{fig:twist}
\end{figure*}

The disk surfaces are characterized by the disk opening angle from the equatorial plane, $\Delta \theta_{\rm d}=\arctan{(H/R)}$, where $H$ is the gas pressure scale height. $\Delta \theta_{\rm d}\approx 8^\circ$.
The measured twist is temporally averaged over the period of $t=157.6$-$204.0$ day. We varied the period for the time average, but the general structures of the solutions are unchanged.

Figure~\ref{fig:twist} displays the radial profiles of the magnetic twist measured at both disk surfaces (red dashed and dotted lines). Note the difference in the vertical range between the left and right panels. The left panels compare the numerical solutions and the analytic prescriptions.
The results of Equations (\ref{eq:twist_an1}) and (\ref{eq:twist_an2}) are shown as the solid and dashed blue lines, respectively (they can be compared with the numerical data measured at the upper surface).
The magnetic twist is of the order of unity around $R=r_{\rm m}$ if we average the values at the upper and lower disk surfaces.
We recall that a very similar result holds for the field around the midplane (Section \ref{subsec:Bfield_magnetosphere}).
However, the magnetic twist at the disk surfaces diverges just outside it because $\langle B_z \rangle_{\rm s}$ approaches to zero. 
The analytic functions largely deviate from the numerical solutions. The analytic functions are zero at $r=r_{\rm cor}$, but the numerical solutions are not. Ignoring the sign of the twist, we also find that the twist inside the magnetosphere in Model B and C is much smaller than the analytic prediction, which indicates the stellar field is almost not twisted at all. These results suggest that the simple assumption about the generation of the magnetic twist is invalid. Considering the above results, we question the picture that the ratio $|B_\varphi/B_z|$ is regulated by the inflation of the twisted stellar field.

The sign of the twist in the magnetosphere is explained as follows. In all the models, $B_z <0$ inside the magnetospheres around the equatorial plane. In Model A and C, magnetospheric accretion occurs mainly in the northern hemisphere. As the accreting flows increase their rotation speed during the infall, they twist up the magnetosphere and develop the toroidal field with the negative sign ($B_\varphi < 0$). Therefore, $\langle B_\varphi\rangle/\langle B_z\rangle > 0$ inside the magnetospheres in Model A and C. On the other hand, magnetospheric accretion occurs in the southern hemisphere in Model B, which results in the generation of a positive $B_\varphi$. For this reason, $\langle B_\varphi\rangle/\langle B_z\rangle < 0$ inside the magnetosphere in Model B. Note that we cannot discuss the stellar spin-up/down only using the magnetic twist measured near the equatorial plane. The spin-up/down torque will be analyzed in Section \ref{subsec:ang}.

We study the twist profiles in more detail. Looking at the right panels of Figure~\ref{fig:twist}, one will notice that the twist shows peaky structures and changes signs at different radii.
The peaky structures just outside $R=r_{\rm m}$ are formed at the boundary between the coherent magnetic field (stellar magnetosphere) and the turbulent disk field. Away from the magnetospheric boundary, the change in sign results from the disk dynamo and turbulence. $\langle B_z\rangle_{\rm s}$ largely fluctuates in response to the turbulence. In addition, $\langle B_\varphi \rangle_{\rm s}$ changes its sign as a result of the disk dynamo \citep[e.g.][]{Flock2011ApJ,Takasao_Tomida_Iwasaki_Suzuki_2018}. The combination of the two processes results in the complex twist profiles.

In previous analytical studies and 2D MHD simulations, the toroidal field is generated via the twisting of the stellar magnetic field. 
Therefore, the twist distribution is smooth in the radial direction. 
However, in our 3D simulations, the magnetic field amplification outside the magnetosphere is governed by the disk processes such as MRI turbulence. The strong toroidal field can be generated without significant twisting of the stellar magnetic field.
Many 3D simulations demonstrated that the toroidal field dominates the other components if the disk vertical field is sufficiently weak \citep[e.g.][]{Flock2011ApJ,Suzuki_etal_2014ApJ,Takasao_Tomida_Iwasaki_Suzuki_2018}. As the strong vertical field is confined approximately inside the magnetospheric radius (Figure~\ref{fig:Bfield}), the disk dynamo outside the magnetosphere should be similar to such a case.
Inside the magnetosphere, the magnetic twist is generated by patchy accretion flows (Figures \ref{fig:interchange} and \ref{fig:midplane_bfield_amplification}).
In previous theories, the angular momentum exchange between the star and the disk has been considered on the basis of a smooth twist profile. However, our results indicate that the previous picture needs to be updated.

It has been assumed that $|B_{\varphi}/B_z|$ at the magnetospheric boundary depends on the stellar spin \citep[e.g.][]{Ghosh_Lamb1979_paperIII,Wang_YM1987}.
\citet{Dangelo_Spurit2010MNRAS} adopted $|B_\varphi/B_z|=0.1$ as a fiducial value.
However, our simulations demonstrate that $|B_{\varphi}/B_z|\approx 1$ at the disk surfaces for all the three models, although Model A shows a larger value. In addition, $\sqrt{\langle B_{\varphi}^2 \rangle}\approx \sqrt{\langle B_{z}^2 \rangle}$ (or more specifically, $|\langle B_{\varphi}\rangle|\approx |\langle B_{z} \rangle|$) at the equatorial plane (Figure~\ref{fig:Bfield}). These results suggest that the ratio only weakly depends on the stellar spin.

\subsection{Velocity Profile of Conical Winds}\label{subsec:conical-wind}

\begin{figure}
    \centering
    \includegraphics[width=0.9\columnwidth]{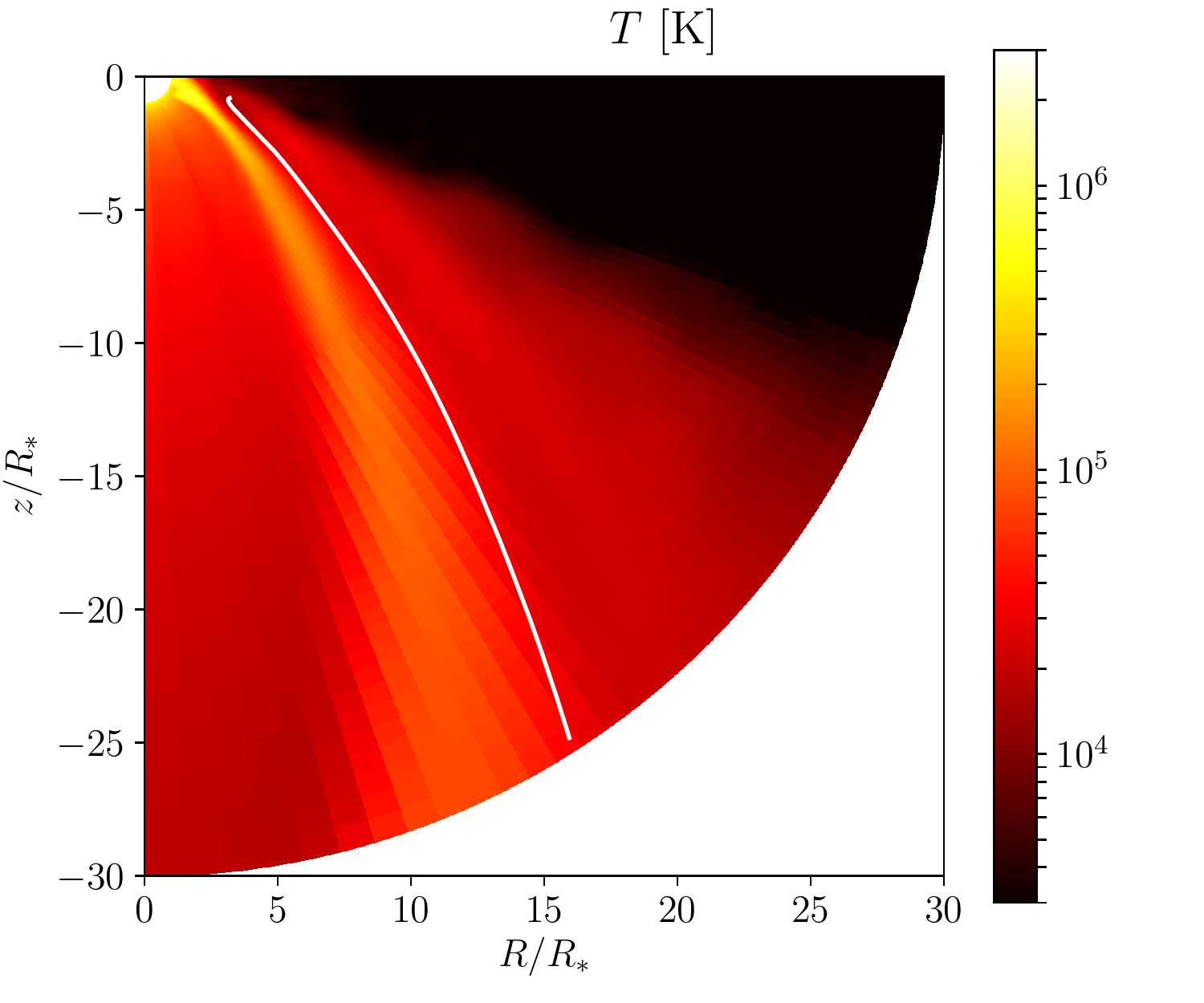}
    \includegraphics[width=0.9\columnwidth]{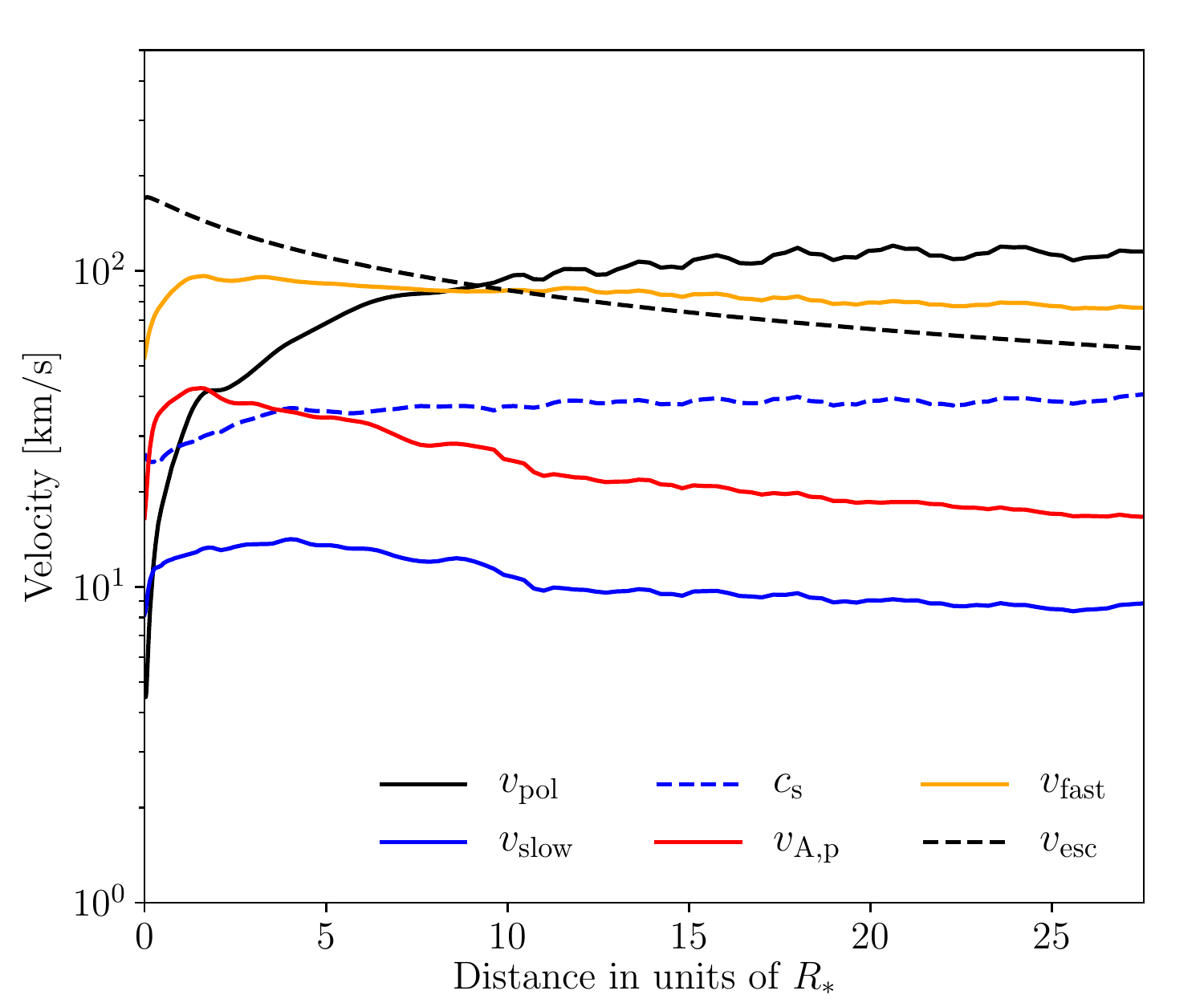}
    \caption{The analysis of the conical wind of Model A. The top panel shows the azimuthally and temporally averaged temperature. The white solid line is the streamline analyzed. The bottom panel displays the profiles of various speeds against the distance from the start point of the streamline, $(R,z)=(3.2R_*,-0.8R_*)$. The black solid line denotes the poloidal speed of the wind, which exceeds the local escape velocity (black dashed line) inside the numerical domain. The blue, red, and orange solid lines show the slow-mode, poloidal Alfv\'en, and the fast-mode speeds, respectively. The blue dashed line denotes the sound speed. The data are averaged over the period of $t=$162.3-166.9~day.}
    \label{fig:conical_wind_streamline}
\end{figure}

To study the acceleration of the conical winds, we measure the velocity profile along a streamline inside the conical wind. 
Figure \ref{fig:conical_wind_streamline} demonstrates an example of Model A. The analyzed streamline is indicated as the white line in the temperature map. The bottom panel shows the profiles of various speeds against the distance from the start point of the streamline. 
We find that the wind is accelerated through two steps: the rapid acceleration near the disk and the subsequent gradual acceleration. During the gradual acceleration, the wind velocity exceeds the local escape velocity (black dashed line). Therefore, this wind will escape from the stellar gravity. Both the density and the magnetic field strength decrease with distance. As a result, the poloidal Alfv\'en (red solid) and fast magnetosonic speeds (orange solid) do not significantly change along the streamline. 
The conical wind is accelerated by the magnetic pressure gradient force of the toroidal field (see also Figure~\ref{fig:conical_wind}), which means that the poloidal field is much weaker than the toroidal field. Indeed, the poloidal Alfv\'en speed is much smaller than the fast magnetosonic speed.

The wind passes through the fast magnetosonic point around the distance of $10R_*$. The fast magnetosonic speed at this point is approximately 60~km~s$^{-1}$. The wind speed finally reaches approximately $120$~km~s$^{-1}$ in the simulation domain.
This velocity is comparable to but smaller than the local escape velocity $v_{\rm esc}$ at the wind base ($\sim$200~km~s$^{-1}$). 
This is because the fast magnetosonic speed at the fast magnetosonic point is smaller than the wind-base escape velocity.

We see a similar two-step acceleration of the conical winds in the other models, although the conical winds exhibit transient behaviors and their terminal velocities are below the local escape velocity during some periods. 
The conical wind is mainly driven by the magnetic pressure gradient force of the toroidal fields, as argued in previous studies \citep[e.g.][]{Romanova2009MNRAS}. 
Therefore, the strength and the stability of the conical wind will depend on the toroidal field strength. 
Model A, whose stellar spin is highest, shows the most stable conical disk wind. The magnetospheric boundary in Model A is subject to the instability that produces the rotating spiral patterns, which efficiently increases the coherent toroidal field (Figures \ref{fig:Bfield} and \ref{fig:angular_momentum_flux_mlwd}).
We consider that the onset of the instability contributes to the production of the strong conical wind.
More detailed investigations about the wind acceleration will be given in our future papers.

\subsection{Mass and Angular Momentum Transfer}\label{subsec:ang}
We investigate the mass and angular momentum transfer processes around the star. We particularly focus on the following two points; 1. the driving mechanism of accretion at different radii and 2. the angular momentum extraction from and injection to the star. 

\begin{figure*}
    \centering
    \includegraphics[width=1.\columnwidth]{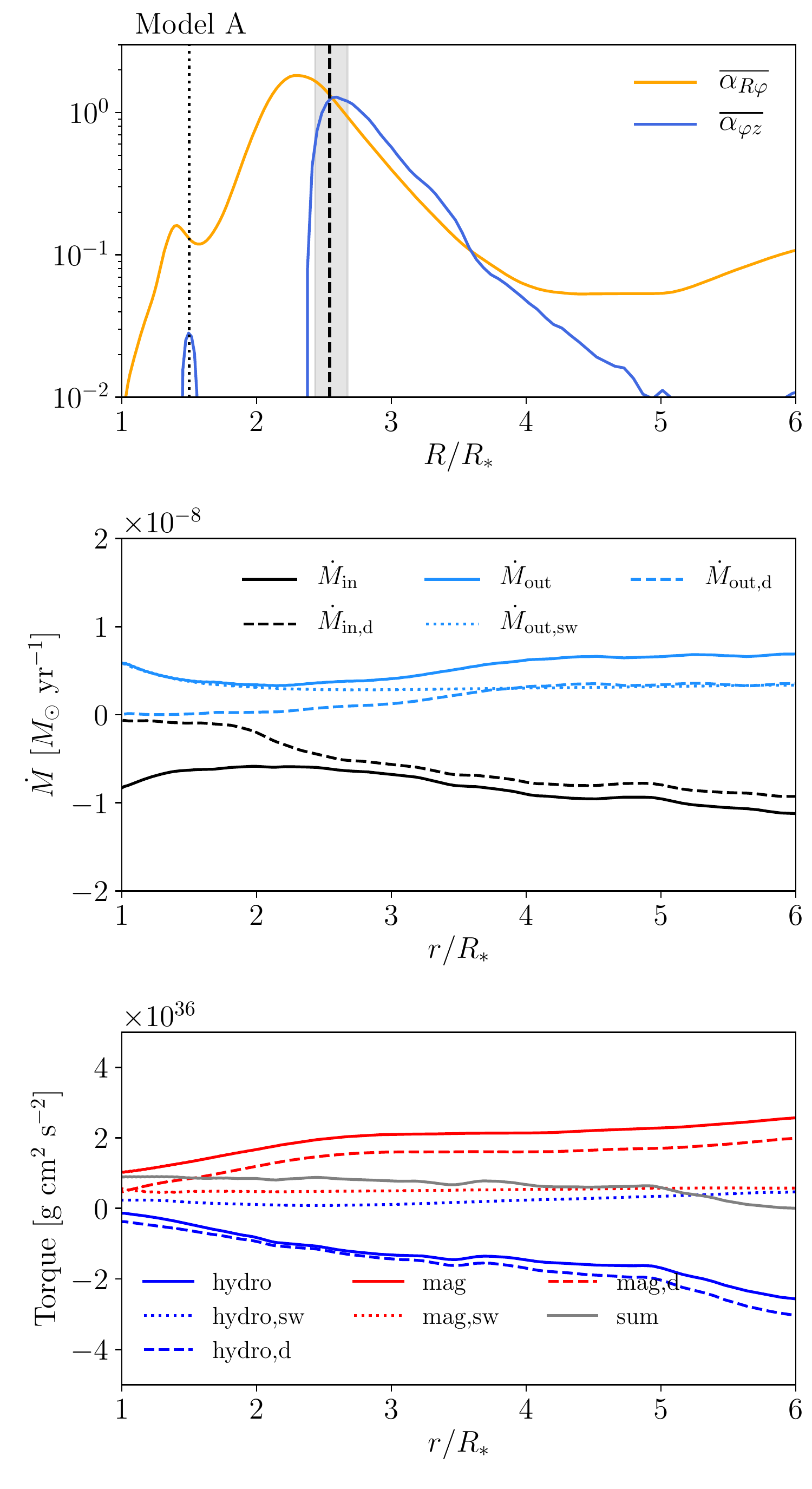}
    \includegraphics[width=1.\columnwidth]{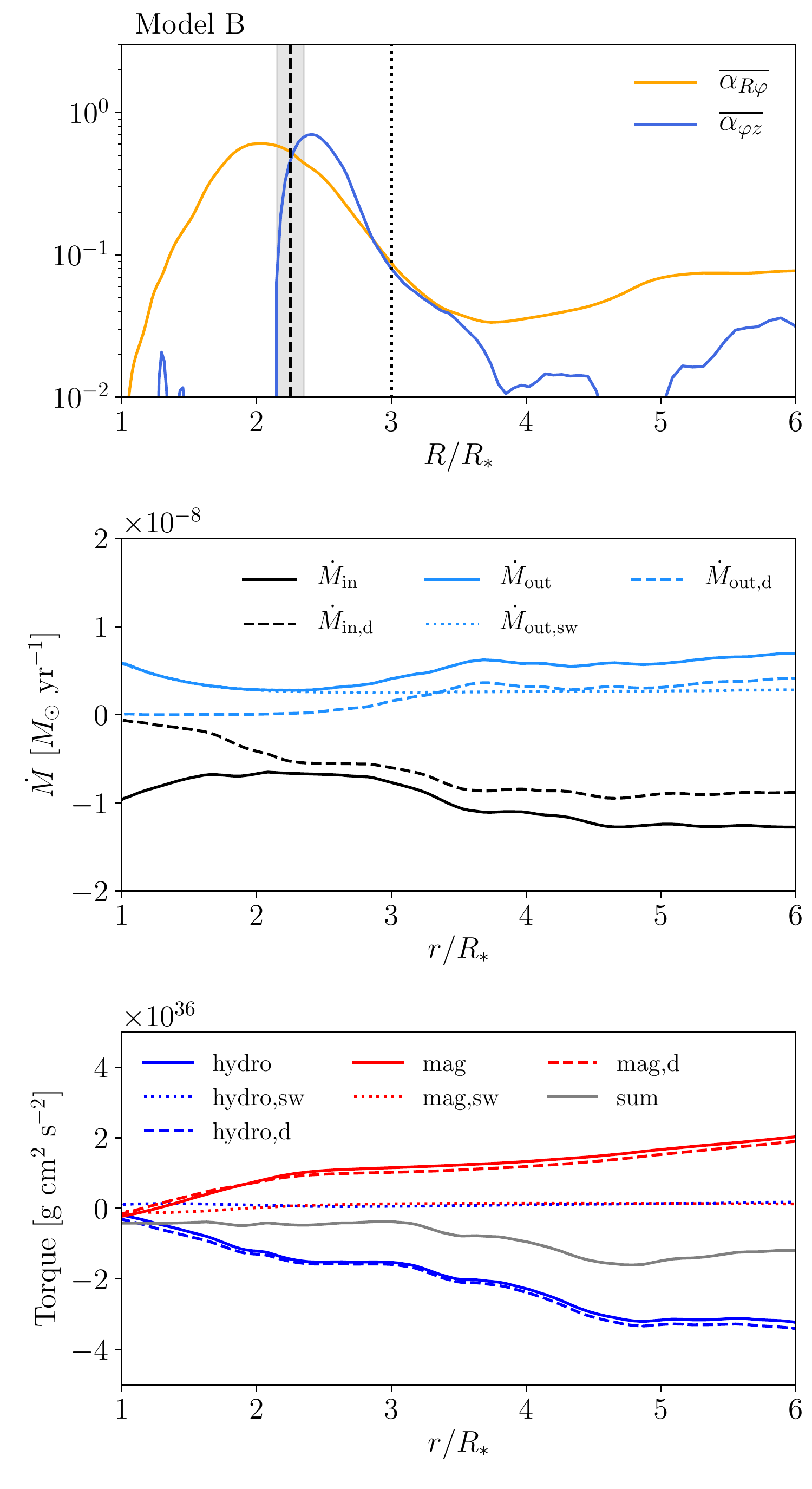}
    \caption{Top panels show the viscous parameters $\overline{\alpha_{m,R\varphi}}$ and $\overline{\alpha_{m,\varphi z}}$. Middle panels display the mass accretion and ejection rates in spherical coordinates. $\dot{M}_{\rm in}$ (black solid) and $\dot{M}_{\rm in,disk}$ (black dashed) show the total accretion rate and the accretion rate inside the disk, respectively. $\dot{M}_{\rm out}$ (blue solid), $\dot{M}_{\rm out,sw}$ (blue dotted), and $\dot{M}_{\rm out,d}$ (blue dashed) denote the total mass outflow rate, the mass outflow rate of the stellar wind, and the mass outflow rate of the disk gas including the disk wind. The bottom panels show the torque profile in spherical coordinates. The Reynolds and Maxwell stress contributions of the stellar wind are shown as the dotted blue and red lines, respectively. Those of the non-stellar wind gas (including the disk wind) are denoted as the dashed blue and red lines, respectively. The sums of the Reynolds and Maxwell stress contributions are shown as the solid blue and red lines, respectively. The gray solid line indicates the total torque.
    Left and right panels are for Model A and B, respectively. The data are averaged between $t=$190.1-199.4~day.
    In the top panels, the vertical dashed lines denote the time averaged values of the magnetospheric radii during the period of $t=$190.1-199.4~day. The gray bands indicate the ranges between the minimum and maximum values of the magnetospheric radii. The vertical dotted lines denote the locations of the corotation radii.}
    \label{fig:alpha_mdot_torque}
\end{figure*}

Top panels of Figure \ref{fig:alpha_mdot_torque} show the midplane radial profiles of the viscous parameters resulting from the Maxwell stress only.
We define the following two quantities:
\begin{align}
    \overline{\alpha_{m,R\varphi}}(R)&=\frac{\int_{-H(R)}^{H(R)} dz \rho (-B_R B_\varphi/(4\pi p))}{\int_{-H(R)}^{H(R)} \rho dz}\\
    \overline{\alpha_{m,\varphi z}}(R)&=\frac{\left[-B_\varphi B_z/4\pi \right]_{-H(R)}^{H(R)}}{p_{\rm mid}}
\end{align}
where $[Q]_{-H}^{H}\equiv Q(H)-Q(-H)$, where $H$ is the pressure scale height, and $H/R=0.14$ for the disk temperature of our models. $p_{\rm mid}=p_{\rm mid}(R)$ denotes the gas pressure at the equatorial plane. The definitions of $\overline{\alpha_{m,R\varphi}}$ and $\overline{\alpha_{m,\varphi z}}$ are the same as those for the Maxwell stress in \citet{Suzuki_et_al2016A&A}, where the $1+1$D ($t$-$R$) MHD disk model is formulated.

In Model A and B, $\overline{\alpha_{m,R\varphi}}$ and $\overline{\alpha_{m,\varphi z}}$ take their peaks around the magnetospheric radii. In each model, the peak of $\overline{\alpha_{m,\varphi z}}$ is located at a larger distance from the center than the peak of $\overline{\alpha_{m,R\varphi}}$. The local enhancement of the viscous parameters is a result of the magnetic field amplification around the magnetospheric boundary (see Figures \ref{fig:Bfield} and \ref{fig:midplane_bfield_amplification}).
The large $\overline{\alpha_{m,\varphi z}}$ around the disk-magnetosphere boundary indicates an efficient angular momentum extraction by the vertical magnetic field associated with the disk wind.

According to the $1+1$D disk model \citep{Suzuki_et_al2016A&A,Tabone2022MNRAS}, the accretion rates by the disk turbulence ($\dot{M}_{\rm acc}^{\rm visc}$) and by the disk wind ($\dot{M}_{\rm acc}^{\rm DW}$) can be respectively expressed as
\begin{align}
    \dot{M}_{\rm acc}^{\rm visc}(R)\sim \overline{\alpha_{m,R\varphi}} \frac{\Sigma c_s^2}{\Omega_{\rm K}} ,\\
    \dot{M}_{\rm acc}^{\rm DW}(R)\sim \overline{\alpha_{m,\varphi z}} \frac{R}{H}\frac{\Sigma c_s^2}{\Omega_{\rm K}}
\end{align}
(note the factor of $R/H (> 1)$ in $\dot{M}_{\rm acc}^{\rm DW}$). $\Sigma(R) = \int_{-H}^{H}\rho dz$ is the disk surface density.
Around the disk-magnetosphere boundary, $(R/H)\overline{\alpha_{m,\varphi z}}$ is larger than $\overline{\alpha_{m,R\varphi}}$. Therefore, accretion is mainly driven by the magnetic torque associated with the disk wind. In the outer region, the MRI turbulence dominates the wind contribution, suggesting that the major driver of accretion is MRI turbulence. We also note that $\overline{\alpha_{m,R\varphi}}$ remains larger than 0.1 in a large range of the magnetosphere, demonstrating that the $R\varphi$ component of the Maxwell stress cannot be ignored (see Section \ref{subsec:rotation}).

In previous 2D models, the effective kinematic viscosity is given as a constant against the radius and time, and the difference between $\overline{\alpha_{m,R\varphi}}$ and $\overline{\alpha_{m,\varphi z}}$ is ignored \citep[e.g.][]{Romanova2009MNRAS,Ustyugova2006ApJ,Zanni2013A&A}. 
However, $\overline{\alpha_{m,R\varphi}}$ and $\overline{\alpha_{m,\varphi z}}$ behave differently, because the importance of the disk wind varies with radius. 
As the magnetospheric boundary fluctuates in response to the instabilities (Section \ref{subsec:magnetosphere-boundary}), the Maxwell stress there is intrinsically time variable.
In addition, their values change by an order of magnitude within the width of $\sim R_*$, as a result of the local field amplification (Figures \ref{fig:Bfield} and \ref{fig:midplane_bfield_amplification}). Our results demonstrate the importance of 3D modeling.

We investigate the radial profiles of the mass accretion and outflow rates measured in spherical coordinates. 
We calculate the total accretion and outflow rates ($\dot{M}_{\rm in}$ and $\dot{M}_{\rm out}$, respectively) as follows:
\begin{align}
    \dot{M}_{\rm in}(r)&=\int_{0}^{\pi/2}\langle \rho v_r\rangle_{-}2\pi r^2 \sin\theta d\theta,\\
    \dot{M}_{\rm out}(r)&=\int_{0}^{\pi/2}\langle \rho v_r\rangle_{+}2\pi r^2 \sin\theta d\theta,
\end{align}
where the subscript $+$ and $-$ indicates the regions with the positive and negative $v_r$, respectively. We also separately measure the accretion rate within the disk $\dot{M}_{\rm in,d}$, the outflow rate of the stellar wind $\dot{M}_{\rm out,sw}$, and the outflow rate of the disk gas $\dot{M}_{\rm out,d}$. These are defined as follows:
\begin{align}
    \dot{M}_{\rm in,d}(r)&=\int_{\pi/2-\Delta \theta_{\rm d}}^{\pi/2+\Delta \theta_{\rm d}}\langle \rho v_r\rangle_{-}2\pi r^2 \sin\theta d\theta\\
    \dot{M}_{\rm out,sw}(r)&=\int_{0}^{\pi/2}\langle \rho v_r\rangle_{\rm sw}2\pi r^2 \sin\theta d\theta\\
    \dot{M}_{\rm out,d}(r)&=\dot{M}_{\rm out}-\dot{M}_{\rm out,sw}.
\end{align}
The subscript ``sw" indicates the values for the stellar wind regions. We define the stellar wind as the outflowing gas ($v_r>0$) with a specific entropy larger than a threshold.
As the specific entropy significantly differs between the stellar wind and the disk gas within several stellar radii, the results shown here are insensitive to the choice of the threshold value. 
The non-stellar-wind region is defined as the disk gas including the disk wind (denoted by the subscript ``d"). We note that the specific entropy of the disk winds increases as they propagate probably because of the turbulent mixing. This makes it difficult to define the disk-origin gas far away from the star.

The middle panels of Figure \ref{fig:alpha_mdot_torque} display the radial profiles of the mass accretion and outflow rates.
The black solid lines show $\dot{M}_{\rm in}$, and the black dashed lines indicate $\dot{M}_{\rm in,d}$. The difference between the two denotes the contribution of the disk surface accretion or coronal accretion including the failed disk wind \citep[e.g.][]{Takasao_Tomida_Iwasaki_Suzuki_2018,Zhu_Stone2018ApJ,Jacquemin-Ide2021A&A}. Our result demonstrates that the contribution of the coronal accretion is approximately a few 10\%. As magnetospheric accretion occurs within the radius of $\sim 2.5 R_*$, a large amount of the disk gas is lifted from the equatorial plane, which results in a reduction in $\dot{M}_{\rm in,d}$.

\begin{figure*}
    \centering
    \includegraphics[width=0.67\columnwidth]{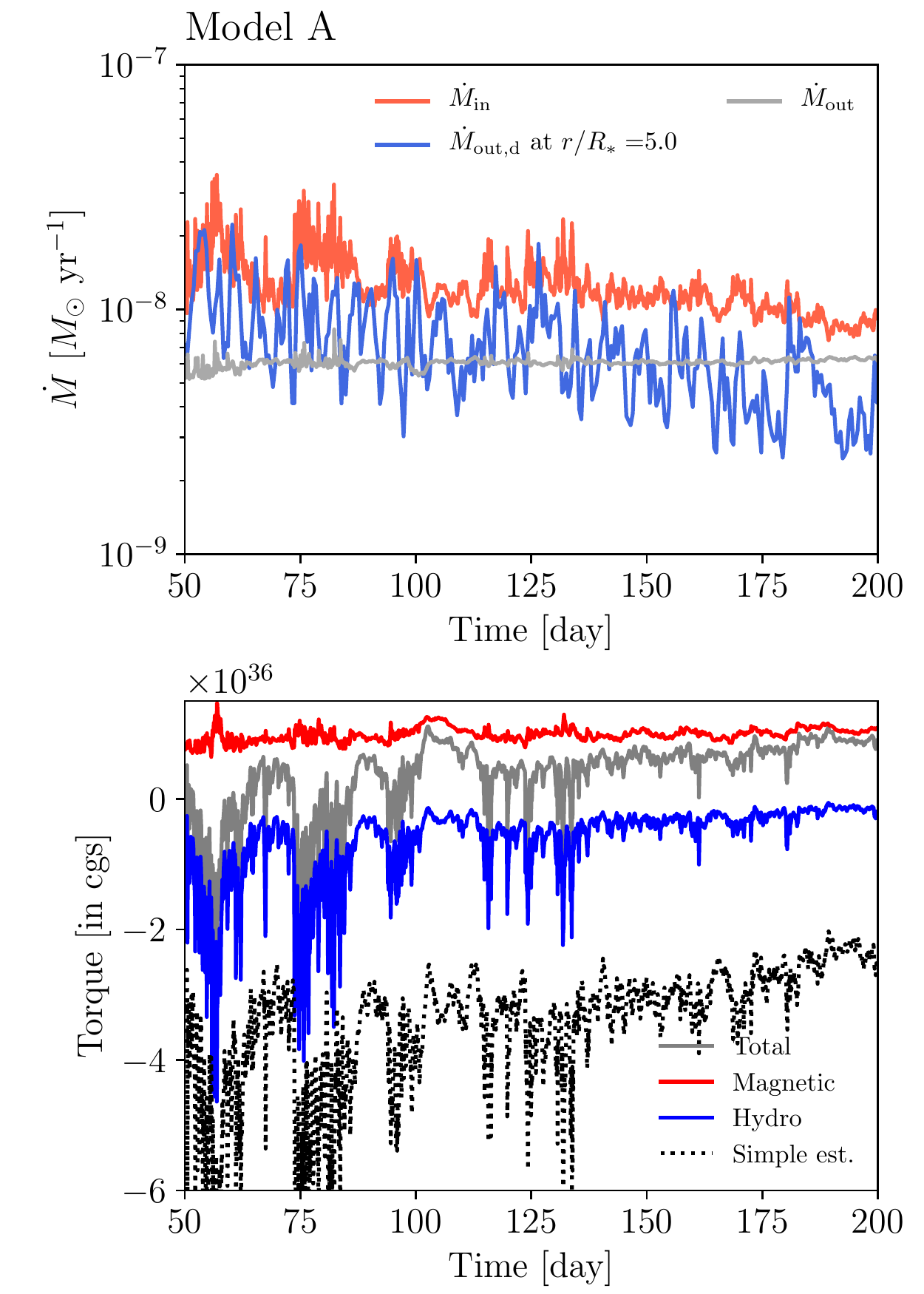}
    \includegraphics[width=0.67\columnwidth]{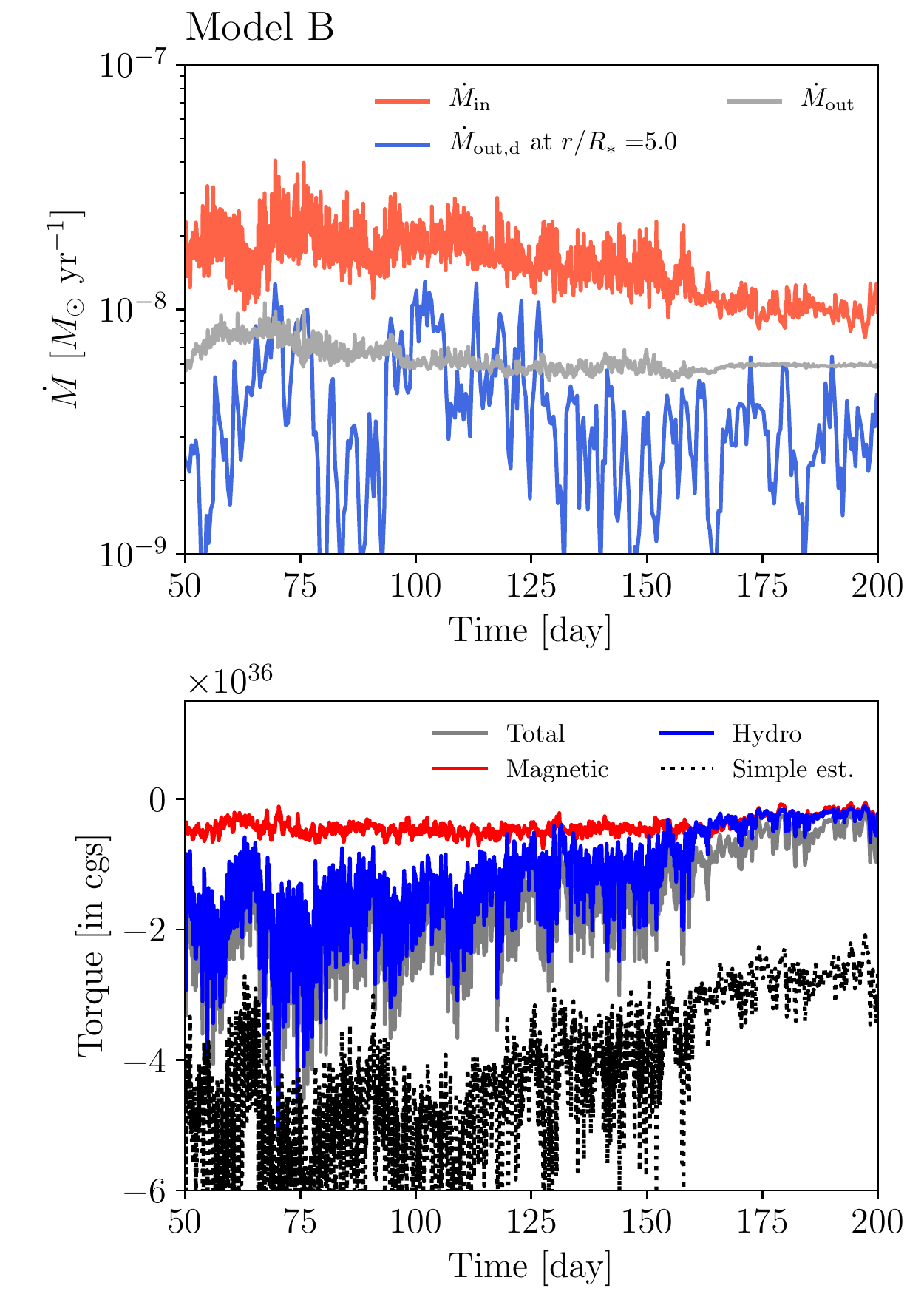}
    \includegraphics[width=0.67\columnwidth]{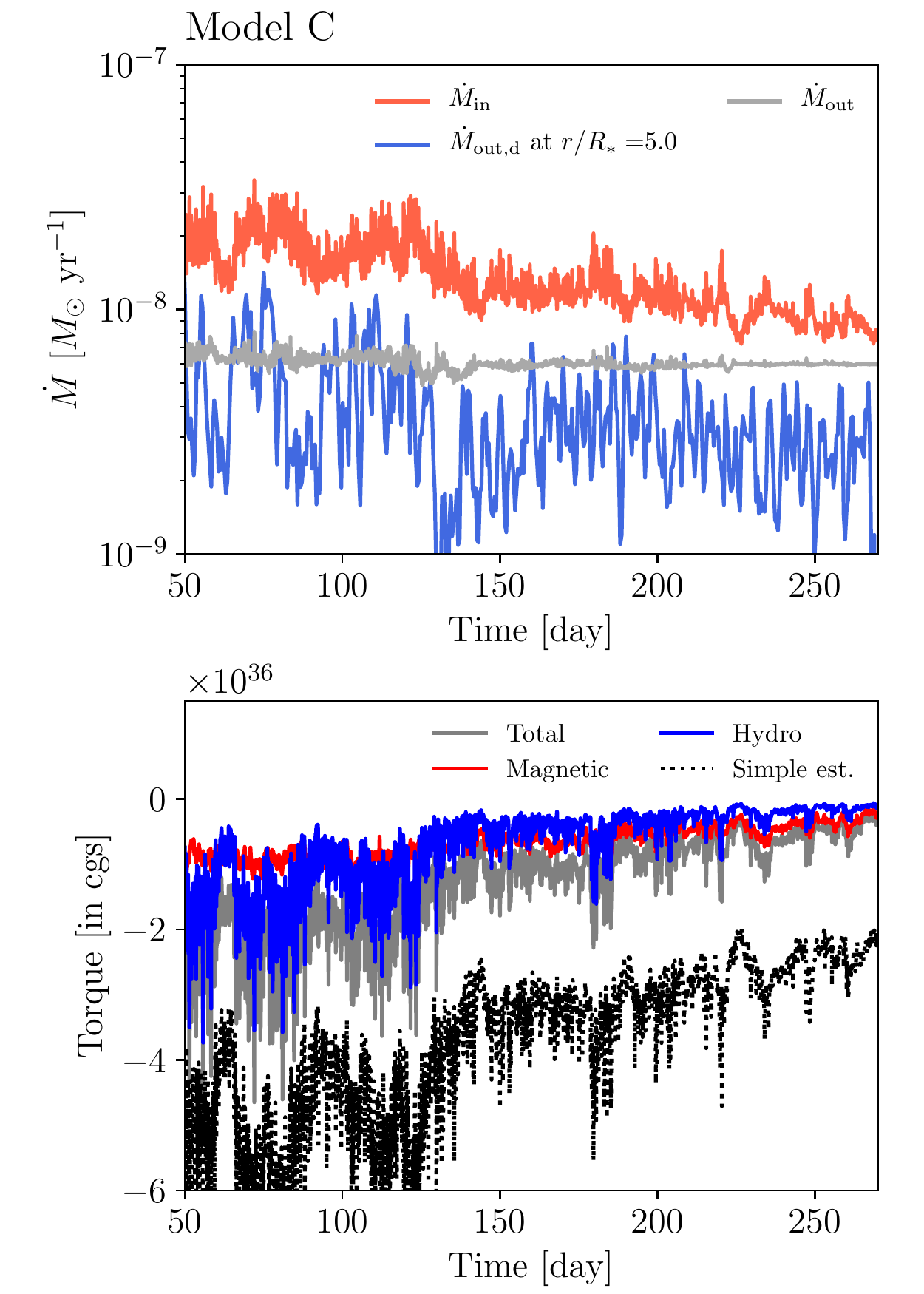}
    \caption{Mass accretion and outflow rates (top) and the torque exerting the star (bottom). The left, middle, and right panels are for Model A, B, and C, respectively. In the top panels, the accretion rate $\dot{M}_{\rm in}$ and the outflow rate $\dot{M}_{\rm out}$ are  measured at $r=R_*$. The outflow rate of the disk gas $\dot{M}_{\rm out,d}$ is measured at $r=5R_*$.
    In the bottom panels, the torques are measured at $r=R_*$. the hydrodynamic and magnetic contributions are shown as blue and red lines, respectively. The gray solid line denotes the total torque. The black dashed line shows $-\dot{M}_{\rm in}\sqrt{GM r_{\rm m}}$, where the magnetospheric radius is approximated as $r_{\rm m}=2R_*$ for all the models.}
    \label{fig:mdot_angmomdot}
\end{figure*}

The outflow rates of the disk gas $\dot{M}_{\rm out,d}$ and the stellar wind $\dot{M}_{\rm out,sw}$ are shown as blue dashed and dotted lines in the middle panels of Figure \ref{fig:alpha_mdot_torque}, respectively. The plot shows that the increase in the outflow rate of the disk gas $\dot{M}_{\rm out,d}$ within $r\approx 3\mbox{-}4R_*$, indicating the launching of the strong disk wind from the magnetospheric boundary.
We note that the mass loss rate of the stellar wind depends on the coronal density we adopt. Therefore, the mass loss rate of the stellar wind cannot be directly compared to actual classical T Tauri stars. It is possible that the value is much larger than the realistic value. Nevertheless, we consider that the stellar wind has a weak impact on the overall structure. A brief discussion is given in Appendix~\ref{appendix:stellar-wind}.

The top panels of Figure \ref{fig:mdot_angmomdot} display the temporal evolution of the mass accretion and outflow rates measured at $r=R_*$.
The blue lines denote the mass outflow rate of the disk gas measured at $r=5R_*$. As expected, Model A shows the most powerful disk wind.
The time variability in the accretion rate is not prominent in all the models, which is different from the results of 2D models \citep[e.g.][]{Zanni2013A&A,Lii2014MNRAS}. 
In 2D models, magnetospheric ejections commonly cause highly variable stellar accretion. Magnetospheric ejections in 2D models may be regarded as huge explosions with superhot ($\gg 10^{6}$ K) plasmas \citep[e.g.][]{Hayashi_etal_1996}.
Although our 3D models show magnetospheric ejections, we do not find significant time variability in the accretion rate and explosions accompanied by superhot plasmas.
The major reasons will be discussed in Section \ref{sec:discussion}.

We study the radial profiles of the torques by different components. The hydrodynamic and magnetic torques at a spherical radius $r$, $\dot{J}_{\rm hydro}$ and $\dot{J}_{\rm mag}$, respectively, are calculated as follows:
\begin{align}
    \dot{J}_{\rm hydro} &= \int_{0}^{\pi/2}R\langle \rho v_r v_\varphi\rangle 2\pi r^2 \sin\theta d\theta\\
    \dot{J}_{\rm mag} &= \int_{0}^{\pi/2}R\frac{\langle -B_r B_\varphi\rangle}{4\pi} 2\pi r^2 \sin\theta d\theta.
\end{align}
The positive and negative torques correspond to the outward and inward transport of the angular momentum, respectively.
Here, we do not subtract the term related to the average azimuthal velocity from the hydrodynamic torque.
The hydrodynamic and magnetic torques in the stellar wind, $\dot{J}_{\rm hydro,sw}$ and $\dot{J}_{\rm mag,sw}$, respectively, are defined as follows:
\begin{align}
\dot{J}_{\rm hydro,sw} &= \int_{0}^{\pi/2}R\langle \rho v_r v_\varphi\rangle_{\rm sw} 2\pi r^2 \sin\theta d\theta\\
    \dot{J}_{\rm mag,sw} &= \int_{0}^{\pi/2}R\frac{\langle -B_r B_\varphi\rangle_{\rm sw}}{4\pi} 2\pi r^2 \sin\theta d\theta.
\end{align}
The hydrodynamic and magnetic torques in the disk gas, $\dot{J}_{\rm hydro,d}$ and $\dot{J}_{\rm mag,d}$, are then calculated as
\begin{align}
    \dot{J}_{\rm hydro,d} & = \dot{J}_{\rm hydro}-\dot{J}_{\rm hydro,sw}\\
    \dot{J}_{\rm mag,d} & = \dot{J}_{\rm mag}-\dot{J}_{\rm mag,sw},
\end{align}
respectively. As a reference, we also define the simple estimation of the accretion torque from the accretion rate as
\begin{align}
    \dot{J}_{\rm acc}' &= \dot{M}_{\rm in}\sqrt{GM_*r_{\rm m}}.
\end{align}
$\sqrt{GM_* r_{\rm m}}$ denotes the specific angular momentum of the Keplerian rotation at the magnetospheric radius $r_{\rm m}$.
This estimation has been widely used to study the spin evolution of the central objects \citep[e.g.][]{Matt&Pudritz2005ApJ}.
Considering the result shown in Figure \ref{fig:rmag}, we use $r_{\rm m}=2R_*$ as a representative value.

The bottom panels of Figure \ref{fig:alpha_mdot_torque} display the radial profiles of the torques of the different components. The blue and red lines denote $\dot{J}_{\rm hydro}$ and $\dot{J}_{\rm mag}$, respectively.
The gray lines show the total torque $\dot{J}_{\rm hydro}+\dot{J}_{\rm mag}$, indicating that the star in Model A is spinning-down (positive torque at $r=R_*$), while the star in Model B is spinning-up (negative torque at $r=R_*$). 
An important result is a significant reduction of the accretion torque $\dot{J}_{\rm hydro,d}$ inside the magnetospheric radius (see the blue dashed lines). 
The angular momentum of the accreting flows is efficiently extracted by the magnetic torque in the disk gas, particularly in the form of the disk winds (See red dashed lines, which indicate $\dot{J}_{\rm mag,d}$). The stellar wind contributions $\dot{J}_{\rm hydro,sw}$ (blue dotted) and $\dot{J}_{\rm mag,sw}$ (red dotted) are negligibly small, as the stellar wind is confined in the polar regions.
One may notice that $\dot{J}_{\rm mag,sw}$ has a small negative value (spin-up torque) at the stellar surface, which is counter-intuitive. The reason will be shown in Section~\ref{subsec:ang_flux}.

The bottom panels of Figure \ref{fig:mdot_angmomdot} show the temporal evolution of the torques measured at $r=R_*$.
Our results demonstrate that the simple torque estimation based on the accretion rate $\dot{J}_{\rm acc}'$ significantly overestimates the actual injection rate of the angular momentum (compare the gray and black dotted lines). This is mainly because the significant reduction of $\dot{J}_{\rm hydro}$ by the turbulent magnetospheric wind and the conical wind (if present). Therefore, the central star is not spinning up as predicted by the simple estimation because the angular momentum of the accreting flow is partially removed inside the magnetospheric radius.

The direct magnetic torque will also be important in the stellar spin evolution, depending on the initial spin.
In Model A, the magnetic torque spins down the star. However, the magnetic torque in Model B and C is negative and spins up the stars. As we have seen in Figure \ref{fig:alpha_mdot_torque}, the magnetic torque carries away a large fraction of the angular momentum of accreting flows inside the magnetosphere. Nevertheless, the accreting flows rotate faster than the stellar surface during the infall (see also Figure~\ref{fig:Omega}). As a result, the rapidly rotating flows drag the stellar magnetic fields, producing negative magnetic torque.

\subsection{Angular Momentum Flux Distribution}\label{subsec:ang_flux}
\begin{figure*}
    \centering
    \includegraphics[width=1.0\columnwidth]{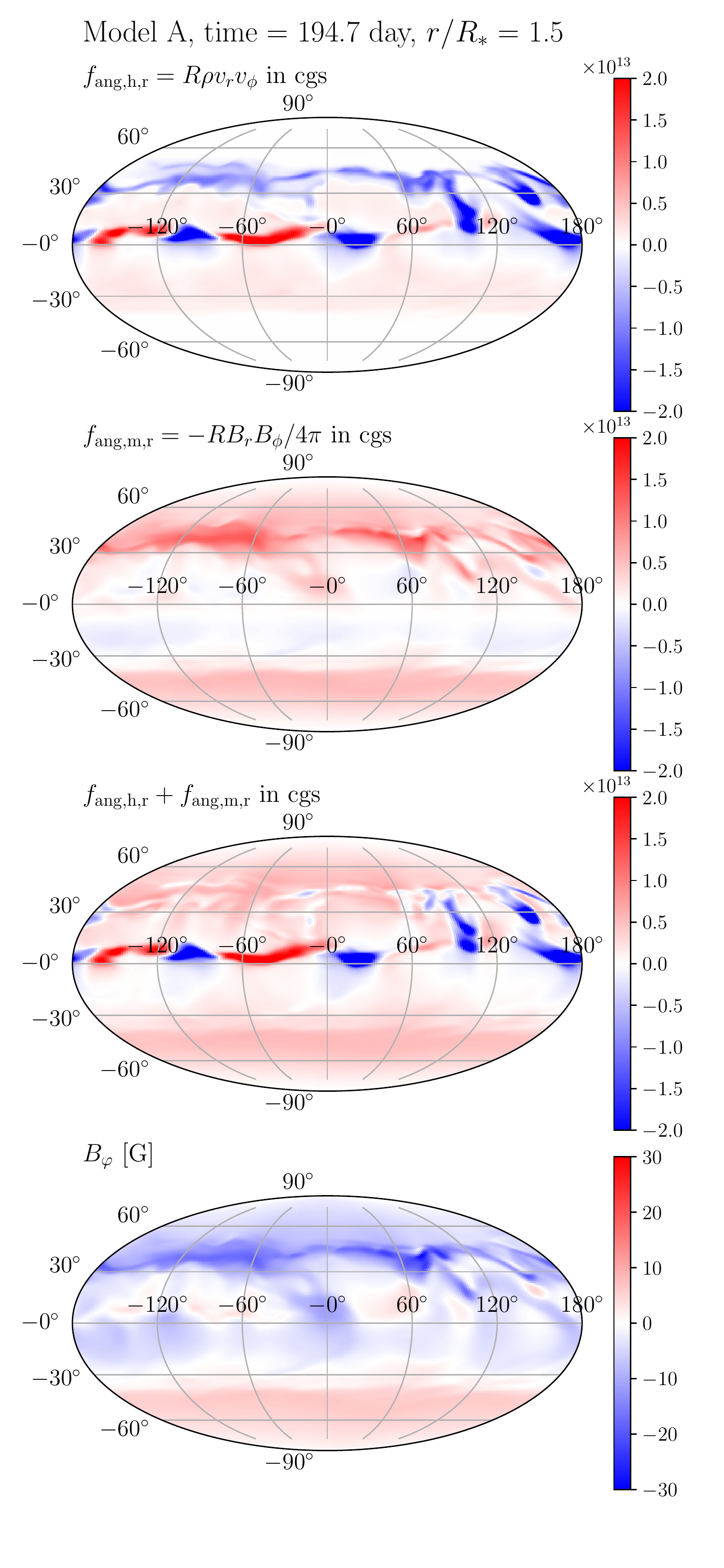}
    \includegraphics[width=1.0\columnwidth]{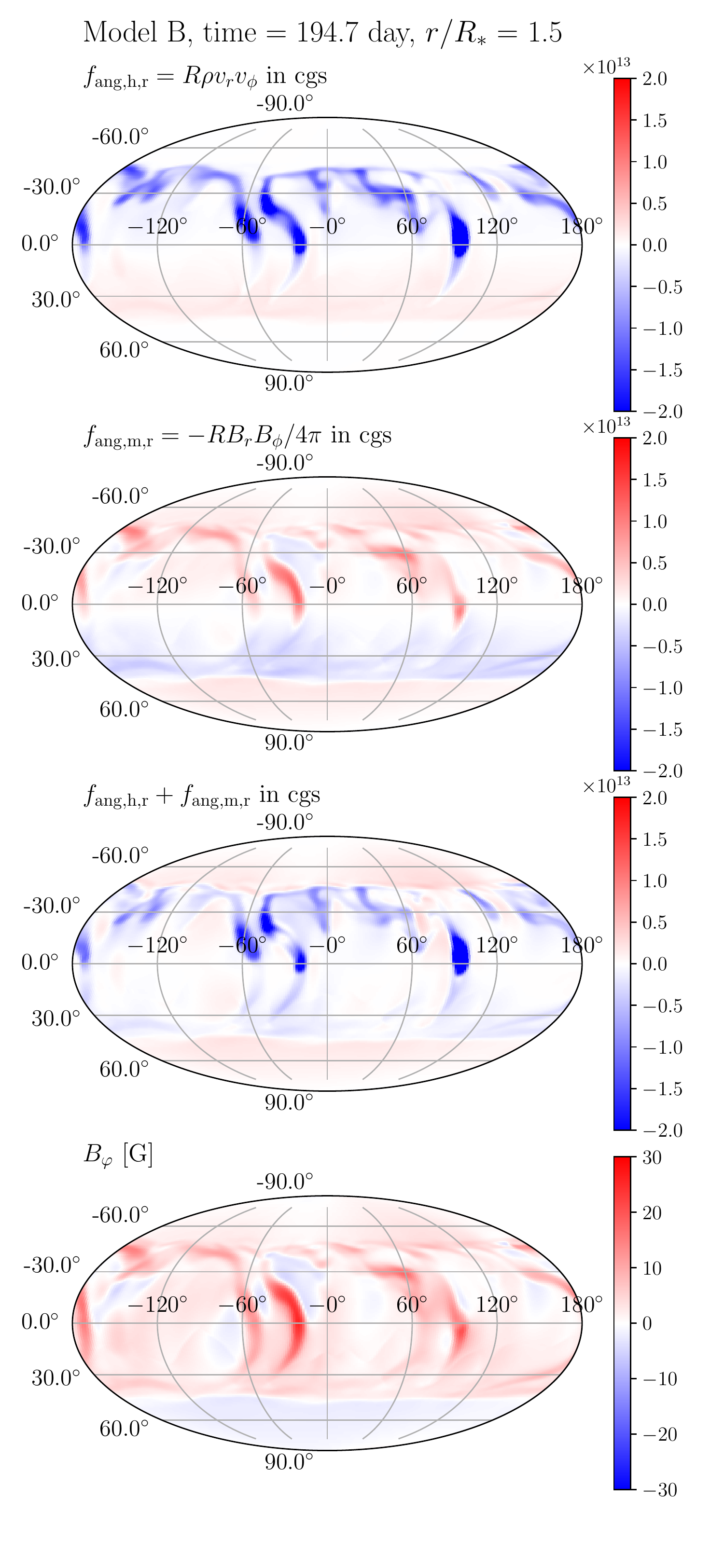}
    \caption{Angular momentum flux distributions at $r=1.5R_*$ for Model A (left) and Model B (right). From top to bottom, $f_{{\rm ang, h},r}=R\rho v_r v_\varphi$, $f_{{\rm ang, m},r}=-RB_r B_\varphi/4\pi$, $f_{{\rm ang, h},r}+f_{{\rm ang, m},r}$, and $B_\varphi$. Note that accretion mainly occurs in the northern and southern hemispheres in Model A and B, respectively. The images of Model B are flipped in the $\theta$ direction.}
    \label{fig:angular_momentum_flux_mlwd}
\end{figure*}

The angular momentum flux distributions on the sphere at $r=1.5 R_*$ (within the magnetospheric radius) for Model A and B are shown in Figure \ref{fig:angular_momentum_flux_mlwd}. As indicated in the top panels ($f_{{\rm ang,h},r}=R\rho v_rv_\varphi$), accretion mainly occurs in the northern and southern hemispheres in Model A and B, respectively (note that the images of Model B are flipped vertically). The outward angular momentum flux by the magnetic fields is also prominent in the accreting regions in both models. In Model A, the outward transport by magnetic fields dominates the inward transport by accreting flows in a large area. However, in Model B, inward transport dominates outward transport. 
The difference between the two models originates from the different stellar spins and the different $B_\varphi$ distributions.
Model A shows strong $B_\varphi$ regions that elongate in the azimuthal direction, increasing the outward angular momentum flux in the large area. However, in Model B, the enhancement of $B_\varphi$ is relatively localized in the azimuthal direction, leading to the localized enhancement of the outward flux.

The different $B_\varphi$ structures arise from different mechanisms that destabilize the magnetospheric boundary. As shown in Section~\ref{subsec:magnetosphere-boundary}, the boundary in Model A is unstable to the instability that creates the spiral pattern. Such spiraling flows efficiently increase the coherent component of $B_\varphi$ around the boundary.
On the other hand, in Model B, the interchange instability creates finger-like structures that mainly extend in the radial direction. Such radial penetration can amplify $B_\varphi$ only locally in the azimuthal direction. Therefore, the mechanisms that perturb the magnetospheric boundary (namely, the mechanisms of mass loading from the disk to the magnetosphere) affect the resulting angular momentum transport process.

\begin{figure*}
    \centering
    \includegraphics[width=2.1\columnwidth]{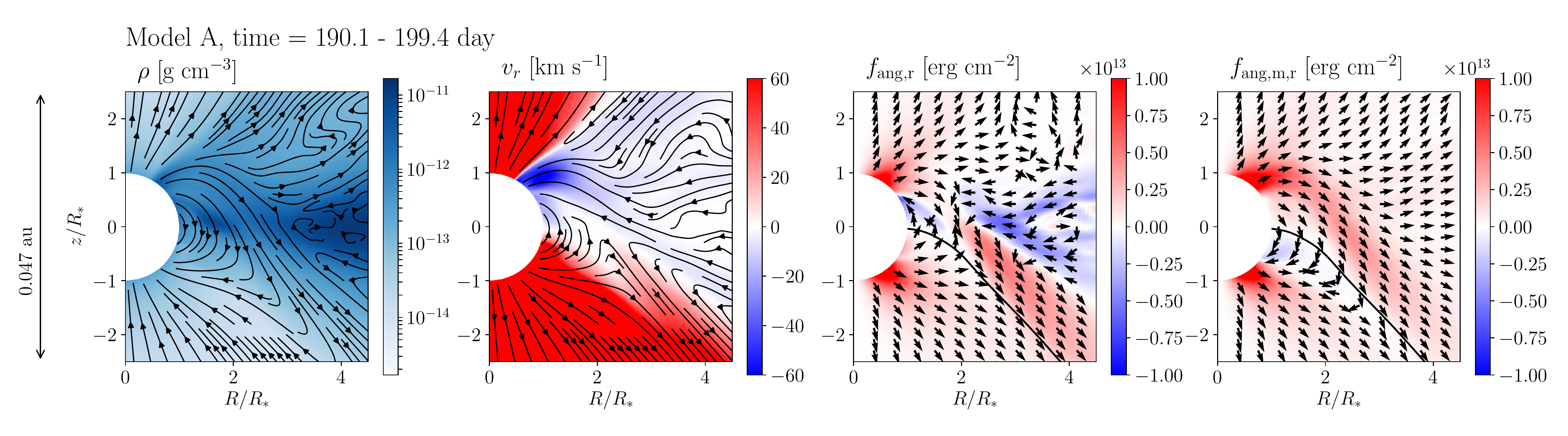}
    \includegraphics[width=2.1\columnwidth]{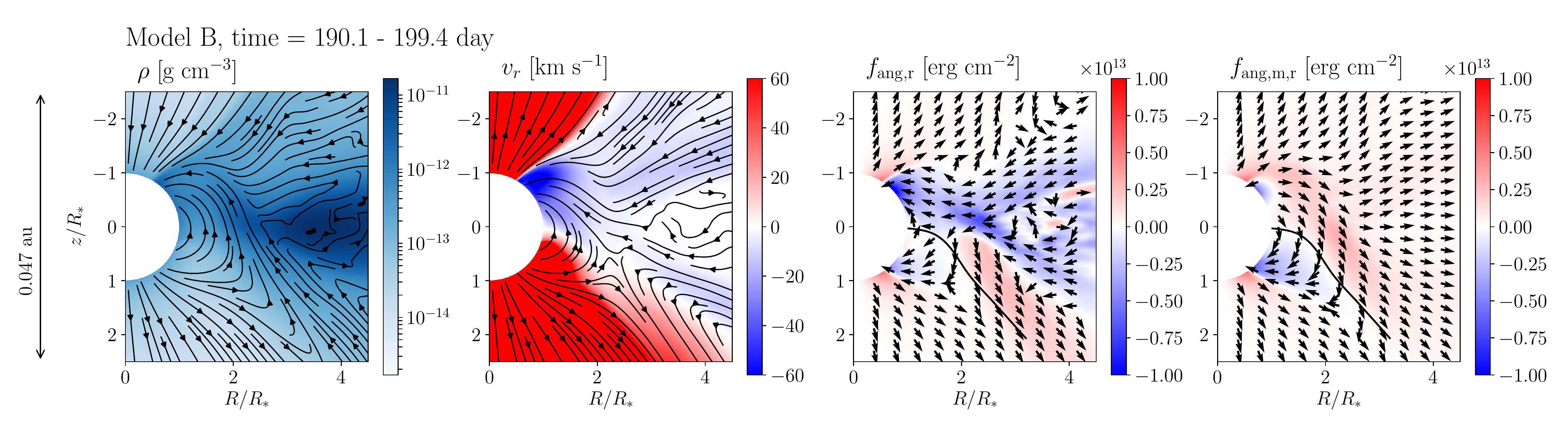}
    \includegraphics[width=2.1\columnwidth]{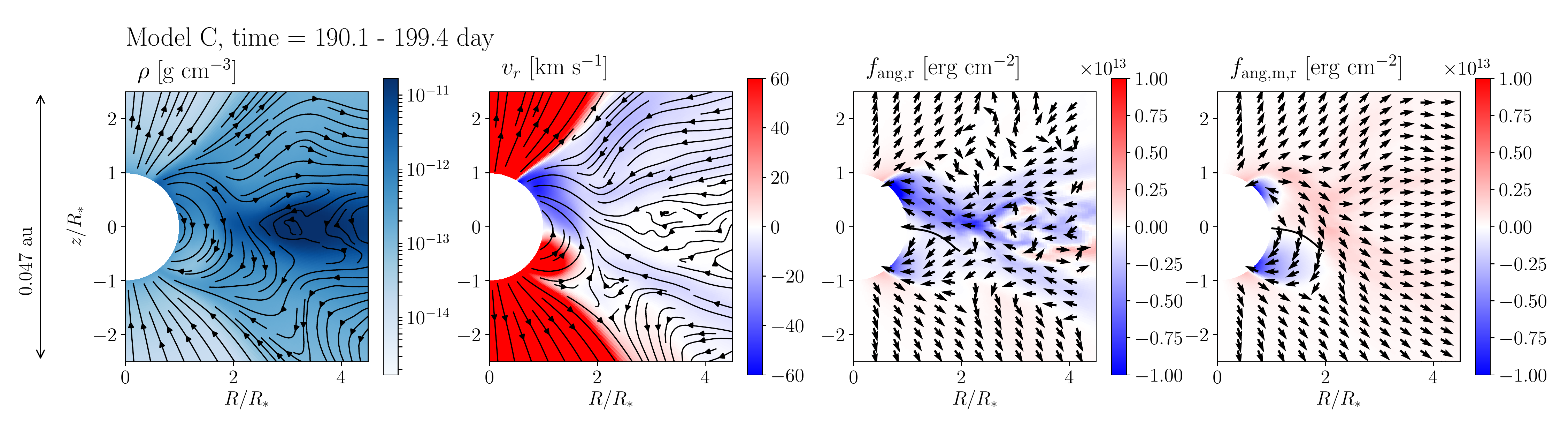}
    \caption{Angular momentum flux distributions of Model A (top), B (middle), and C (bottom). The first column shows the density and the poloidal magnetic field structures. 
    The second column displays the $r$ component of the velocity, $v_r$. Solid lines with arrows denote streamlines. 
    The third column displays the $r$ component of the total angular momentum flux, $f_{{\rm ang},r}$. Arrows denote the direction of the total angular momentum flux. The solid lines indicate the positions of $B_r=0$ in the expanding magnetospheres. The solid lines approximately denote the locations of the current sheets in the expanding magnetospheres. The fourth column shows the $r$ component of the angular momentum flux from the Maxwell stress, $f_{{\rm ang,m},r}$. Arrows denote the direction of the magnetic angular momentum flux. 
    The data are azimuthally averaged and temporally averaged between $t$=190.1-199.4 day. }
    \label{fig:angular_momentum_flux}
\end{figure*}

Figure \ref{fig:angular_momentum_flux} shows the structure of the angular momentum transport in the poloidal plane. The figure compares the results of Model A, B, and C. The first column displays the density and poloidal magnetic field structures. 
The second column shows the $v_r$ maps with streamlines. 
The color in the third column indicates the $r$ component of the total angular momentum flux, and arrows show the direction of the total angular momentum flux vectors. The inward flux by the funnel accretion is prominent in the northern and southern hemispheres in Model A and B, respectively (note that the image for Model B is vertically flipped). The fourth column shows the $r$ component of the angular momentum flux by the magnetic fields. The fourth column indicates that in Model A and B the magnetic field transports a large amount of the angular momentum in the form of the conical wind.

In Model C, a stable conical wind appears only in the late phase.
Nevertheless, Figure \ref{fig:mdot_angmomdot} indicates that the angular momentum is extracted from the accreting flows even in the absence of the conical wind. By investigating this phase, we can highlight the roles of turbulent winds emanating from the magnetosphere.
The bottom panels of Figure \ref{fig:angular_momentum_flux} displays the structure before the stable conical wind appears. The angular momentum flux from the Maxwell stress takes a large value along the disk surfaces. The outward angular momentum transport at high latitudes is mediated by the turbulent magnetospheric wind. A very similar process is also reported in ST18, although there are some differences (see Section~\ref{subsec:accretion_structure} for the comments about this point). A large fraction of the turbulent wind falls back to the disk. Therefore, the wind mass circulates but the angular momentum is removed from the vicinity of the star.
When the stable conical wind is established in the southern hemisphere, the accretion structure is similar to that of Model B.

Figure~\ref{fig:alpha_mdot_torque} shows a small negative $\dot{J}_{\rm mag,sw}$ at the stellar surface for Model B.
The small spin-up torque is caused by the stellar corona confined in the magnetosphere (magnetospheric plasma). Figure \ref{fig:angular_momentum_flux_mlwd} indicates that a large area around at a latitude of 30~degree has a negative value. This region corresponds to the root of the magnetospheric plasma. When the magnetospheric plasma moves outward in response to the magnetospheric expansion, for instance, this region is categorized as the stellar wind region under our definition (see Section \ref{subsec:ang}) even though this region is inside the magnetosphere. Therefore, the spin-up torque is not given by the polar stellar wind but by the magnetospheric plasma. The negative torque in the magnetospheric region is a result of back-reaction of the magnetospheric accretion. The accretion flows can produce such a negative torque by dragging the stellar field.

\subsection{Density Structure around the Star}\label{subsec:shielding}
\begin{figure*}
    \centering
    \includegraphics[width=0.68\columnwidth]{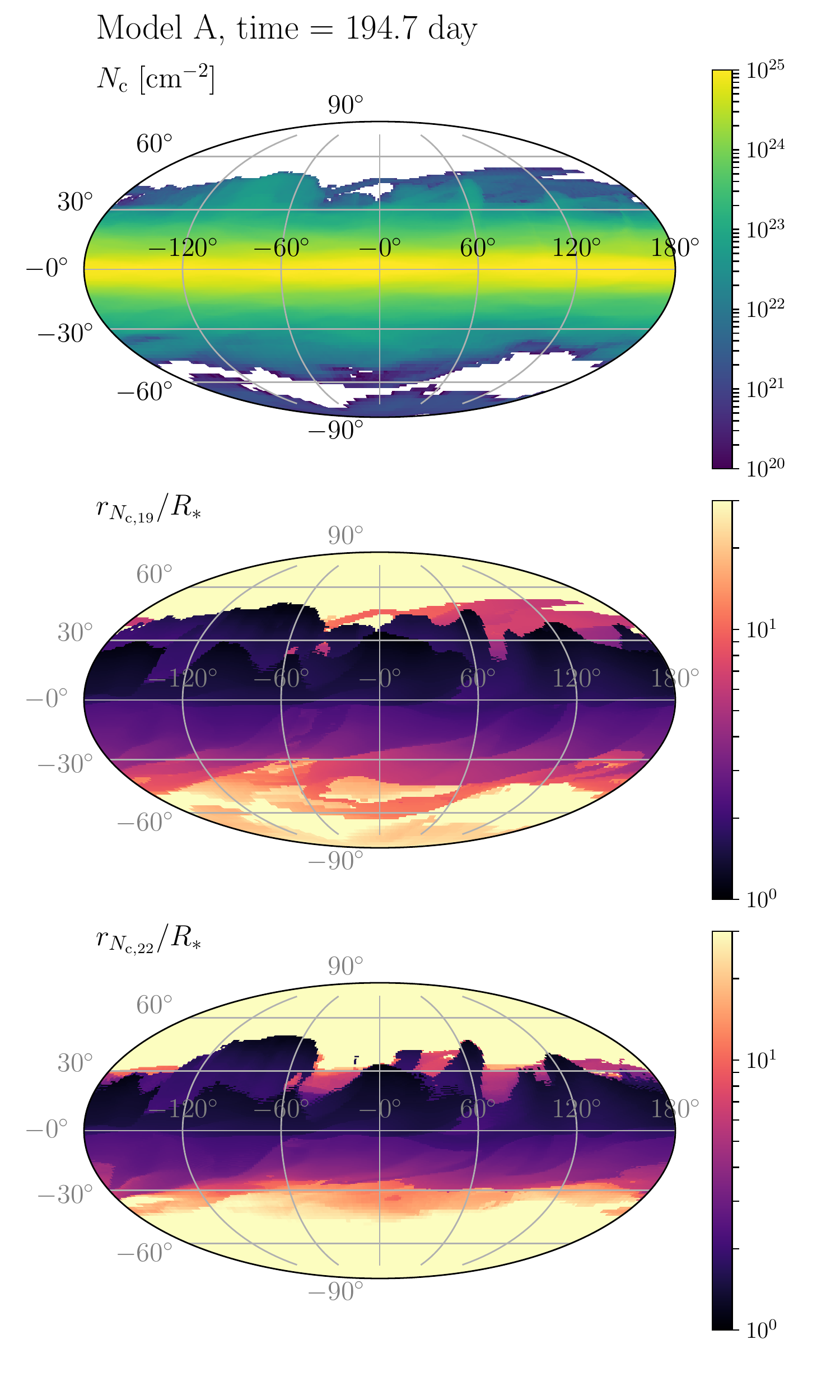}
    \includegraphics[width=0.68\columnwidth]{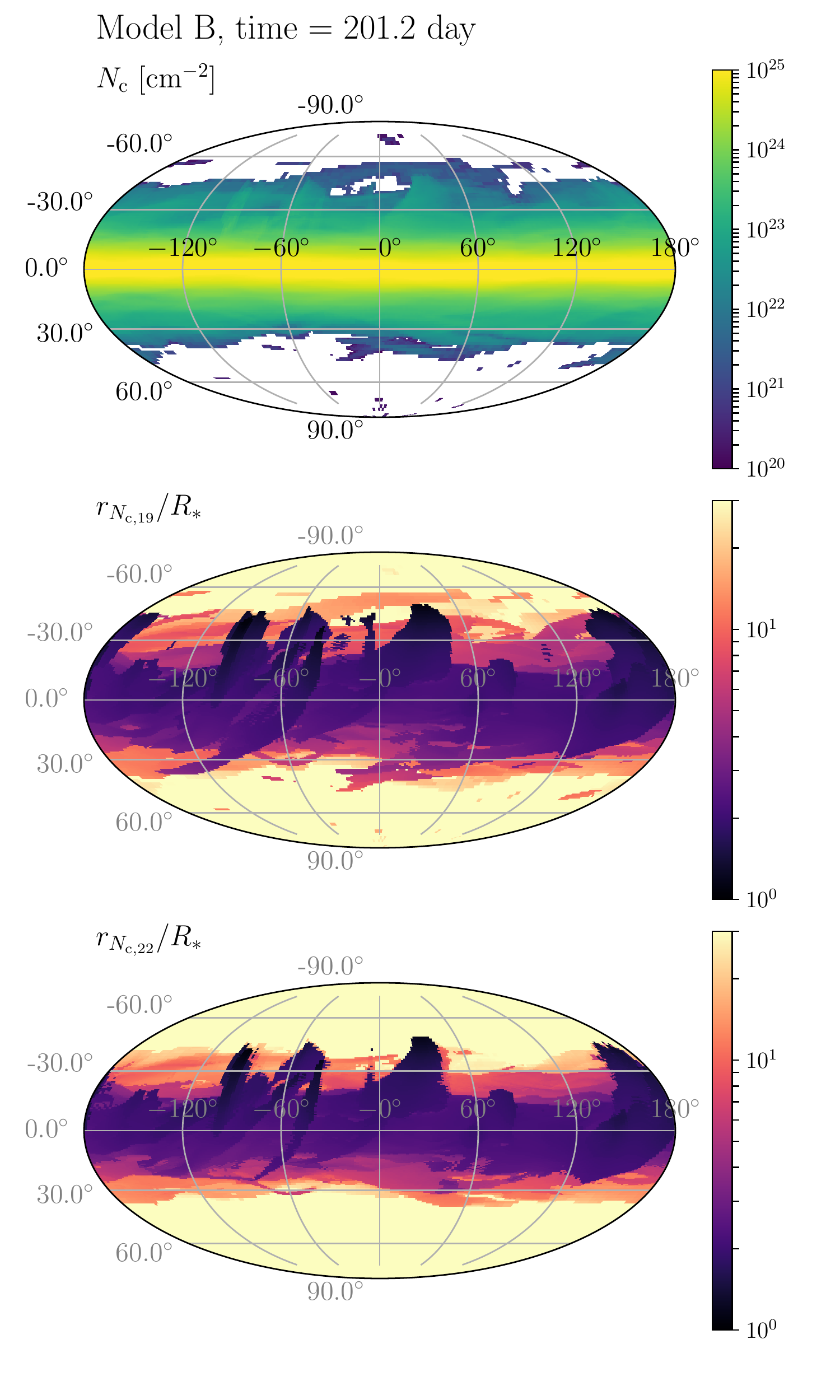}
    \includegraphics[width=0.68\columnwidth]{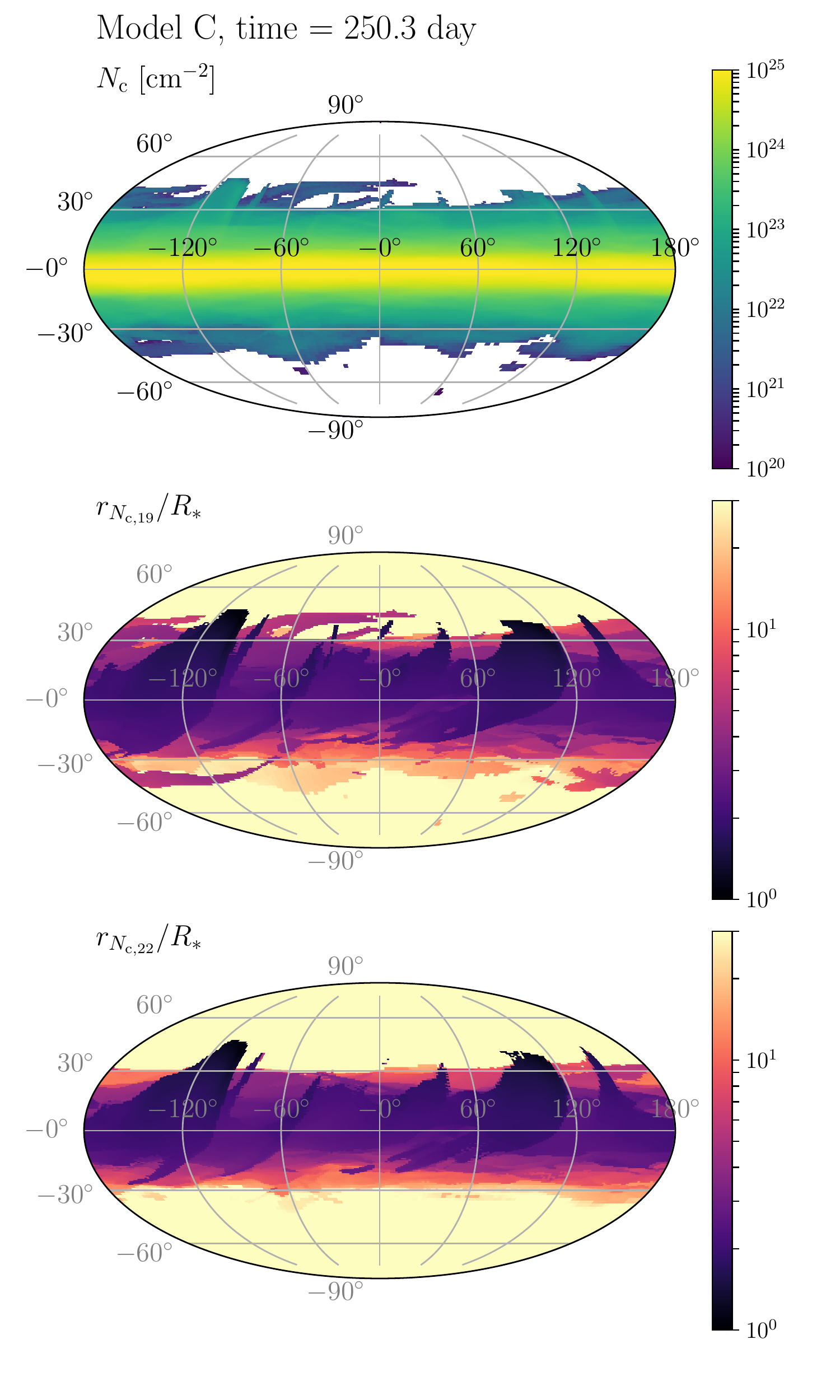}
    \caption{Density structures around the star for Model A (left), B (middle), and C (right). The top panels show the column density $N_{\rm c}$ measured from $r=R_*$. The middle and bottom panels display the radii (normalized by $R_*$) at which the column density becomes $10^{19}$ cm$^{-2}$ and $10^{22}$ cm$^{-2}$, respectively. These radii are expressed as $r_{N_{\rm c,19}}$ and $r_{N_{\rm c,22}}$, respectively.}
    \label{fig:column_density}
\end{figure*}

The density structure around the star controls the amount of the stellar radiation that can reach the outer disk. Figure \ref{fig:column_density} shows the results for the three models. The top panels display the column density $N_{\rm c}$ calculated by integrating the density in the radial direction from the stellar surface. We only count the plasma with a temperature lower than $\sim 2\times 10^4$~K in this plot by simply assuming that the hotter plasma will not contribute to the blocking of the stellar FUV, EUV, and X-rays.

The hydrogen atom column densities for shielding EUV and X-ray are expected to be $10^{19}$ cm$^{-2}$ and $10^{22}$ cm$^{-2}$, respectively \citep[e.g.][]{Ercolano2009ApJ,Owen2010MNRAS}. Considering this, we investigate the shielding region for these wavelengths in our models. 
The middle and bottom panels of Figure \ref{fig:column_density} indicate the radii at which the column density exceeds $10^{19}$ cm$^{-2}$ and $10^{22}$ cm$^{-2}$, respectively. These radii are expressed as $r_{N_{\rm c,19}}$ and $r_{N_{\rm c,22}}$, and they are normalized by the stellar radius $R_*$ in the figure. In all the models, the EUV shielding regions extend broader than the X-ray shielding regions. As EUV cannot penetrate at low latitudes, EUV may generally irradiate a more outer part of the disk than X-ray. Therefore, the main wavelength range that drives the photoevaporation is expected to change with radius. We also find filamentary structures which extend in the latitudinal direction. These correspond to the multiple accretion columns (Figures \ref{fig:3D_modelB}, \ref{fig:interchange} and \ref{fig:angular_momentum_flux_mlwd}). EUV and X-rays will reach the outer disks through the windows between the multiple accretion columns.

We have to note that the above estimate about the photon shielding is incomplete in the sense that our models treat the thermal evolution only in a very simplified manner. We plan to update our models so that the radiative transfer and chemical reactions are solved to obtain a more realistic thermochemical structure. The result shown here is the first step for this direction of study.

\section{Discussion} \label{sec:discussion}
\subsection{Short summary}
We presented the initial results of MHD simulations of magnetospheric accretion in a T Tauri star and studied the accretion and wind structures within a few 10 $R_*$. We analyzed three models with different stellar spins. 
Multi accretion columns are produced in all models (Figures \ref{fig:3D_modelB}, \ref{fig:interchange} and \ref{fig:angular_momentum_flux_mlwd}). It is found that accretion flows in the magnetosphere originate not only from the magnetospheric boundary but also from a broader region of the inner disk. The failed disk wind from the inner disk becomes a part of the magnetospheric accretion flows (Figures \ref{fig:4phys_modelA}, \ref{fig:4phys_modelB}, \ref{fig:4phys_modelC}, and \ref{fig:alpha_mdot_torque}).

Our models demonstrate that a large amount of the angular momentum of accreting flows is extracted by conical winds and failed turbulent disk winds.
As a result, the spin-up accretion torque is smaller than the simple estimation based on the accretion rate and the magnetospheric radius, $\dot{J}_{\rm acc}'=\dot{M}_{\rm acc}\sqrt{GM r_{\rm m}}$ \citep[e.g.][]{Matt&Pudritz2005ApJ}.
A qualitatively similar result was found in 2D simulations by \citet{Zanni2013A&A}. We confirmed this behavior using 3D models.
The spin-up torque is substantially smaller than $\dot{J}_{\rm acc}'$ even during the period when a stable conical wind is absent (before $t\approx 250$ day in Model C. See Figure \ref{fig:mdot_angmomdot}). The angular momentum of accreting flows is extracted by turbulent, weak winds emanating from the magnetosphere. Similar angular momentum transport is also discussed in ST18.
As most of the weak wind falls back to the disk, a part of the mass seems to circulate around the star. 
This result indicates that disk winds play essential roles in extracting the angular momentum from accreting flows regardless of whether the winds can escape from the stellar gravity and extend to a much larger scale or not.
The angular momentum is continuously extracted from the failed disk winds and is deposited somewhere in the disk atmospheres, as found in ST18.

\subsection{Magnetosphere-disk boundary \\
and magnetospheric radius}\label{subsec:summary_magnetospheric_boundary}
Many previous theories assume that the accreting gas rotates rigidly with the magnetic field corotating with the star after it penetrates into the magnetosphere. In addition, the strength of the toroidal magnetic field has been thought to depend on the stellar spin. However, our simulations disagree with these assumptions (Figures~\ref{fig:Bfield}, \ref{fig:Omega}, and \ref{fig:twist}). 
A large volume of the magnetosphere is forced to rotate nearly at the Keplerian velocity by the penetrating accreting gas. Around the magnetospheric boundary, the toroidal field becomes as strong as the vertical field, regardless of the stellar spin in our parameter range.
The detailed theoretical model by \citet{Kluzniak2007ApJ} explains a smooth transition in the angular velocity. However, we demonstrated the mismatch between the 3D simulations and the prediction of the theory (Section \ref{subsec:rotation}). The breakdown of model assumptions are found.

As described in Section~\ref{subsec:magnetosphere-boundary}, the three-dimensional and two-dimensional interactions are very different. In three-dimension, the stellar field is continuously extruded to the disk via the instabilities at the magnetospheric boundary. This leads to an efficient production of the toroidal field there (Figure~\ref{fig:midplane_bfield_amplification}). When the plasma $\beta$ becomes close to unity, the amplified toroidal field escapes from the disk because of magnetic buoyancy (see also ST18). The balance between the field amplification and escape regulates the toroidal field strength. The MRI turbulent disk also produces the toroidal field, which helps the innermost disk to keep the strong toroidal field against diffusion around the boundary. These processes are not present in the 2D models.

There are many theoretical predictions about the magnetospheric radius. Although only three models of ours are insufficient to completely check the scaling relations of the theories, we briefly compare some predictions with our numerical results.

By considering the balance between the magnetic pressure and the ram pressure for the spherically accreting gas, \citet{Ghosh_Lamb1979_paperIII} and \citet{Koenigl1991ApJ} estimated the magnetospheric radius as 
\begin{align}
    r_{\rm m,GL79}\approx k_{\rm GL} \left(\frac{\mu_*^4}{2GM_*\dot{M}^2}\right)^{1/7}
\end{align}
or
\begin{align}
    \frac{r_{\rm m,GL79}}{R_*}&\approx 2.5 \left(\frac{k_{\rm GL}}{1}\right)\left( \frac{B_*}{160~{\rm G}}\right)^{4/7}\left(\frac{R_*}{2R_\odot}\right)^{5/7}\nonumber \\
    &\times \left(\frac{M_*}{0.5M_\odot}\right)^{-1/7}\left(\frac{\dot{M}}{10^{-8}M_\odot~{\rm yr^{-1}}}\right)^{-2/7},
\end{align}
where $\mu_*=B_* R_*^3$. The estimated value may be consistent with the numerical result if we take the numerical factor $k_{\rm GL}\approx 1$. The numerical factor has not been determined in their papers. In addition, the above argument assumes the free-falling, spherical accretion, which is inconsistent with the disk accretion. Although some theories derived similar relations for the disk accretion \citep[e.g.][]{Shu1994ApJ,Ostriker1995ApJ}, the discussions are limited only to the case $r_{\rm m}\approx r_{\rm cor}$.

\citet{Bessolaz2008A&A} derived another relation by assuming that the magnetic pressure resulting from the poloidal field balances the ram pressure of the accreting gas in a disk:
\begin{align}
    \frac{r_{\rm m,B08}}{R_*} &\approx 1.5 \left(\frac{m_{\rm s}}{0.2}\right)^{2/7}\left(\frac{B_*}{160~{\rm G}}\right)^{4/7}\left(\frac{\dot{M}}{10^{-8}~M_{\odot}~{\rm yr^{-1}}}\right)^{-2/7} \nonumber \\
    &\times \left(\frac{M_*}{0.5M_\odot}\right)^{-1/7}\left(\frac{R_*}{2R_\odot}\right)^{5/7},
\end{align}
where $m_{\rm s}$ is the acoustic Mach number of the accretion speed at the magnetospheric boundary. 
It is not clear if this relation is applicable to a wide range of the fastness parameter.
They consider that $m_{\rm s}\approx 1$ at the boundary, but our simulation suggests that the azimuthally averaged $m_{\rm s}$ is typically $\mathcal{O}(0.1)$ and at most 0.2. The estimated magnetospheric radius is considerably smaller than the numerical result. They assumed that the accreting gas at the magnetospheric boundary is lifted from the midplane mainly by the gas pressure. However, this is not necessarily true when the magnetosphere becomes asymmetric about the equator. In addition, the magnetic force in the disk significantly contributes to the lifting (e.g. the magnetic pressure gradient force by the toroidal field, the MRI-driven wind). For these reasons, $m_{\rm s}\approx 1$ will not be generally required for the magnetospheric boundary.

\citet{Dangelo_Spurit2010MNRAS} estimated the magnetospheric radius by using the angular momentum transport equation \citep[see also][]{Spruit1993ApJ}. The result is
\begin{align}
    r_{\rm m,DS10} &= \left(\frac{\eta'\mu_*^2}{2\Omega_* \dot{M}}\right)^{1/5} \label{eq:rmag_DS10} \\
    \frac{r_{\rm m,DS10}}{R_*}&\approx 1.6\left( \frac{\eta'}{0.1}\right)^{1/5}\left(\frac{B_*}{160~{\rm G}}\right)^{2/5}\left( \frac{R_*}{2R_\odot}\right)^{6/5} \nonumber \\
    &\left(\frac{\dot{M}}{10^{-8}~{M_\odot~{\rm yr}^{-1}}}\right)^{-1/5}\left( \frac{P_*}{3~{\rm day}}\right)^{1/5},
\end{align}
where $\eta'=|B_\varphi/B_z|$ is the ratio of the toroidal field strength to the vertical field strength at the boundary (the numerical factor is slightly different from their papers, but here we consider the magnetic torques exerting both disk surfaces). $P_*=2\pi/\Omega_*$ is the stellar rotational period. This original estimates gives a smaller value than the numerical result.

Here, we update Equation (\ref{eq:rmag_DS10}) considering our 3D simulations. Our models indicate that $\eta'= 1$ (Figures \ref{fig:Bfield} and \ref{fig:twist}) and that the magnetospheric spin rate is not $\Omega_*$ but $\Omega_{\rm K}(r_{\rm m})$. Therefore, we modify Equation (\ref{eq:rmag_DS10}) as follows:
\begin{align}
    r_{\rm m, mod} = \left(\frac{\mu_*^2}{2\Omega_{\rm K}(r_{\rm m,mod})\dot{M}}\right)^{1/5}
\end{align}
or 
\begin{align}
    r_{\rm m,mod}= \left(\frac{\mu_*^4}{4GM_*\dot{M}^2}\right)^{1/7}
\end{align}
This expression is equivalent to $r_{\rm m,GL79}$ except for the difference in the numerical factor. That is, our simulations allow us to determine the numerical factor in the expression of $r_{\rm m,GL79}$. $r_{\rm m,mod}$ is consistent with the numerical result (Figure~\ref{fig:rmag}):
\begin{align}
    \frac{r_{\rm m,mod}}{R_*} &\approx 2.3\left( \frac{B_*}{160~{\rm G}}\right)^{4/7}\left(\frac{R_*}{2R_\odot}\right)^{5/7} \nonumber \\
    &\times \left(\frac{M_*}{0.5M_\odot}\right)^{-1/7}\left(\frac{\dot{M}}{10^{-8}M_\odot~{\rm yr^{-1}}}\right)^{-2/7}.
\end{align}

As mentioned above, the magnetospheric radius has been estimated in different ways. One is based on the pressure balance \citep[e.g.][]{Ghosh_Lamb1979_paperIII,Koenigl1991ApJ} and another is based on the angular momentum transfer equilibrium \citep{Spruit1993ApJ,Dangelo_Spurit2010MNRAS}. These two approaches originally predict different scaling relations. However, we find that they give the same scaling relation if the rotation of the magnetosphere is not governed by the stellar spin but is controlled by the rotation of the accreting flows that penetrate the magnetosphere. Our models suggest that $r_{\rm m}$ only weakly depends on the stellar spin, although the simulated cases are quite limited. For the magnetosphere-disk interaction, the estimation based on the angular momentum transfer equilibrium seems more reasonable than the estimation based on the pressure balance. \citet{Ostriker1995ApJ} also arrived at a very similar result by considering the angular momentum transfer, although they assumed the situation of $r_{\rm m}\approx r_{\rm cor}$.
Most of the theoretical studies consider the case where $r_{\rm m}\lesssim r_{\rm cor}$, and the applicability of the Ghosh \& Lamb relation to the propeller regime has remained unclear \citep[e.g.][]{Blinova2016MNRAS}. However, we show that the magnetospheric radii in the three models are approximately consistent with the estimation, which suggests that the modified expression will be applicable to both the slow rotator and the propeller regimes.
We note that the accretion can be quenched if the stellar wind expels the accreting gas \citep[e.g.][]{Parfrey2017ApJ}. This may occur if the accretion rate decreases to some value.

Previous 3D simulations by \citet{Kulkarni_Romanova2013MNRAS} and \citet{Blinova2016MNRAS} suggest that the scaling of $r_{\rm m}$ with the parameter $\mu_*^2/\dot{M}$ depends on the stability of the magnetospheric boundary.
A slightly flatter scaling is found for unstable cases (the power of 0.22, while the power of $r_{\rm m,GL79}$ is $2/7\approx 0.286$.
If the magnetospheric boundary is stable against the interchange instability, a more flatter scaling is found.
The direct comparison between our simulations and the previous simulations by \citet{Kulkarni_Romanova2013MNRAS} and \citet{Blinova2016MNRAS} is not straightforward because their models adopt an explicit viscosity using the so-called $\alpha$-model even though the magnetic field is solved. As the main driver of the disk accretion in such a model is difficult to interpret, the outcome of the combination of the $\alpha$ viscosity and the Maxwell stress is also unclear.

\citet{Blinova2016MNRAS} argue that the magnetospheric boundary will be unstable to the interchange instability when $\omega_{\rm s}\lesssim 0.6$. Our Model B and C are generally consistent with the result.
However, we find a qualitatively different behavior for Model A.
Our Model A, which has $\omega_{\rm s}\approx 2.2$, displays the highly unstable boundary. Therefore, we infer that unstable magnetospheric boundaries would be found in a wider range of the fastness parameter.
Larger parameter surveys without using the $\alpha$ viscosity would be required to clarify a more detailed dependence. This would be a future task.
Particularly, investigating rapid rotators with $\omega_{\rm s}>1$ is important. Our Model A is one example.

\clearpage
\subsection{Comparison with previous 2D axisymmetric models}\label{subsec:comparison_2D}
Previous 2D models \citep[e.g.][]{Ustyugova2006ApJ,Lii2014MNRAS} indicate that in the propeller regime, the stellar accretion is largely suppressed, and a large fraction of accreting gas is ejected away by the spinning magnetosphere. In terms of the fastness parameter $\omega_{\rm s}$, Model A is expected to be in the propeller regime ($\omega_{\rm s}\approx 2.2$).
However, the accretion rate in our Model A is similar to those of Model B and C (Figure \ref{fig:mdot_angmomdot}), and a significant reduction in the accretion rate is not found. 
\citet{Ustyugova2006ApJ} discussed that the outflow efficiency is a decreasing function of the effective magnetic diffusivity when the diffusivity is too high ($>0.2$ in the nondimensional form). This is a direct consequence that strong magnetic diffusion weakens the magnetic coupling and the disk wind. Therefore, most of the accreting gas will fall onto the star in the strong diffusion case. Although we do not show the effective magnetic diffusion, the viscous parameters $\overline{\alpha_{m,R\varphi}}$ and $\overline{\alpha_{m,\varphi z}}$ in our models are of the order of unity around the magnetospheric boundary. If the effective magnetic diffusivity is on the same order of magnitude as the viscous parameters, the discussion based on 2D models seems to be consistent with our 3D results.

In previous 2D models \citep{Hayashi_etal_1996,Zanni2013A&A,Lii2014MNRAS}, powerful magnetospheric ejections are intermittently driven as the rotating disk gas twists up the stellar magnetic field and efficiently increases the free magnetic energy. However, although our simulations show magnetospheric ejections, ejections do not induce significant variability in the accretion rate (Figure \ref{fig:mdot_angmomdot}). In other words, magnetospheric ejections in three-dimension are not as powerful as expected from 2D models.

We find that there are mainly three reasons why the 3D models are less time-variable.
The first reason is, as demonstrated in our models, that accreting flows penetrating the magnetosphere are fragmented (Figure~\ref{fig:interchange}). It is difficult for such fragmented flows to coherently twist the stellar magnetosphere.

The second reason is that the penetrating flows generally infall onto the star within one or two rotations. As a result, accreting flows can only twist the outer part of the stellar magnetic field, as shown in Figures \ref{fig:Bfield}, \ref{fig:midplane_bfield_amplification} and \ref{fig:twist}. As the magnetic field strength around the magnetospheric radius is much weaker than that near the stellar surface, the magnetic energy built-up by shearing motions of disk gas is less efficient than expected in 2D models.

The third reason is that the plasma condition around the electric current sheet of the inflating magnetosphere is not suitable for violent magnetic reconnection. For magnetic reconnection to produce a very hot plasma that can be observed as an X-ray flare, the plasma $\beta$ around the current sheet should be much smaller than unity. For the relation between the plasma $\beta$ and the temperature of the plasma heated by magnetic reconnection, see e.g., \citet{Takasao_Shibata2016ApJ}. However, the azimuthally averaged plasma $\beta$ around the current sheets is approximately 0.3-1 in our models (Figures \ref{fig:4phys_modelA}, \ref{fig:4phys_modelB}, and \ref{fig:4phys_modelC}) because of mass loading and heating by the MRI-driven wind (see also ST18). Therefore, violent magnetic reconnection that can produce strong X-ray flares is suppressed in our 3D models.
Our results are qualitatively consistent with X-ray observations. \citet{Getman2008ApJ} found no clear evidence that the disk-magnetosphere interaction in pre-MS stars produces powerful flares \citep[see also][]{Getman_Feigelson2021ApJ}. However, we note that powerful flares can be driven by accretion in young protostars which are surrounded by disks with a high accretion rate and a strong poloidal field \citep{Takasao_Tomida_Iwasaki_Suzuki_2019}.

\subsection{Comments on the stellar spin-down}
\citet{Ireland2021ApJ} argued using 2D models that massive stellar winds with a mass loss rate of a few 10\% would be required to balance the spin-up torque due to accretion \citep[see also][]{Pantolmos2020A&A}. 
However, such a strong stellar wind would be difficult to realize.
\citet{Shoda2020ApJ} performed a series of 1D MHD simulations for solar-type main-sequence stars by considering the effect of the stellar rotation and the detailed process of the coronal heating based on the modern understanding. Note that they solve the stellar winds along fixed average magnetic flux tubes based on the observation of solar-type stars \citep[e.g.][]{See2019ApJ,See2020ApJ}. They found that the mass loss rate would saturate around $\sim 3\times 10^{-14}~M_\odot~{\rm yr^{-1}}$ because the Alfv\'en waves, the main energy carriers, are subject to the strong reflection and dissipation in the chromosphere. 
Observations suggest somewhat higher mass-loss rates for active solar-type and lower-mass stars; $\sim 10^{-12}M_{\odot}$ yr$^{-1}$ from the observations of astrospheres \citep{Wood2005ApJ,Wood2021ApJ} and $\sim 10^{-11}M_{\odot}$ yr$^{-1}$ from the observations of slingshot prominences \citep{Jardine2019MNRAS,Waugh2021MNRAS}.
Although there are some differences in the stellar parameters between the MS solar-type stars and T Tauri stars, these theoretical and observational values would be reference values we have to remember. Considering the range of the accretion rate of classical T Tauri stars ($10^{-9}$-$10^{-7}~M_\odot~{\rm yr^{-1}}$), the above reference values are much smaller than required in the stellar wind paradigm.
Accretion may be able to power such a massive stellar wind from the viewpoint of energetics \citep[e.g.][]{Matt2005bApJ,Cranmer2008ApJ}, but there is no established theory that explains such a very large mass loss rate.

Although predicting the stellar spin evolution is still challenging, our simulations demonstrate the importance of turbulent magnetospheric and MRI-driven winds on the stellar spin evolution. The magnetospheric winds can significantly reduce the angular momentum injection to the star. This finding therefore suggests that the required mass loss in the stellar wind paradigm will be smaller than previously expected. 
On the other hand, the MRI-driven winds can have negative effects on the stellar spin-down, because the MRI-driven winds collimate the stellar winds and reduce the Alfv\'en radius (see also Appendix~\ref{appendix:stellar-wind}). 
The interplay of these effects will be important. To improve our understanding, a detailed modeling of the thermal structure is required, as the property of these turbulent winds can depend on the details of the thermal structure. In addition, more realistic modeling of the stellar winds is also demanded.

\subsection{Possible destabilizing mechanism for the case of the rapidly rotating star}\label{subsec:intability_modelA}
Although the destabilizing mechanism for Model A remains unresolved,
the instability seems to be relevant to the magneto-gradient driven instability proposed by \citet{Hirabayashi2016ApJ}. This instability is driven by the expansive nature of the magnetic pressure gradient force arising from a nonuniform toroidal field. The modes with the wavelength larger than $2\pi L_B$ can become unstable, where $L_B$ is the thickness of the toroidal field in the radial direction. The growth rate is comparable to the local Keplerian spin frequency when the plasma $\beta=1$. In fact, the top panel of Figure \ref{fig:Bfield} (the solid blue line) indicates that the thickness of the coherent magnetic flux bundle is a few $R_*$, which will allow the $m=1$ mode or the spiral pattern to grow at $r=r_{\rm m}$. As the plasma $\beta$ is comparable to unity at the boundary, we expect the rapid growth on the orbital timescale.

The velocity shear around the magnetospheric boundary naturally amplifies the toroidal field from the magnetospheric field. Therefore, the magnetospheric boundary will be a suitable location for this instability.
As similar situations are also realized in Model B and C (the middle and bottom panels of Figure \ref{fig:Bfield}), it is possible that both the interchange instability and the magneto-gradient driven instability operate in the two models. However, the interchange modes will dominate in the two cases because the growth rate of the shorter-wavelength modes is larger.

\subsection{Implications for dipper phenomena}\label{subsec:dipper}
It is likely that partial occultation of the stellar surface found in our simulations is relevant to the dipper phenomena. 
The column density distribution is highly time variable because the accreting and ejected flows are significantly inhomogeneous, which may explain the behavior of aperiodic dippers.
In addition, the failed MRI-driven disk winds could be involved with dipper phenomena caused by dusty materials \citep[e.g.][]{Bodman2017MNRAS} because the failed disk winds can bring the dusty materials close to the star from the outer region where the temperature is smaller than the dust sublimation temperature ($\sim 1,000-2,000$ K).

We note that some other processes need to be considered as well.
A warped inner disk formed by a rotating star with a dipole magnetic field misaligned from the stellar rotation axis can results in the dipper phenomena. \citet{Romanova2013MNRAS} showed that the density waves excited by the inclined, rotating magnetosphere perturb the disk surfaces \citep[also see observations of, e.g.,][]{McGinnis2015A&A}. The dust grains may come with the funnel accretion flows from the truncation radius \citep{Nagel2020A&A}, but detailed modeling of the temperature around the inner disk is necessary. It is possible that the inner disk gas temperature is well above the sublimation temperature because of viscous heating. Indeed, the funnel accretion is often identified in emission lines from warm gas such as H$\alpha$ and Br$\gamma$ \citep[e.g.][]{Hartmann2016ARA&A}.
In our simulations, multiple accretion columns consist of both warm and cool materials. The former comes from the truncation radius, while the latter originates from the outer disk. We will investigate the observable properties in more detail in future studies.

\subsection{Heating of the accreting gas\\
around the magnetospheric boundary}\label{subsec:heating}
The hydrogen atomic line observations suggest the formation of the warm accreting gas with the temperature of approximately $10^4$~K \citep[regarding Br$\gamma$ observations, see, e.g.][]{Gravity_Collaboration2017A&A, Eisner2009ApJ,Gravity_Collaboration2020Natur}. As the disk temperature is expected to be a few 1,000~K, the accreting gas should experience some heating \citep[e.g.][]{Hartmann1994ApJ,Muzerolle2001ApJ}.
However, the gas heating mechanism remains unresolved \citep[for a brief summary, see][]{Hartmann2016ARA&A}.

Although our models do not solve the thermal structures in detail, we find indications of magnetic heating around the base of the magnetospheric accretion flows. Figures~\ref{fig:4phys_modelA}, \ref{fig:4phys_modelB}, and \ref{fig:4phys_modelC} demonstrate that the temperature around the base is locally enhanced (see also Figure~\ref{fig:schematic_diagram}). This hot innermost disk may be regarded as a hot rotating ring. As the plasma $\beta$ around the base is close to or smaller than unity (see the plasma $\beta$ images in those figures), the dissipation of magnetic energy can lead to significant gas heating. We have seen that magnetospheric fields are tangled by instabilities at the magnetospheric boundary (Figure \ref{fig:interchange}). The magnetosphere-disk interaction produces the fluctuating magnetic fields around the base of the magnetospheric accretion flows (Figure~\ref{fig:Bfield}). Therefore, magnetic heating such as small-scale reconnection of the fluctuating magnetic fields can be important.
The presence of the hot innermost disk may be supported by spectroastrometry of Br$\gamma$ emission toward TW Hya \citep{Goto2012ApJ}, although more detailed considerations are required to interpret the observed line width.

\subsection{Future prospects}\label{subsec:future}
This study focused on the star with a dipole magnetic field aligned with the star's rotation axis. Magnetospheres in our models show asymmetric structures about the equatorial plane at least during some periods. Such asymmetric accretion is commonly seen in 2D axisymmetric models \citep{Lii2014MNRAS,Romanova2018NewA}, probably because one-sided accretion can prevent accumulation of mass at the magnetospheric boundary and efficiently release the gravitational energy of accreting gas. 
However, there are T Tauri stars with an inclined magnetosphere to the rotation axis \citep[e.g.][]{Bouvier2007A&A,McGinnis2015A&A}. \citet{Romanova2003ApJ} have performed 3D simulations of accretion to an inclined dipole magnetosphere and showed the funnel accretion at both hemispheres and the formation of bipolar winds \citep[see also][]{Romanova2009MNRAS}.
We will also investigate the inclined field cases in future papers to find the relation between the stellar magnetic field structure and the wind driving.

We are updating our model to improve the treatment of the thermal structure of the disk. The temperature structure controls the disk thickness and the ionization degree, affecting the optical depth for the high energy radiation. We will implement the radiation transfer and the chemical reactions for more realistic modeling.

Our models suggest that the turbulent magnetospheric winds are always present even when stable conical winds are absent. As the failed winds (turbulent MRI-driven and magnetospheric winds) play roles in the circulation of the mass and the removal of the angular momentum in the innermost disk (Figure \ref{fig:angular_momentum_flux}), observations that investigate the innermost structure smaller than 0.1 au are highly important. The HI Br$\gamma$ line will be particularly useful for this purpose. Indeed, near-infrared interferometric observations using GRAVITY at the Very Large Telescope Interferometer (VLTI) succeeded in probing the wind in the innermost region for some pre-MS stars \citep[e.g.][]{Gravity_Collaboration2017A&A}. Increasing the number of samples will enable us to relate the wind property to physical quantities of accretion.

\section{Summary}\label{sec:summary}
We list the key findings from three simulations with different stellar spins.
\begin{itemize}
    \item The accretion flows onto the star consist of two components; the gas infalling from the magnetospheric boundary and the failed disk winds. The failed disk winds are turbulent winds that fail to escape from the stellar gravity. 
    Both flows are fluctuating, which results in the formation of multi-accretion columns  (Section~\ref{subsec:accretion_structure}). 
    \item Our models show various outflows (Section~\ref{subsec:accretion_structure}). Turbulent failed winds emanate from both the disk and the magnetosphere, which are absent in 2D models. Even when the escaping conical wind is absent, the failed disk winds are always present around the magnetosphere. They are important in the mass circulation and the angular momentum transfer in the innermost region (Section~\ref{subsec:ang_flux}).
    \item The magnetospheric ejections resulting from magnetic reconnection occur in 3D, but they are not as powerful as expected from 2D models, which is consistent with the X-ray observations (Section~\ref{subsec:comparison_2D}). 
    \item The accretion torque exerting on the star is significantly smaller than the simple estimation based on the accretion rate (Section~\ref{subsec:ang}). A large amount of the angular momentum is extracted by conical disk winds (if present) and turbulent magnetospheric winds (Section~\ref{subsec:ang_flux}). 
    \item Previous theories expect that the ratio of the toroidal to the poloidal field strengths at the magnetospheric boundary depends on the stellar spin. However, our simulations show that the ratio $\approx \mathcal{O}(1)$ is insensitive to the spin (Sections \ref{subsec:Bfield_magnetosphere} and \ref{subsec:magnetic_twist}). 
    \item We compare the rotation profile around the magnetosphere with the analytical prediction by \citet{Kluzniak2007ApJ} and find a significant difference. We point out the breakdown of some assumptions in the theory (Section \ref{subsec:rotation}).
    \item Considering the 3D effects found in this study, we demonstrate that the relation very similar to the Ghosh \& Lamb relation is obtained from the steady angular momentum transport equation (Section~\ref{subsec:summary_magnetospheric_boundary}). The theoretical relation is also consistent with our numerical results. 
    Although the number of the models are quite limited, this study suggests that the relation will be applicable to a wide range of the fastness parameter, unlike the previous expectation.
    \item The magnetospheric boundary is unstable not only in the slow rotator cases but also in the case with a large fastness parameter ($\approx 2.2$ in Model A) (Section~\ref{subsec:magnetosphere-boundary}). The destabilization in the case with such a large fastness parameter was not found in previous 3D simulations of \citet{Blinova2016MNRAS}. 
    \item In our models, the stellar spin changes the types of the instabilities at the magnetospheric boundary (Section \ref{subsec:magnetosphere-boundary}). The mechanisms that destabilize the magnetospheric boundary are found to affect the level of the time variability of the conical wind. In addition, the angular momentum extraction from the magnetosphere depends on the property of the instabilities because the toroidal field is amplified differently (Section \ref{subsec:ang_flux}).
    \item We investigated the X-ray and UV shielding. As expected, our models suggest that the EUV shielding regions extend broader than the X-ray shielding regions (Section~\ref{subsec:shielding}). Therefore, the main wavelength range that drives the photoevaporation will change with radius.
    \item The magnetosphere-disk interaction produces the fluctuating magnetic fields (Section \ref{subsec:Bfield_magnetosphere}). We found that magnetic heating such as small scale magnetic reconnection of the fluctuating magnetic fields is important around the base of the magnetospheric accretion flows. Such magnetic heating should be responsible for determining the temperature of the accretion flows (see Section~\ref{subsec:heating}).
\end{itemize}

%\begin{acknowledgments}
\vspace{10pt}
We thank Riouhei Nakatani and Yuhiko Aoyama for fruitful comments. S.T. was supported by the JSPS KAKENHI grant Nos. JP18K13579, JP21H04487, and JP22K14074. K.T was supported by the JSPS KAKENHI grant Nos. JP16H05998 and JP21H04487.
K.I. was supported by the JSPS KAKENHI grant No. JP21H00056.
T.K.S. was supported by the JSPS KAKENHI grant Nos. JP17H01105, JP21H00033, and JP22H01263.
Numerical computations were carried out on Cray XC50 at Center for Computational Astrophysics, National Astronomical Observatory of Japan.
Test calculations in this work were in part carried out at the 
Yukawa Institute Computer Facility.
This work was supported by MEXT as a Program for Promoting Researches on the Supercomputer Fugaku by the RIKEN Center for Computational Science (Toward a unified view of the universe: from large-scale structures to planets, grant no. 20351188(PI J. Makino)).
%\end{acknowledgments}

%\software{aaa}




\appendix
\section{Estimation of the impacts of our stellar wind}\label{appendix:stellar-wind}
Our stellar wind model is based on some simplifications. We briefly discuss the robust structures and the model limitations.

\begin{figure*}
    \centering
    \includegraphics[width=\columnwidth]{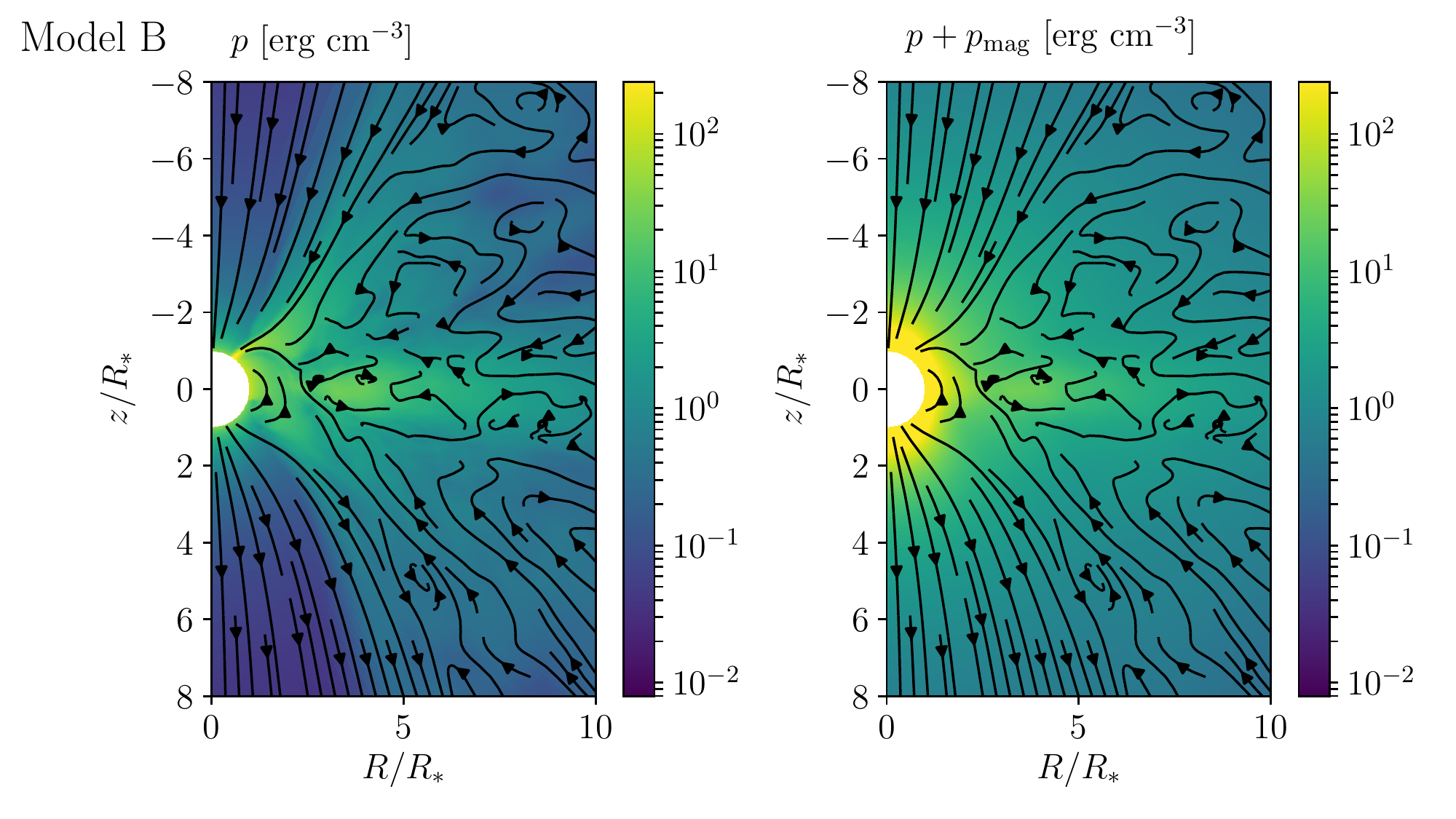}
    \caption{Gas pressure (left) and the sum of the gas and magnetic pressures (right) of Model B at $t=194.7$ day. Lines with arrows display the magnetic structures. The data are averaged in the azimuthal direction. }
    \label{fig:stellar_wind_pressure}
\end{figure*}

We show that the turbulent MRI-driven winds play a role in confining the stellar winds in the polar regions. The left panel of Figure~\ref{fig:stellar_wind_pressure} displays the gas pressure distribution of Model B. The gas pressure in the stellar winds is approximately an order of magnitude smaller than that in the turbulent MRI-driven winds (also see Figure \ref{fig:4phys_modelB} for more details about the structure of the MRI-driven winds). The right panel of the figure shows the sum of the gas and magnetic pressures. 
The sum show no discontinuities in the image, which indicates that the gas pressure of the MRI-driven winds confines the stellar wind. The absence of the discontinuities also demonstrates that the ram pressure of the stellar wind has a minor influence on this plasma structure. 
As long as the gas pressure of the turbulent MRI-driven winds is much larger than that of the stellar wind, the overall structure will be the same.
Considering that the ram pressure of the wind only work in its direction, the collimated stellar winds have minor effects on the accretion taking place at lower latitudes.

Even if the stellar winds try to compete the accretion, they do not significantly quench the accretion.
To clarify this point, we compare the gas pressure of our stellar corona ($p_*$) and the ram pressure of the accretion columns ($p_{\rm ram}$).
Using our model parameters, $p_*$ is estimated to be
\begin{align}
p_{*}\approx 8.0\times 10 ~{\rm erg~cm^{-3}}\left( \frac{\rho_*}{5.1\times 10^{-13}~{\rm g~cm^{-3}}}\right)\left( \frac{T_*}{0.87~{\rm MK}}\right).
\end{align}
Accretion columns typically have a density of $10^{-11}~{\rm g~cm^{-3}}$ or larger (see Figure \ref{fig:rho_ekinflux_mollweide}) and the velocity of 100-150~${\rm km~s^{-1}}$. Therefore, the ram pressure of an accretion column with the density $\rho$ and the velocity $v$ is estimated to be
\begin{align}
    p_{\rm ram}\sim 10^3~{\rm erg~cm^{-3}}\left(\frac{\rho}{10^{-11}~{\rm g~cm^{-3}}}\right)\left(\frac{v}{150~{\rm km~s^{-1}}}\right)^2
\end{align}
Therefore, the coronal gas pressure cannot stop the accretion flows with such parameters. The accretion flows with a much smaller density can be quenched, but such flows should have minor impacts on the accretion rate.

Our coronal temperature $T_*$ would be comparable to that of the classical T Tauri stars, but our coronal density $\rho_*$ may be much larger than the realistic value. Adopting a much smaller value is difficult because of the numerical limitation. The numerical time step of our simulations is determined by the Alfv\'en speed in the polar regions, which means that the computational time increases if we adopt a smaller coronal density. On the other hand, using a much larger coronal density is undesirable because the coronal plasma in the magnetosphere can be high-$\beta$ plasma. To study the dynamics of the typical classical T Tauri stars, the coronal region in the magnetosphere should be low-$\beta$ and confined by the stellar magnetic field unless the magnetic structure is largely modified by the accretion (the low-$\beta$ condition is mostly satisfied for our magnetospheric plasma. See Figures \ref{fig:4phys_modelA}-\ref{fig:4phys_modelC}). The coronal density of this study is chosen by considering these points.
As discussed above, our stellar winds are expected to have minor impacts on the accretion. This will be particularly true around the midplane. Therefore, our stellar winds should have a weak influence on the magnetosphere-disk interaction. 
However, the mass loss rate of our stellar wind may be much larger than the realistic value.

\bibliography{main_arxiv_ver1}{}
\bibliographystyle{aasjournal}

\listofchanges
\end{document}